\documentclass[twocolumn]{aastex701}

\usepackage{amsmath,amsmath,amssymb}
\usepackage{mathrsfs}
\usepackage{subcaption}
\usepackage{xcolor}
\usepackage[normalem]{ulem}

\begin{document}

\title{Constraining the Binary Fraction and Mass-Ratio Distribution of FGK Dwarfs with Gaia XP Spectra: Metallicity, $\alpha$-Element, and Age Dependencies}

\correspondingauthor{Zexi Niu}
\email{niuzexi@ucas.ac.cn}

\author[0009-0005-7351-1092]{Yinuo Zhang}
\affiliation{National Astronomical Observatories, Chinese Academy of Sciences, Beijing 100101, People's Republic of China}
\affiliation{School of Astronomy and Space Science, University of Chinese Academy of Sciences, Beijing 100049, People's Republic of China}
\email{zhangyinuo@bao.ac.cn}

\author[0000-0002-3651-0681]{Zexi Niu}
\affiliation{School of Astronomy and Space Science, University of Chinese Academy of Sciences, Beijing 100049, People's Republic of China}
\affiliation{National Astronomical Observatories, Chinese Academy of Sciences, Beijing 100101, People's Republic of China}
\email{niuzexi@ucas.ac.cn}

\author[0000-0003-2471-2363]{Haibo Yuan}
\affiliation{Department of Astronomy, Beijing Normal University, 19th Xinjiekouwai Street, Haidian District, Beijing, People's Republic of China}
\email{yuanhb@bnu.edu.cn}

\author[0000-0002-4731-9698]{Ning-Chen Sun}
\affiliation{School of Astronomy and Space Science, University of Chinese Academy of Sciences, Beijing 100049, People's Republic of China}
\affiliation{National Astronomical Observatories, Chinese Academy of Sciences, Beijing 100101, People's Republic of China}
\affiliation{Institute for Frontiers in Astronomy and Astrophysics, Beijing Normal University, Beijing 102206, People's Republic of China}
\email{sunnc@ucas.ac.cn}

\author[0009-0009-3372-5271]{Jia-Hui Wang}
\affiliation{National Astronomical Observatories, Chinese Academy of Sciences, Beijing 100101, People's Republic of China}
\affiliation{School of Astronomy and Space Science, University of Chinese Academy of Sciences, Beijing 100049, People's Republic of China}
\email{wangjh@bao.ac.cn}

\author[0000-0002-7559-315X]{César Rojas-Bravo}
\affiliation{School of Astronomy and Space Science, University of Chinese Academy of Sciences, Beijing 100049, People's Republic of China}
\affiliation{National Astronomical Observatories, Chinese Academy of Sciences, Beijing 100101, People's Republic of China}
\email{cesar.rojasbravo@ucas.ac.cn}

\author{Jifeng Liu}
\affiliation{National Astronomical Observatories, Chinese Academy of Sciences, Beijing 100101, People's Republic of China}
\affiliation{School of Astronomy and Space Science, University of Chinese Academy of Sciences, Beijing 100049, People's Republic of China}
\affiliation{Institute for Frontiers in Astronomy and Astrophysics, Beijing Normal University, Beijing 102206, People's Republic of China}
\affiliation{New Cornerstone Science Laboratory, National Astronomical Observatories, Chinese Academy of Sciences, Beijing 100012, People's Republic of China}
\email{jfliu@nao.cas.cn}

\begin{abstract}

We present a study of the binary fraction and mass-ratio distribution for near 200,000 FGK dwarf stars within 1\,kpc, utilizing the Stellar Locus Outlier (SLOT) method applied to synthetic Gaia XP photometry and stellar parameters from LAMOST DR9. By modeling the asymmetric color residuals relative to a metallicity-dependent single-star locus, we measure the binary fraction $f_b$ and constrain the power-law slope $\gamma_{\rm large\,q}$ of the mass-ratio distribution at high mass ratio $q$. Higher metallicity is associated with a steeper (more negative) $\gamma_{{\rm large}q}$, indicating a larger proportion of low-mass-ratio systems. For $\alpha$-elements, [Mg/H] and [Si/H] are anticorrelated with $f_b$ in thin-disk-dominated subsamples; conversely, [Ca/H] shows a positive correlation with $f_b$ in thin-disk-dominated subsamples. No correlation is found between $f_b$ and $\alpha$-element abundances in thick-disk-dominated subsamples.
The binary fraction increases modestly toward older ages in the thin-disk population, a trend that likely reflects underlying correlations with chemistry. Our results shed light on how chemical composition and dynamical processes shape the binary properties of solar-type stars.

\end{abstract}

\section{Introduction} \label{sec:introduction}

The vast majority of stars reside in binary or higher-order multiple systems \citep[e.g.,][]{2005Goodwin,Duchene2013multipl}. A comprehensive understanding of binary systems is fundamental to many fields of astrophysics. Binary interactions alter the evolutionary pathways of component stars and give rise to a rich diversity of exotic outcomes (\citealt{HanZ2020binarypopsyn,ChenX2024newMil,Sana2012BInteract} and references therein; quantitative results for O stars cannot be directly transferred to the FGK population considered here), including Type~Ia supernovae \citep[e.g.,][]{2023Liu}, core-collapse supernovae \citep[e.g.,][]{2026Souropanis,2026Niu}, kilonovae \citep[e.g.,][]{Metzger2020KNe}, short-duration gamma-ray bursts \citep[e.g.,][]{2014Berger}, X-ray binaries \citep[e.g.,][]{Tetarenko2016BHXB}, cataclysmic variables \citep[e.g.,][]{Warner1995cataclysmic} and many others. They also play an essential role in galactic chemical evolution \citep[e.g.,][]{Arcones2023eleorigin} and gravitational waves sources \citep[e.g.,][]{Abbott2023GWTC3}.

The characterization of binary systems relies on a set of fundamental statistical properties: the binary fraction $f_b$, the distributions of orbital period $P$, mass ratio $q$, and orbital eccentricity $e$ \citep{Duchene2013multipl}. Among these, the binary fraction is a key diagnostic for star formation and evolution, providing critical constraints for stellar population synthesis models \citep{Moe2017PandQ, HanZ2020binarypopsyn, Offner2022OriginEvoMultiplestar, ChenX2024newMil}. The mass-ratio distribution, on the other hand, probes the pairing mechanisms and subsequent dynamical disruptions as well as binary interactions \citep{1995Kroupa,Tokovinin2000origin, Sana2012BInteract, ElBadry2019twinbinary}. Correlations between $q$ and stellar type, chemical composition, or other binary parameters thus provide valuable clues to the evolutionary history of binary populations.

Over the past decades, binary surveys have expanded dramatically thanks to a variety of detection techniques. Classical visual methods \citep{Eggleton1989vbs, Hartkopf1996orbits} have been complemented by spectroscopic monitoring of radial velocity (RV) variations \citep{Badenes2018APOGEEview, TianZ2018lmdr4, Mazzola2020closebinary} and by the identification of secondary component through spectral fitting \citep{RM2007PCEB, ElBadry2018Signatures, Roulston2020classify, Xiang2021DDSp, Lijd2025MSbinary}. Photometric surveys have exploited eclipsing binaries (EBs) \citep{Moe2013MWandMC, Pawlak2016OGLEMCs, Soszynski2016OGLEbulge, Drake2017catalina, Heinze2018ATLAS, Jayasinghe2019ASASSN, ChenX2020ZTF}, and common proper motion \citep{ElBadry2019wb, Hartman2020SUPERWIDE}, long-baseline interferometry \citep{Sana2014southernmassive, OT2024DYNAMO}, and deviations on color–magnitude or color–color diagrams \citep{LiuC2019SmokingGun, Yuan2015slot, Niuzx2021FGK} have also proven effective. These complementary approaches have enabled large-scale statistical studies across different stellar populations.

A consistent picture has emerged from these efforts: the binary fraction increases with primary mass and is anticorrelated with metallicity among solar-type (FGK) stars \citep{Raghavan2010survey, Yuan2015slot, Badenes2018APOGEEview, Moe2019STclosebinary, Mazzola2020closebinary, Niuzx2021FGK}. The mass-ratio distribution is often modeled either by a broken power law with an excess of near-equal-mass (``twin") systems \citep{Moe2017PandQ, ElBadry2019twinbinary} or by a single power law. There is growing evidence that $q$ depends on primary mass, orbital period, and separation, and \citet{LiuC2019SmokingGun} reported a possible non-monotonic relation between $q$ and [Fe/H].

Despite this progress, several important questions remain. The connection between binary fraction and $\alpha$-element abundances is relatively unexplored, while the existing studies do not reach a consensus. \citet{Mazzola2020closebinary} found a significant negative correlation between close binary fraction and $\alpha$ abundance, while \citet{Niuzx2021FGK} reported a positive trend with their averaged [$\alpha$/Fe] that disappears for thick-disk stars. The relationship between binary fraction and stellar age is also poorly investigated, as obtaining reliable ages for large, homogeneous binary samples has long been challenging. Furthermore, previous mass-ratio analyses have relied mainly on spectroscopy or cluster members, limiting both sample size and population coverage; whether $q$ correlates with metallicity needs more follow-up confirmations.

We address these issues by building on the Stellar Locus Outlier (SLOT) method \citep{Yuan2015slot, Niuzx2021FGK}. The SLOT technique infers the binary fraction from the asymmetric color residuals induced by unresolved multiples relative to a single-star locus, and is particularly well-suited for large photometric surveys because it is relatively insensitive to the distributions of orbital period, eccentricity, and viewing angle, and requires no multi-epoch observations. In this work, we combine stellar parameters derived from LAMOST low-resolution spectroscopy \citep{Cuix2012LAMOST,Zhaog2012LAMOSToverview,Dengl2012LAMOSTsp,Luoa2015firstlm} with the state-of-the-art photometric precision of Gaia XP synthetic photometry \citep{GaiaCollab2023Summary} to conduct a systematic study of binary fractions and mass-ratio distributions among FGK field dwarfs. Compared to previous implementations, our analysis covers a larger sample volume, achieves higher photometric precision, and simultaneously explores dependencies on metallicity, individual $\alpha$ elements, and age.

This paper is organized as follows. In Section\,\ref{sec:data}, we describe the datasets used in this work, with the procedure of data selection. In Section\,\ref{sec:method}, we describe the method in details. In Section\,\ref{sec:results}, we present the results for various subsamples. In Section\,\ref{sec:discussion}, we compare our results with previous works, discuss the implication and possible caveats. In Section\,\ref{sec:summary}, we conclude by summarizing the main findings and discussing future work.

\section{Data} \label{sec:data}

\subsection{Gaia DR3} \label{sec:gaiadr3}

Gaia is a global space astrometry mission operated by European Space Agency (ESA) \citep{GaiaCollab2016mission}. The third data release 3 (DR3) was released in June 2022 \citep{GaiaCollab2023Summary}.  In addition to astrometric solutions (celestial positions, parallaxes, and proper motions) for $1.46$ billion sources, Gaia DR3 also provides broad $G$-band photometry for over $1.8$ billion sources, with $1.54$ billion of them having  broad-band $BP$ magnitude and $1.55$ billion have  broad-band $RP$ magnitude, reaching a limiting magnitude of $G\,\approx 21$\,mag.  A revolutionary feature of DR3 is the low-resolution ($R\,\approx\,50$) spectra from the Blue and Red Photometers ($BP$/$RP$), collectively referred to as XP spectra\footnote{\url{https://gaia.aip.de/}}. These spectra cover a wavelength range of $330-1020\,{\rm nm}$ and are released for approximately 220 million sources. A detailed description of the XP spectra and their practical usage can be found in \citet{GaiaCollab2023Summary}. 

We utilize the photometric calibration for Gaia XP spectra provided by \citet{Huangbw2024xp}\footnote{\url{https://github.com/HiromonGON/GaiaXPcorrection/blob/main/GaiaDR3XPspectracorrectionV1/readme}}, which corrects the systematic errors in the Gaia XP spectra with an internal precision of 1\%--2\%. The applied correction depends on $BP-RP$, $G$, and reddening $E(B-V)$, and is validated within $-0.5\,<\,BP-RP\,<\,2\,{\rm mag}$, $3\,<\,G\,<\,17.5\,{\rm mag}$, and $E(B-V)\,<\,0.8\,\rm mag$.  
For each star, we synthesize the $BP$, $G$, and $RP$ magnitudes from the calibrated XP spectra. We restrict our sample to sources with the quality flag {\tt\string Caution} being {\tt\string [0 0 0]}, {\tt\string [0 0 1]} or {\tt\string [0 0 2]}. According to \citet{Huangbw2024xp}, these values indicate that the correction is within the normal range and the resulting photometry is reliable.

We adopt the extinction coefficients from \citet{Zhangry2023dered}\footnote{\url{https://github.com/vnohhf/XP_Extinction_Toolkit}}, which presents an empirical median reddening model constructed by combining multi-band photometry from the far-ultraviolet to mid-infrared: GALEX \citep{Martin2005GALEX}, PanSTARRS \citep{Chambers2018Panstarrs}, Gaia EDR3 \citep{GaiaCollab2022edr3}, SDSS \citep{Alam2015sdss}, 2MASS \citep{Skrutskie20062MASS}, and WISE \citep{Wright2010WISE}, with LAMOST DR7 spectroscopy\footnote{\url{https://www.lamost.org/dr7/v2.0/}}. The resulting extinction coefficients depends on the effective temperature $T_{\rm eff}$ and $E(B-V)$ and is mostly valid in the $E(B-V)$ range of 0.0 -- 0.5\,mag and the temperature range of $4000$ -- $10,000\,\rm K$.

In our work, we generate the synthetic $BP/G/RP$-band magnitude by 
convolving the calibrated and dereddened XP spectra with the Gaia filter profiles\footnote{Gaia DR3 passbands: \url{https://www.cosmos.esa.int/web/gaia/edr3-passbands\#}}. All Gaia photometric magnitudes ($BP$, $G$, $RP$) and corresponding errors ($\sigma_{BP}$, $\sigma_{G}$, $\sigma_{RP}$) used in this paper refer to the data synthesized from Gaia XP spectra.

We adopt the distances from \citet{BJ2021distances}, who derives distance posteriors for 1.47 billion stars by combining Gaia EDR3 \citep{GaiaCollab2022edr3} parallaxes with a three-dimensional Galactic model. Throughout this work, we use the median geometric distance estimates from this catalog. 
The distance uncertainties are computed as half the difference between the 84th and 16th percentiles of the posterior distribution provided by \citet{BJ2021distances}. Following their Figure\,18, we retain sources with $d / \sigma_{d}\,>\,20$.

\subsection{LAMOST Low-resolution Spectroscopy} 
\label{sec:lamost_lowres_sp}

The Large Sky Area Multi-Object Fiber Spectroscopic Telescope (LAMOST) is a 4\,m reflecting Schmidt telescope with a 5$^\circ$ field of view, capable of simultaneously acquiring spectra for 4,000 targets \citep{Cuix2012LAMOST,Zhaog2012LAMOSToverview,Dengl2012LAMOSTsp,Luoa2015firstlm}. Its ninth data release (DR9)\footnote{\url{https://www.lamost.org/dr9/v1.0/}} comprises about 10 million low-resolution spectra obtained between 2012 and 2021. 

We employ the value-added catalog of stellar labels recently published by \citet{Zhangm2025elements}, which was constructed using the data-driven Payne (DD-Payne) algorithm applied to LAMOST DR9 low-resolution spectra. This catalog provides atmospheric parameters ($T_{\rm eff}$, $\log g$, and [Fe/H]) for $6.4$ million FGK stars, along with [X/Fe] for 22 elements for a subset of 3.6 million stars. The underlying model was trained on LAMOST spectra with precise labels drawn from APOGEE DR17 or GALAH DR3, yielding typical uncertainties of $\sim 30$\,K in $T_{\rm eff}$, $\sim 0.07$\,dex in $\log g$, and $\sim 0.05$\,dex in abundances for elements lighter than Sr \citep{Zhangm2025elements}. From this catalog, we adopt $T_{\rm eff}$, $\log g$, and [Fe/H], as well as five $\alpha$-process elements: Mg, Si, Ca, C, and O. The typical uncertainty for the five abundance is $0.03,~0.06,~0.03,~0.11,~0.05 \rm\, dex$, respectively. We further apply the temperature-dependent correction for [Fe/H] \citep{Niuzx2023calibration}. Because [X/Fe] estimates for stars with $T_{\rm eff}\,<\,4800$\,K are subject to larger errors, we exclude these cooler sources from our analysis of $\alpha$-process elements abundances and binarity (see Section\,\ref{sec:alpha_abund}). To exclude stars with likely unreliable [X/Fe] measurements, we applied a $3\sigma$ clipping to the spectral fitting quality parameter  (see Table~3 in \citealp{Zhangm2025elements}).

\citet{Wangjh2025Ages} applies a data-driven model based on the XGBoost algorithm to LAMOST low-resolution spectra to estimate stellar ages for dwarf stars. Their model was trained using wide binaries with known ages, and complemented by open cluster members.
The resulting catalog provides spectroscopic ages for approximately 4 million dwarf stars in LAMOST. We adopt their age estimates in this work.

\subsection{Sample Construction} \label{sec:sample_construction}

The sample construction in this work follows the procedure described in \citet{Niuzx2021FGK}.
We cross-match the aforementioned datasets, and we impose quality constraints as summarized below.

For Gaia data:
\begin{enumerate}
    \item {\tt \string phot\_bp\_mean\_flux\_over\_error} $>\,100$.
    \item {\tt \string phot\_g\_mean\_flux\_over\_error} $>\,100$.
    \item {\tt\string phot\_rp\_mean\_flux\_over\_error} $>\,100$.
    \item {\tt \string phot\_proc\_mode\,=\,0}.
    \item {\tt \string duplicated\_source=False}.
    \item $BP/RP$ photometric excess factor\\
            {\tt phot\_bp\_rp\_excess\_factor}: outliers removed.
\end{enumerate}

For LAMOST data:
\begin{enumerate}
    \item The signal-to-noise ratios (S/N) for the g band ($\mathrm{S/N}_g$) are larger than 20.
    \item Metallicity [Fe/H]\,$\geq\,-1.0\,\rm dex$.
\end{enumerate}

For spatial location:
\begin{enumerate}
    \item $E(B-V)\,<\,0.05\rm\,mag$ according to the \citet{Schlegel1998extin} dust reddening map.
    \item Galactic latitude $|b|\,> 20^\circ$.
    \item Vertical distance to the Galactic disk\\$|Z|\,>\,0.2\,\rm kpc$.
    \item Distance $d\,\leq\,1$\,kpc.
\end{enumerate}

The {\tt phot\_bp\_rp\_excess\_factor} is derived from Gaia DR3 catalog, and we exclude outliers following the same criteria as in \citet{Niuzx2021FGK}.
We restrict the sample to $|Z|~>~0.2$~kpc to minimize extinction-related systematic errors in the \citet{Schlegel1998extin} reddening map, since most dust is confined to the Galactic thin dust disk, which has a scale height of $\sim$ 60~pc in the solar neighborhood \citep[see Table~1 of][]{ZhangRY2026taledustdisksmilky}. 
Because unresolved multiple systems are systematically brighter than single stars of the same color, they can be detected to larger distances in a magnitude-limited sample. To mitigate this bias, we restrict our sample to $d\,\leq\,1$\,kpc. We note, however, that this distance cut reduces but does not eliminate the bias, since our sample remains magnitude-limited (mostly through the Gaia flux-over-error and LAMOST S/N cuts), and unresolved binaries are still preferentially retained near the faint magnitude limit. We therefore assess the possible impact caused by distance-dependent incompleteness in Section\,\ref{sec:potential_caveats}.

From the Kiel Diagram ($\log g$ vs. $T_{\mathrm{eff}}$), we select the main-sequence (MS) stars. The main-sequence turn-off (MSTO) stars are sifted out from the [Fe/H]-binned color-absolute magnitude diagrams (CMDs), following the procedure of \citet{Niuzx2021FGK}.

We utilize PARSEC single-star isochrones\footnote{\url{https://stev.oapd.inaf.it/cgi-bin/cmd}} \citep{Bressan2012PARSEC} for FGK dwarfs to derive metallicity-dependent relations between mass and color, where [Fe/H] is binned into the interval of $0.1\,\rm dex$. Rough estimates of stellar masses for our sample are obtained based on the interpolated relations.
The same interpolations are also used in the simulation step described in Section\,\ref{sec:slot_overview}.

Due to the scarcity of red sources with $BP-RP\,>\,1.4\,\rm mag$ and the removal of the binary branch at $BP-RP\,<\,0.8\,\rm mag$ together with the MSTO stars, we will apply the binary estimation to sample in the color range of $0.8\,\leq\,BP-RP\,<\,1.4\,\rm mag$. This contains 181,011 stars. 

\begin{figure*}[ht!]
    \centering
    \includegraphics[width=\textwidth]{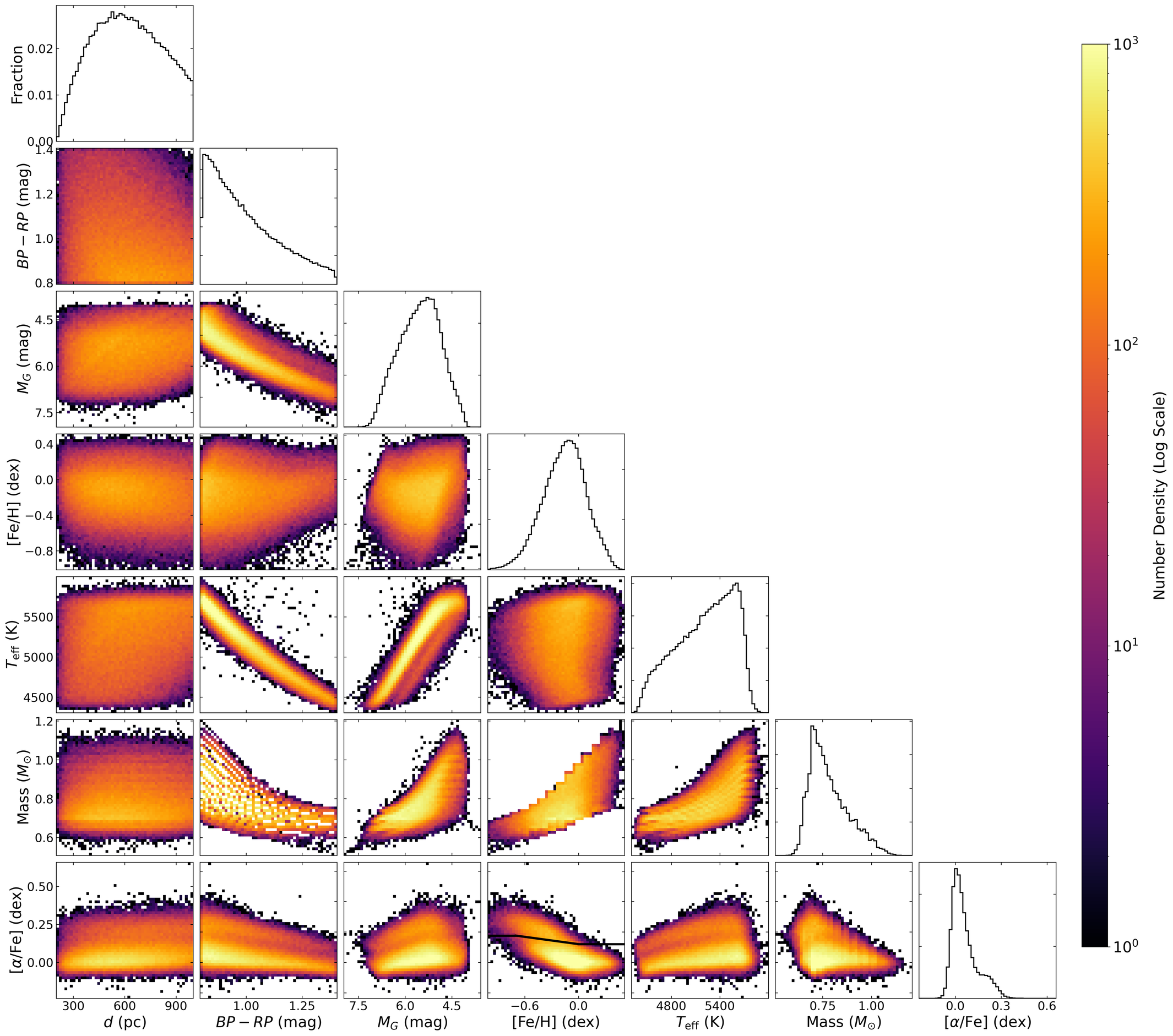}
    \caption{
    Two-dimensional histograms show the joint distributions of stellar parameters for our sample.
    Parameters depicted are: distance, synthetic $BP-RP$ color, absolute $G$ magnitude, [Fe/H], $T_{\mathrm{eff}}$, Mass, [$\alpha$/Fe]. 
    The $\alpha$-process elements abundance [$\alpha$/Fe] is calculated as the weighted arithmetic mean of Mg, Si, Ca, and Ti elements. The separation between thick-disk stars and thin-disk stars is indicated by the black lines in the [Fe/H]-[$\alpha$/Fe] panel.
    }
    \label{fig:stellarparams}
\end{figure*}

The distributions of their stellar parameters of our sample are shown in Figure\,\ref{fig:stellarparams}. By applying an empirical cut plotted as black lines in the [Fe/H]-[$\mathrm{\alpha}$/Fe] plane, thin-disk and thick-disk stars are grouped and labeled in our catalog.

In addition to the sample used for binarity estimation, we construct a training sample for fitting the stellar loci. To make sure that the loci represents the single-star sequence, we roughly eliminate the binary branch on the metallicity-binned CMDs. 
Because the secondary stars in binary systems are systematically redder, we supplement sample with M dwarfs to improve the constraints on the red end of the stellar loci.
The M dwarfs are from the catalog of \citet{Qiud2024Mdwarfs} and processed the same procedure applied to the FGK dwarfs described above, with slight modifications: the S/N cut for the LAMOST spectra is based on the $i$ band instead of the $g$ band, as the $i$-band flux is more representative for M dwarfs. 
The full training sample comprises 164,470 FGK dwarfs and 9,908 M dwarfs.

\section{Method} \label{sec:method}

\subsection{SLOT Overview} \label{sec:slot_overview}

The SLOT method is an approach to identify the binary fraction for unresolved photometrically detectable binaries in large photometric samples \citep{Yuan2015slot}.
For photometrically unresolved binaries, the combined colors are redder than that of single primaries, causing the systems to deviate from the single-star locus in color-color space. Consequently, a sample containing both single and binary systems exhibits an asymmetric distribution of color residuals. The binary fraction can be derived by modeling this asymmetric tail. The SLOT method does not require assumptions about orbital orientation or multi-epoch variability, and it is relatively insensitive to the orbital-period and eccentricity distributions. Moreover, it requires neither multi-epoch nor time-consuming spectroscopic observations, and is therefore highly compatible with large-scale survey data covering wide ranges of stellar parameters.

In \citet{Yuan2015slot}, the SLOT method was applied to a sample of 14,650 stars from the SDSS Stripe 82 region, using SDSS photometric data with the spectroscopic information from the SDSS and LAMOST surveys. \citet{Niuzx2021FGK} subsequently adapted the method for use with Gaia photometric survey data. Using Gaia ERD3 broad-band photometry, they investigated the dependence of the binary fraction on [Fe/H] and $\alpha$-process elements abundance for 86,141 GK dwarfs.
Following them, we take advantage of the precision of Gaia synthetic photometry to examine the binary fraction of an expanded stellar sample.

\begin{figure*}[ht!]
    \centering
    \includegraphics[width=\textwidth]{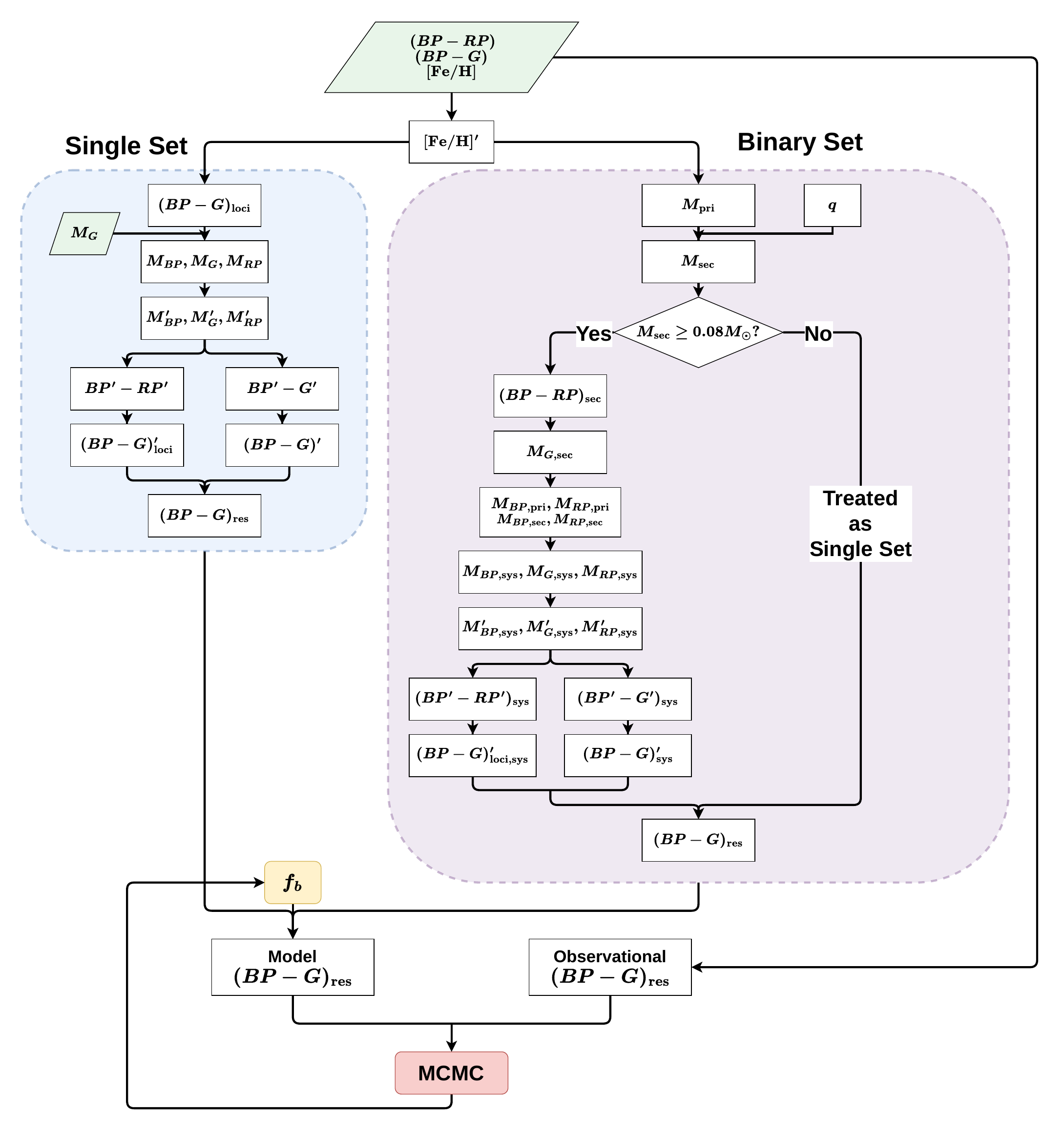}
    \caption{
    Schematic illustration of the SLOT method employed in this work.
    }
    \label{fig:sim_flowchart}
\end{figure*}

Figure\,\ref{fig:sim_flowchart} illustrates the workflow. Generally, we follow that of \citet{Niuzx2021FGK}, with the key difference that we empirically determine the mass-ratio distribution rather than adopting a fixed form.

In brief, we construct two mock datasets: one consisting purely of single stars (``Single Set'') and one consisting purely of binary systems (``Binary Set''). For each object, random errors ($\sigma_{BP}$, $\sigma_{G}$, $\sigma_{RP}$ and $\sigma_{\mathrm{[Fe/H]}}$) are drawn from normal distributions; quantities with added errors are denoted with a superscript $'$.
For the mock Single Set, we apply the single-star, metallicity-dependent stellar loci derived in Section\,\ref{sec:bpg_loci}. 
When generating the mock Binary Set, we adopt the form of the mass-ratio distribution employed in \citet{Moe2017PandQ} and \citet{ElBadry2019twinbinary}. The power-law slope of this distribution is treated as a free parameter and is determined empirically by minimizing $\chi^2$, as detailed in Section\,\ref{sec:q_selection}.
As noted in \citet{Niuzx2021FGK}, the systematic zero-point error in color, $\sigma_{\rm ZP,sys}$, is expected to be significantly smaller than involving $\sigma_{\rm ZP}$ of two bands independently, and it exhibits a dependence on $G$-band magnitude. We follow the same empirical procedure described by \citet{Niuzx2021FGK} to estimate $\sigma_{\rm ZP,sys}$ using red giant branch (RGB) stars. A detailed description of this estimation is provided in Appendix\,\ref{sec:AP_bpg_syserr}.

To determine the binary fraction $f_b$, we fit the observed distribution of color residual by adjusting the relative weights of the mock Single Set and Binary Set. The uncertainty in the observed residual distribution, $\sigma_{{\rm obs}}$, is quantified via bootstrap resampling. We perform the fit using a Markov Chain Monte Carlo (MCMC) approach implemented in the \texttt{emcee} package \citep{FM2013emcee}, optimizing the likelihood defined in Equation\,\eqref{eq:likelihood}. 
Although the binary branch has been largely removed from the training sample used to construct the stellar loci, as described in Section\,\ref{sec:stellar_loci}, contamination from low-mass-ratio binaries still exists and introduces systematic offsets in the fitted stellar loci (seen from Figur\,\ref{fig:bpgfigres_1d}). The MCMC fit therefore involves two free parameters, the $f_b$ and the systematic offset $\mu$.
The prior on $f_b$ is uniform over the interval $(0, 1]$, while $\mu$ is restricted to the range $(-3, 0]$\,mmag. 

\begin{equation}
    \label{eq:likelihood}
    \ln {\mathcal{L}(\theta)} = -\frac12 \sum_i\left({\frac{(N_{\mathrm{i,model}}(\theta)-N_{\mathrm{i,obs}})^2}{\sigma_{\mathrm{i,obs}}^2}} + \ln{(2\pi{\sigma_{\mathrm{i,obs}}^2})}\right)
\end{equation}

We firstly divide the sample in [Fe/H] space, treating the thin-disk and thick-disk populations separately. This is for investigating the variation of the binary fraction with metallicity, and more importantly, to constrain the form of the mass-ratio distribution for different stellar subsamples across [Fe/H].

Having empirically established the appropriate mass-ratio distribution for each bin, we subsequently focus on two subsamples with distinct metallicities: near solar metallicity ([Fe/H] = $0.00 \pm 0.05\,$dex) and moderately metal-poor ([Fe/H] = $-0.5 \pm 0.05\,$dex). For these subsamples, we examine the relationship between $\alpha$-process abundance (Mg, Si, Ca, C, and O) and the binary fraction. By isolating narrow ranges of [Fe/H], the influence of $\alpha$-process elements abundance on binarity from that of overall metallicity is disentangled.

\subsection{Stellar Loci} \label{sec:stellar_loci}

We use the training sample described in Section\,\ref{sec:sample_construction} to derive the metallicity-dependent stellar loci. The color $BP-RP$ and the metallicity [Fe/H] are adopted as the independent variables, and we fit the color $BP-G$ as well as the absolute $G$-band magnitudes $M_G$ as functions of these two parameters. The fitting procedure is performed iteratively on a uniform two-dimensional grid in ($BP-RP$, [Fe/H]), and outliers are rejected at each iteration via $3\sigma$ clipping.

It is noteworthy that not all the binary systems detected by the SLOT method are expected to interact. An estimation of the proportion of potential interacting systems is presented in Section\,\ref{sec:potential_caveats}.

\subsubsection{Metallicity-dependent color-color loci} \label{sec:bpg_loci}

We follow the approach in \citet{Huangbw2025loci} to fit the stellar color-color loci. The form of fitting function is given in Equation\,\eqref{eq:bpg}, where $x$ and $y$ denote $BP-RP$ and [Fe/H], respectively, and $k~=~-5.25$. The fitted coefficients for the terms $f_1$ and $f_3$ are presented in Table\,\ref{tab:bpg_coeff}. As noted by \citet{Huangbw2025loci}, our dwarf sample spans a broad range in color, which makes achieving mmag-level precision with low-order polynomials challenging. We therefore obtain the color-dependent term $f_2$ via interpolation rather than a global polynomial fit; its numerical values are provided in Table~\ref{tab:bpgf2} in Appendix\,\ref{sec:AP_stellarlocifitting}.

\begin{figure*}[ht!]
    \centering
    \includegraphics[width=\textwidth]{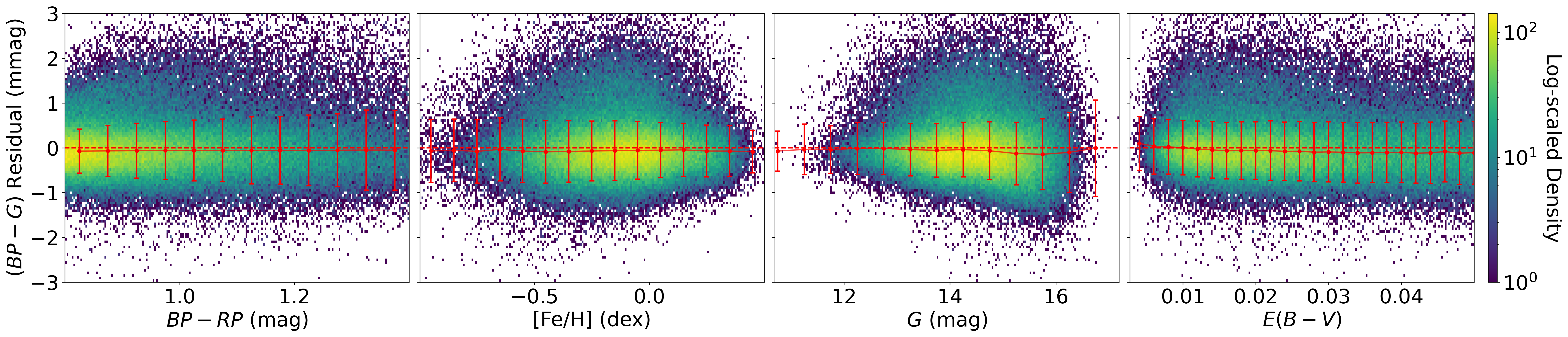}
    \caption{
     Fitting residuals of the color-color loci for FGK dwarfs in the training sample as functions of $BP-RP$, [Fe/H], $G$, and $E(B-V)$. 
     Red points and error bars mark the medians and standard deviations.
     The residuals show no systematic dependence on these stellar parameters and are distributed approximately centering at $0$\,mmag. The extended tails at the red side correspond to unresolved binaries in the training sample.
     The errorbars represent the precision of photometry, typically better than $1$ mmag.
    }
    \label{fig:bpgfigres_1d}
\end{figure*}

\begin{figure*}[ht!]
    \centering
    \includegraphics[width=\textwidth]{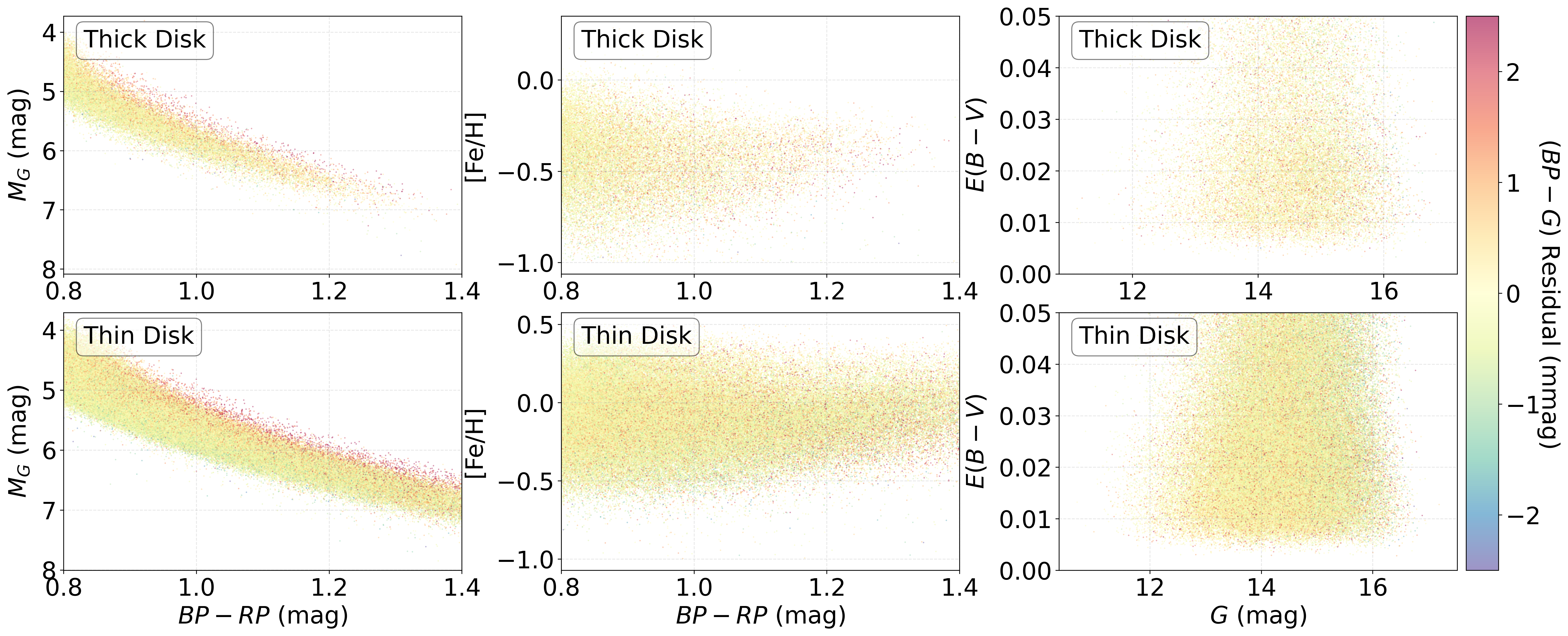}
    \caption{Same as Figure~\ref{fig:bpgfigres_1d}, but showing the fitting residuals in two-dimensional parameter spaces, color-coded by the residual value in mmag. The red-end points are unresolved binaries in the training sample.
    }
    \label{fig:bpgfigres_dcol}
\end{figure*}

\begin{equation}
\label{eq:bpg}
\begin{split}
&f_1=(a_1x^3+a_2x^2+a_3x+a_4)\times (y^3 - 3ky^2 + 3k^2y),\\
&f_2=b_1x^5+b_2x^4+b_3x^3+b_4x^2+b_5x+b_6,\\
&f_3=c_1y^2+c_2y+c_3,\\
&BP-G =f_1 + f_2 + f_3.
\end{split}
\end{equation}

\begin{figure}[ht!]
    \centering
    \includegraphics[width=1.0\columnwidth]{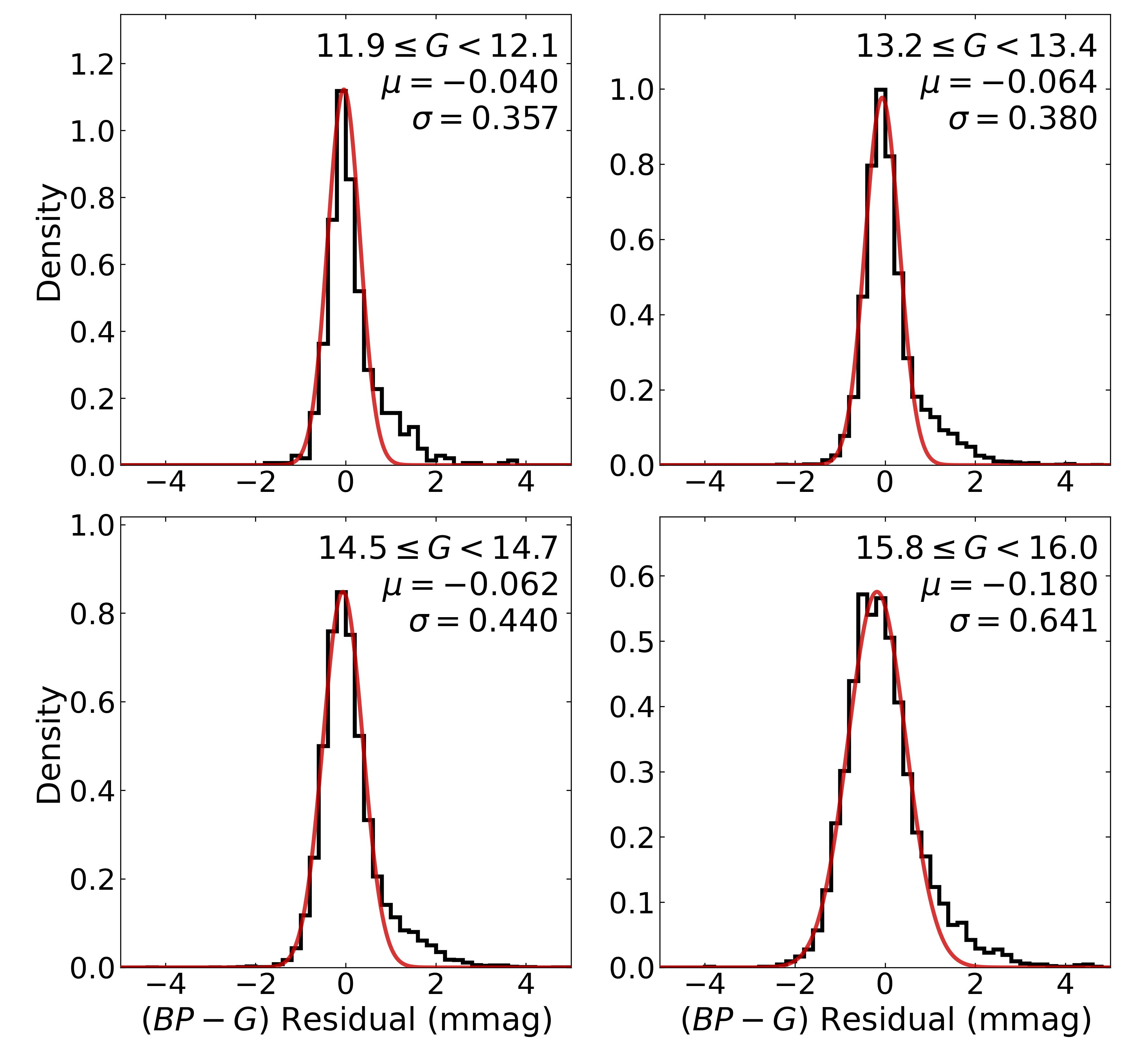}
    \caption{
Distribution of the $BP-G$ residuals for the binarity-estimation sample in four ranges of $G$ magnitude. 
In each panel, the red curve shows the best-fit Gaussian distribution, with the corresponding mean ($\mu$) and standard deviation ($\sigma$) indicated. The extended tail on the red side reflects the contribution of unresolved binaries in the sample.
    \label{fig:resgaussian_scol}
    }
\end{figure}

Figure\,\ref{fig:bpgfigres_1d} and Figure\,\ref{fig:bpgfigres_dcol} demonstrate the fitting residuals of the FGK dwarfs in the training sample as functions of $BP-RP$, [Fe/H], $G$, and $E(B-V)$, in both one-dimensional projections and two-dimensional parameter spaces. Figure~\ref{fig:resgaussian_scol} further shows the residual distributions of the binarity-estimation sample in different bins of $G$-band magnitude. 
The residuals exhibit no systematic dependence on any of these stellar parameters. Although the scatter increases toward fainter magnitudes due to the growing photometric uncertainties, the standard deviation of the residuals remains below $1$\,mmag across all parameter space. This level of precision is consistent with the calibrated photometric accuracy of Gaia XP spectra and represents a significant improvement over the intrinsic widths of stellar loci previously derived from SDSS photometry or Gaia broad-band photometry.  

We note that the residual distributions are not perfectly symmetric and their medians are slightly negative, which are caused by the unresolved binaries. These binary systems manifest as extended red tails in Figures~\ref{fig:bpgfigres_1d} and~\ref{fig:resgaussian_scol}, and as red-end points in Figure~\ref{fig:bpgfigres_dcol}. Consequently, the fitted stellar loci are systematically offset from the true single-star loci by a small negative amount, which we parameterize as $\mu$. As shown in Figures~\ref{fig:bpgfigres_1d} and~\ref{fig:resgaussian_scol}, the magnitude of this offset is significantly smaller than the standard deviations. Because the offset is systematic, it does not bias our subsequent measurement of the binary fraction.

\begin{figure}[ht!]
    \centering
    \includegraphics[width=0.98\columnwidth]{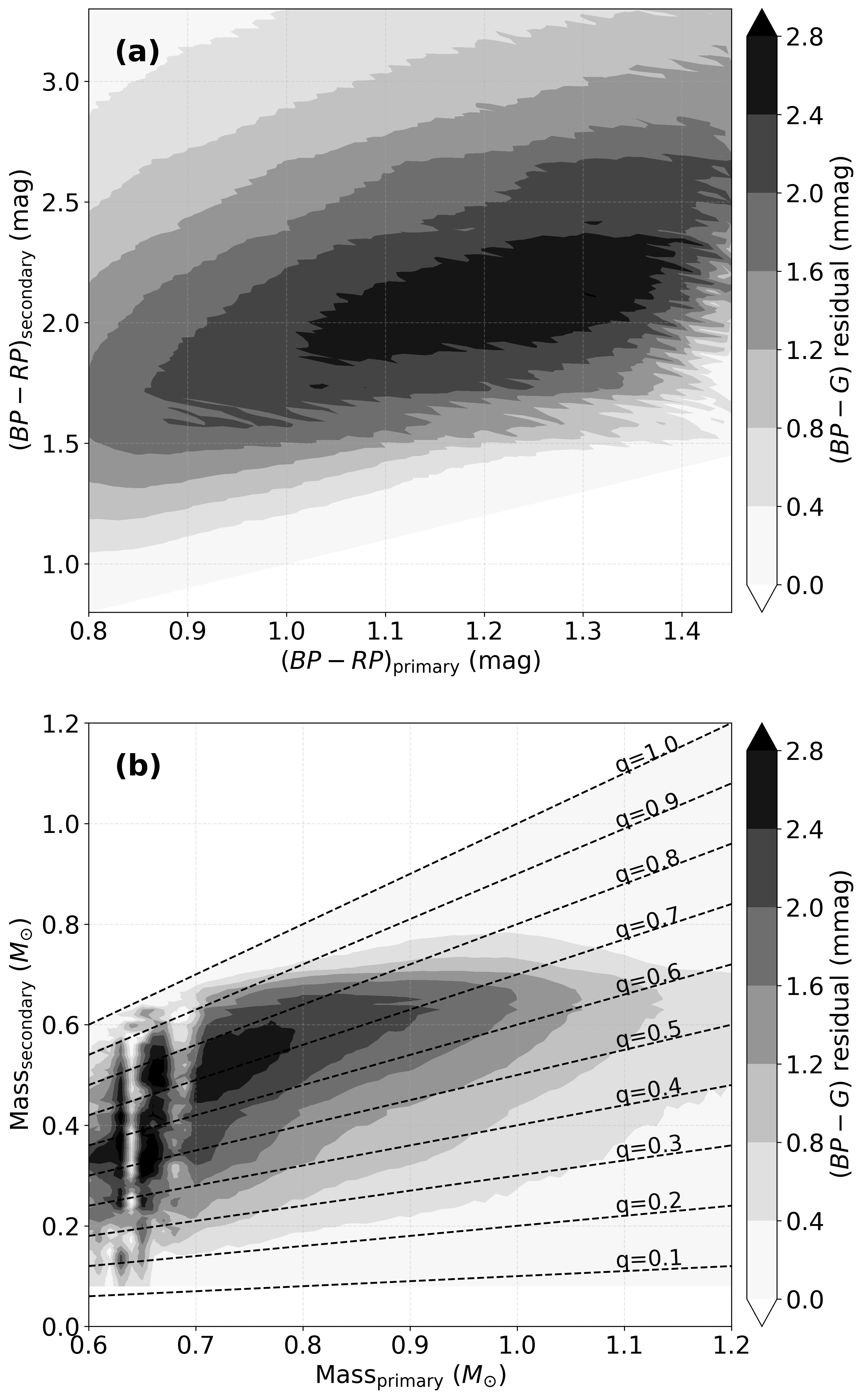}
    \caption{
    Simulated distributions of the $(BP-G)$ residuals, defined as $(BP-G)_{\rm loci}\,-\,(BP-G)_{\rm obs}$, for a fixed metallicity of [Fe/H] $=\,0$~dex. Panel (a) and (b) show distributions color-color plane and mass-mass plane, respectively. Models for different mass ratios $q$ are overplotted in panel (b).
    \label{fig:contour_total}
    }
\end{figure}

The effect of unresolved binaries on the $BP-G$ residual is typically on the order of $1$--$2$\,mmag, which substantially exceeds both the intrinsic width of the loci and the precision of the Gaia synthetic photometry. The maximum influence from the binary system on the $BP-G$ residual about 3~mmag and occurs at intermediate mass ratios, as illustrated in Figure~\ref{fig:contour_total}. The color residuals therefore provide a robust diagnostic for constraining the binary fraction through the fitting procedure described in Section~\ref{sec:slot_overview}.

We noticed that the fitting sample contains few very red, metal-poor stars, and the dependence of the color-color loci on [Fe/H] becomes progressively weaker toward redder colors \citep{Huangbw2025loci,Yuan2015stellarloci1}. For sources with $BP-RP\,>\,1.65$\,mag, we therefore adopt the loci of [Fe/H]$\,=\,0\,\rm dex$. Meanwhile, in a binary system, the secondary star can have a very low mass, such that its $BP-RP$ exceeds the reddest M dwarfs present in our training sample ($BP-RP\,\approx\,3.27\,\rm mag$). For such cases, we adopt the color-color relation interpolated from PARSEC isochrones at [Fe/H]$\,=\,0$\,dex to avoid unreliable extrapolation. We note that at such extreme mass ratios ($q \rightarrow 0$), the contribution from the secondary star to the color residual is also negligible.

\subsubsection{Metallicity-dependent color-absolute magnitude loci}\label{sec:mg_loci}

The absolute magnitude $M_G$ is fitted as a low-order polynomial function of $BP-RP$ and [Fe/H]. Following the approach of Wang et al. (in prep.),  we divide the sample into a blue regime ($BP-RP\,<\,1.8\,\rm mag$) and a red regime  ($BP-RP\,\geq\,1.4\rm\, mag$). The blue side is fitted with a fourth-order polynomial, while the red side is fitted with a third-order polynomial. The resulting coefficients for both regimes are listed in Table\,\ref{tab:mg_coeff} in Appendix\,\ref{sec:AP_stellarlocifitting}.

To ensure a smooth transition across the intermediate color range, we adopt a weighted mean of the blue and red polynomial predictions. The transition function is defined as: 
\begin{equation}
\label{eq:mg_trans}
\begin{split}
f_{\mathrm{mid}} \,&=\, (1-\frac{x-x_{\mathrm{blue}}}{x_{\mathrm{red}}-x_{\mathrm{blue}}}) f_{\mathrm{blue}}
		+ \frac{x-x_{\mathrm{blue}}}{x_{\mathrm{red}}-x_{\mathrm{blue}}} f_{\mathrm{red}}\\
        &(x_{\mathrm{blue}}\,\leq\, x \,<\,x_{\mathrm{red}}).
\end{split}
\end{equation}
where $x$ is $BP-RP$, and we set the boundaries to $x_{\rm blue} = 1.45$\,mag and $x_{\rm red} = 1.75$\,mag.

\begin{figure}[ht!]
    \centering
    \includegraphics[width=0.98\columnwidth]{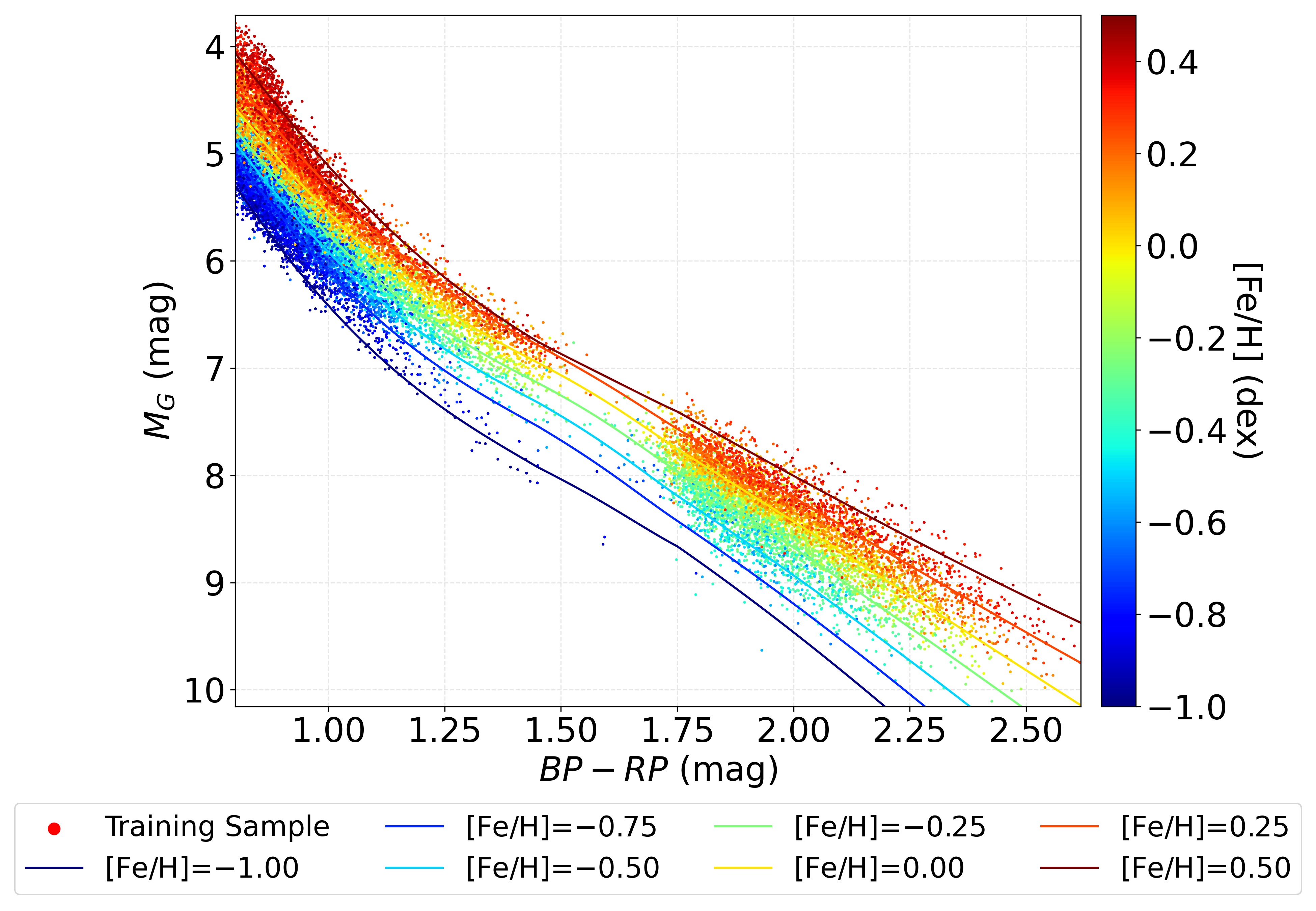}
    \caption{
    CMD of the training sample. The plotted points are uniformly sampled with respect to [Fe/H] for visual clarity and are color-coded by metallicity. The fitted loci are overplotted with the same color map.
    \label{fig:cmdfit}
    }
\end{figure}

Figure\,\ref{fig:cmdfit} displays the CMD of the training sample, color-coded by [Fe/H]. The overplotted curves show the corresponding metallicity-dependent $M_G$ loci derived from our polynomial fits.

\subsection{Mass-ratio distribution} \label{sec:q_selection}

\begin{figure}[ht!]
    \centering
    \includegraphics[width=0.98\columnwidth]{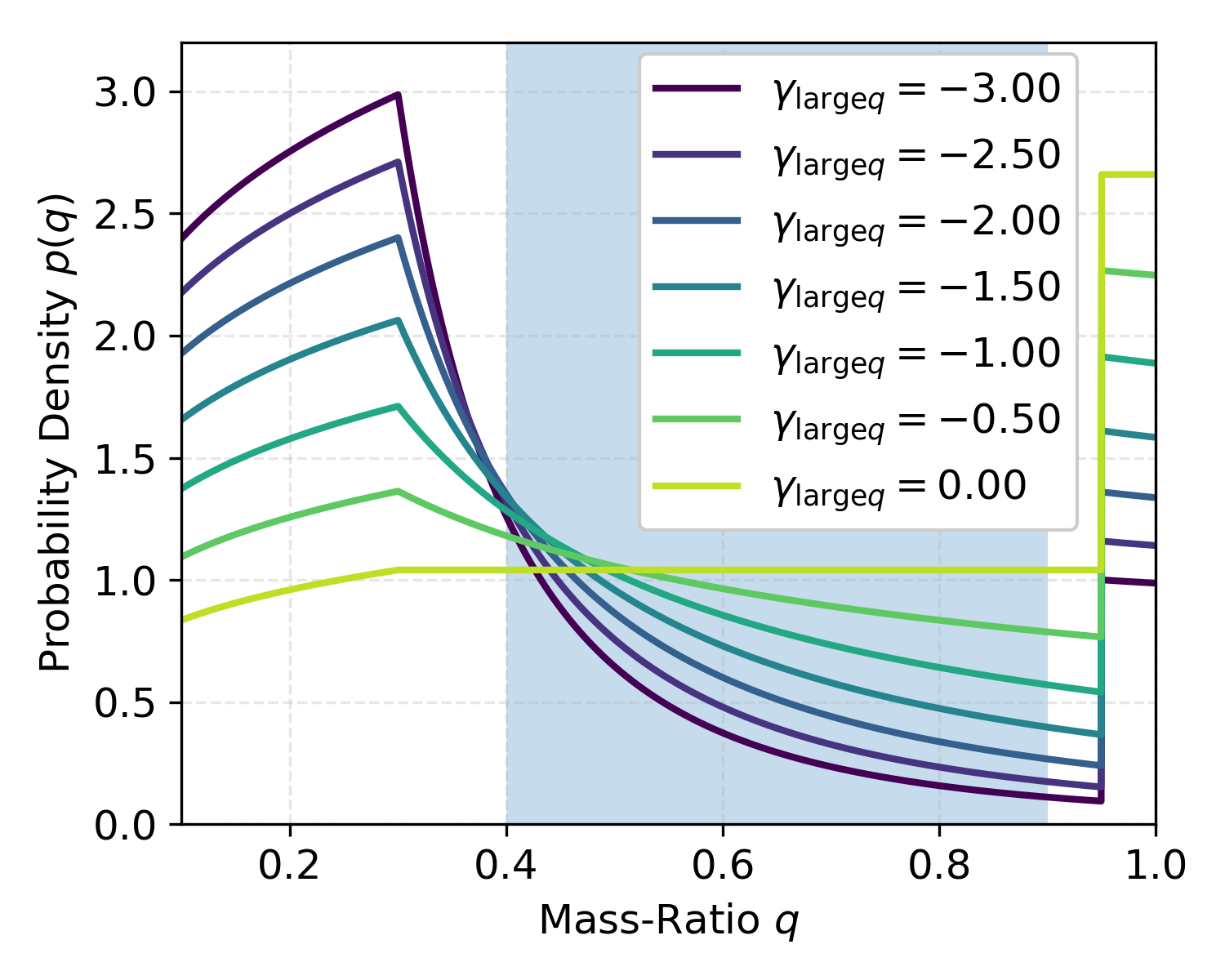}
    \caption{
    Probability density functions of the mass ratio $q$ for seven different values of the large-$q$ power-law index $\gamma_{\mathrm{large}q}$. The shaded region indicates the range of $q$ that produces the largest $BP-G$ residuals.
    \label{fig:qdists}
    }
\end{figure}

For the mass-ratio probability distribution, we adopt the parametric form introduced by \citet{Moe2017PandQ}. This function is characterized by four parameters: the power-law index at small-$q$ regime $\gamma_{\mathrm{small}q}$, the power-law index at large-$q$ regime $\gamma_{\mathrm{large}q}$, the transition mass ratio $q_{\mathrm{split}}$ between the two power-law regimes, and the excess fraction of twin binaries ($q\,>\,0.95$) relative to the underlying power law, $\mathcal{F}_{\mathrm{twin}}$.

With the help of the high precision of the Gaia XP synthetic photometry employed in this work, we find that the SLOT method is sensitive to the shape of the mass-ratio distribution, particularly to the parameter $\gamma_{\mathrm{large}q}$; this sensitivity arises because the largest deviations in the color residual occur at intermediate mass ratios ($0.4 \lesssim q \lesssim 0.9$), where the slope of the $q$ distribution is governed predominantly by the $\gamma_{\mathrm{large}q}$ (see Figure\,\ref{fig:contour_total}). Therefore, our approach offers a degree of empirical leverage on this parameter. 

To assess the influence of the assumed $q$ distribution on the color residual, we fix the remaining three parameters to representative values obtained from \citet{Moe2017PandQ} and \citet{ElBadry2019twinbinary}: $\gamma_{\mathrm{small}q}\,=\,0.2$, $q_{\rm split}\,=\,0.3$, and $\mathcal{F}_{\mathrm{twin}}\,=\,0.1$. We perform the analysis for a grid of seven values of $\gamma_{\mathrm{large}q}$: $-3.0$, $-2.5$, $-2.0$, $-1.5$, $-1.0$, $-0.5$, and $0.0$. The resulting probability density functions are shown in Figure\,\ref{fig:qdists}, with the shaded blue region indicating the range of $q$ that yields the largest $BP-RP$ residuals.

\begin{figure*}[ht!]
    \centering
\includegraphics[width=0.99\textwidth]{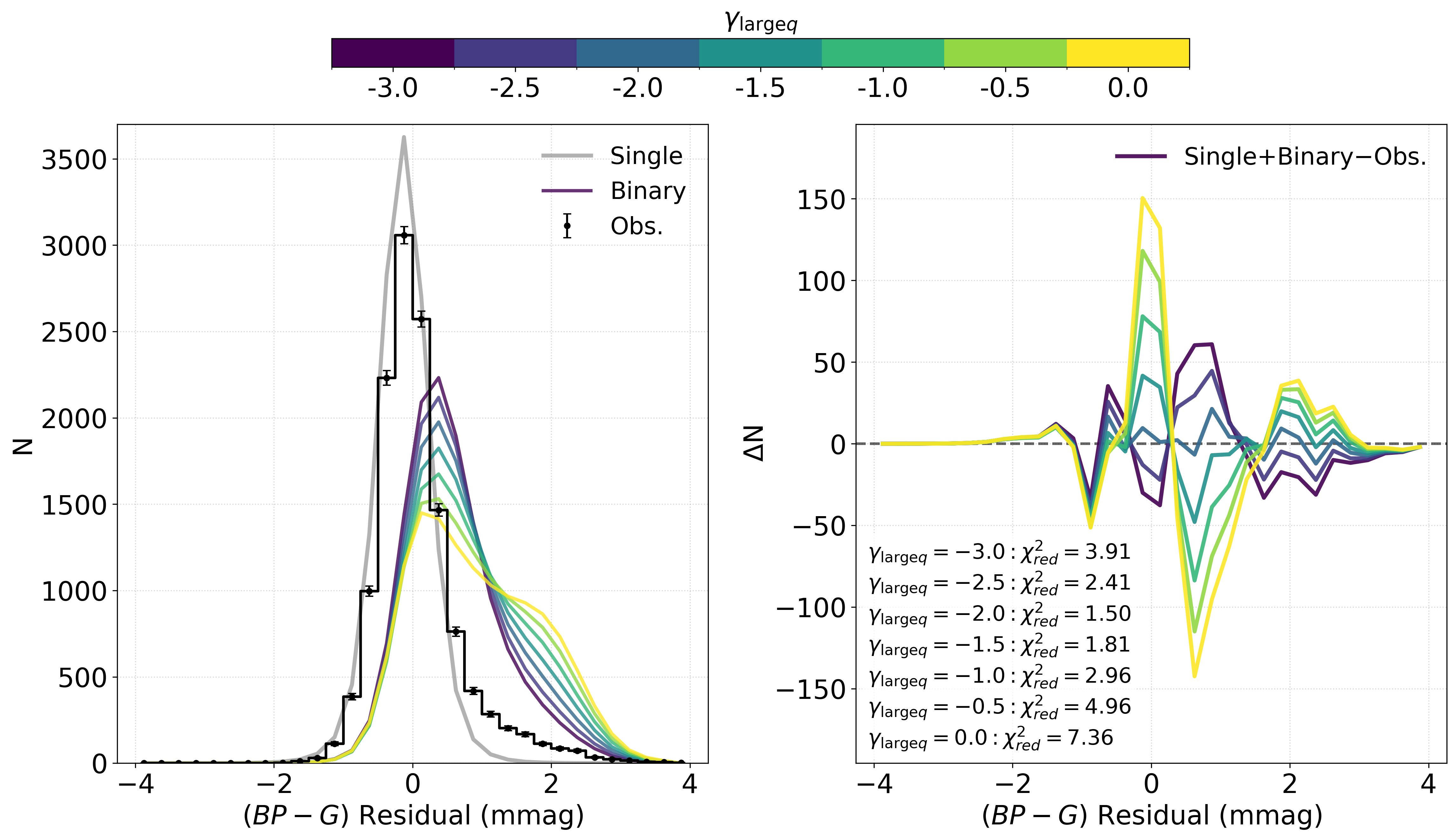}
    \caption{An illustration for mock data sets and SLOT results for a single bin: thin-disk stars within the range $0.1\,\leq\,$[Fe/H]$\,<\,0.2\,\rm dex$. The residual distribution of the mock Single Set is shown in gray, while that of the mock Binary Set is color-coded according to the assumed value of $\gamma_{\mathrm{large}q}$. ``Obs.'' denotes the observed distribution. For this bin, $\gamma_{\mathrm{large}q} = -2.0$ yields the best fit to the data; $\gamma_{\mathrm{large}q} = -1.5$ also has a good fitting effect; shallower slopes (e.g., $\gamma_{\mathrm{large}q} = 0.0$) overpredict the fraction of systems with large color residuals and result in a poorer fitting.
    \label{fig:qfit_example}
    }
\end{figure*}

As suggested by \citet{LiuC2019SmokingGun}, the power-law index of the mass-ratio distribution depends on stellar metallicity. We therefore divide the sample into bins of width of 0.1\,dex in [Fe/H]. For each bin, we perform the fit with seven assumed $q$ distributions (that is, seven models of $q$).
Figure\,\ref{fig:qfit_example} illustrates a representative case for a single bin.

We determine $\gamma_{\mathrm{large}q}$ using a bootstrap-based model-averaging procedure, rather than selecting the single model with the minimum reduced $\chi^2$. This avoids noise-driven choices when several models fit the data almost equally well. For each [Fe/H] bin and each disk subsample, we perform 1,000 resamplings of the observed sample to generate pseudo-datasets. Each pseudo-dataset is fitted with the seven $\gamma_{\mathrm{large}q}$ models, yielding a set of reduced $\chi^2$ values. We then take the 95th percentile of these 1,000 differences as an empirical threshold for assessing whether a model is compatible with the best-fit model. For each bin, we sort the seven models by their reduced $\chi^2$ and include consecutive models from the smallest reduced $\chi^2$ until the adjacent difference exceeds the threshold. We adopt $f_b$, $\mu$, and $\gamma_{\mathrm{large}q}$ for the bin as the uniform average over the models in the compatible group. If all seven models are included, the bin is flagged as unconstrained; otherwise, it is considered constrained. The uncertainty of $f_b$ combines the uncertainty from the individual fits and the scatter among the compatible models; $\mu$ is treated in the same way.

This approach follows the model-averaging rationale of \citet{Buckland1997modelavg}: when several candidate models are comparably supported, committing to a single one overstates the precision and underestimates the uncertainty of the derived parameters.

\section{Results} \label{sec:results}

\subsection{Binarity and Mass-ratio Distribution in Metallicity Space} \label{sec:fb_bprp_feh_space}

\begin{figure}[ht!]
    \centering
    \includegraphics[width=0.98\columnwidth]{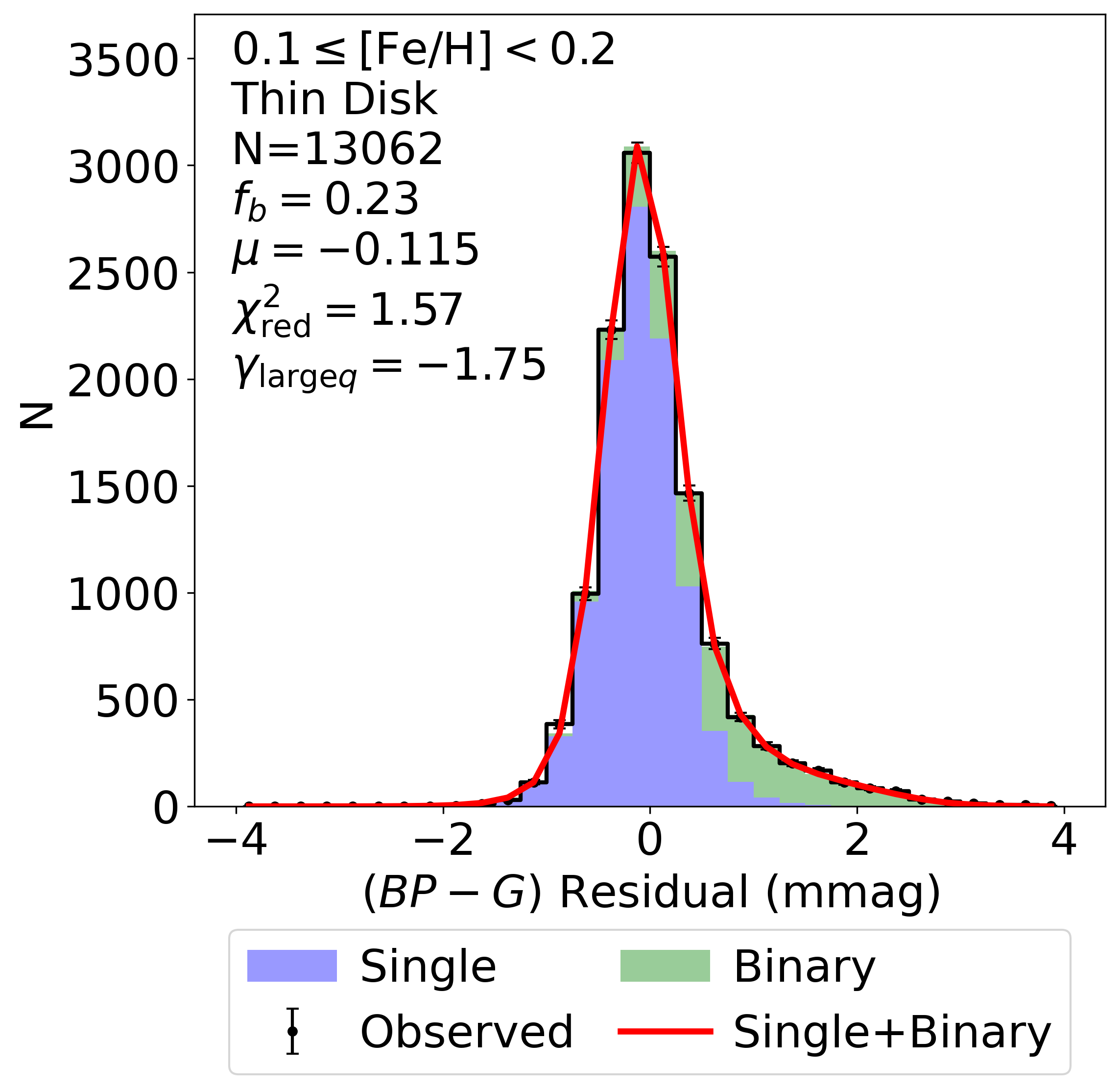}
    \caption{The SLOT fitting result of the thin-disk subsample in
    $0.1~\leq$ [Fe/H] $<~0.2$ dex. The best-fit binary fraction $f_b$, $\mu$,
    reduced $\chi^2$, and the corresponding $\gamma_{{\rm large}q}$ are labeled
    on the upper-left area. The complete figure set (2 images) for the whole
    binarity-estimating sample is available in the online journal, and the data
    used to create these figures are available online as Data behind the Figure.
    For all bins the best-fit $\mu$ lie well inside the prior range, and
    varying the prior boundary does not change the main results.}
    \label{fig:slot_result_example}
    \digitalasset
\end{figure}

\begin{figure*}[htbp]
    \centering
    \includegraphics[width=\textwidth]{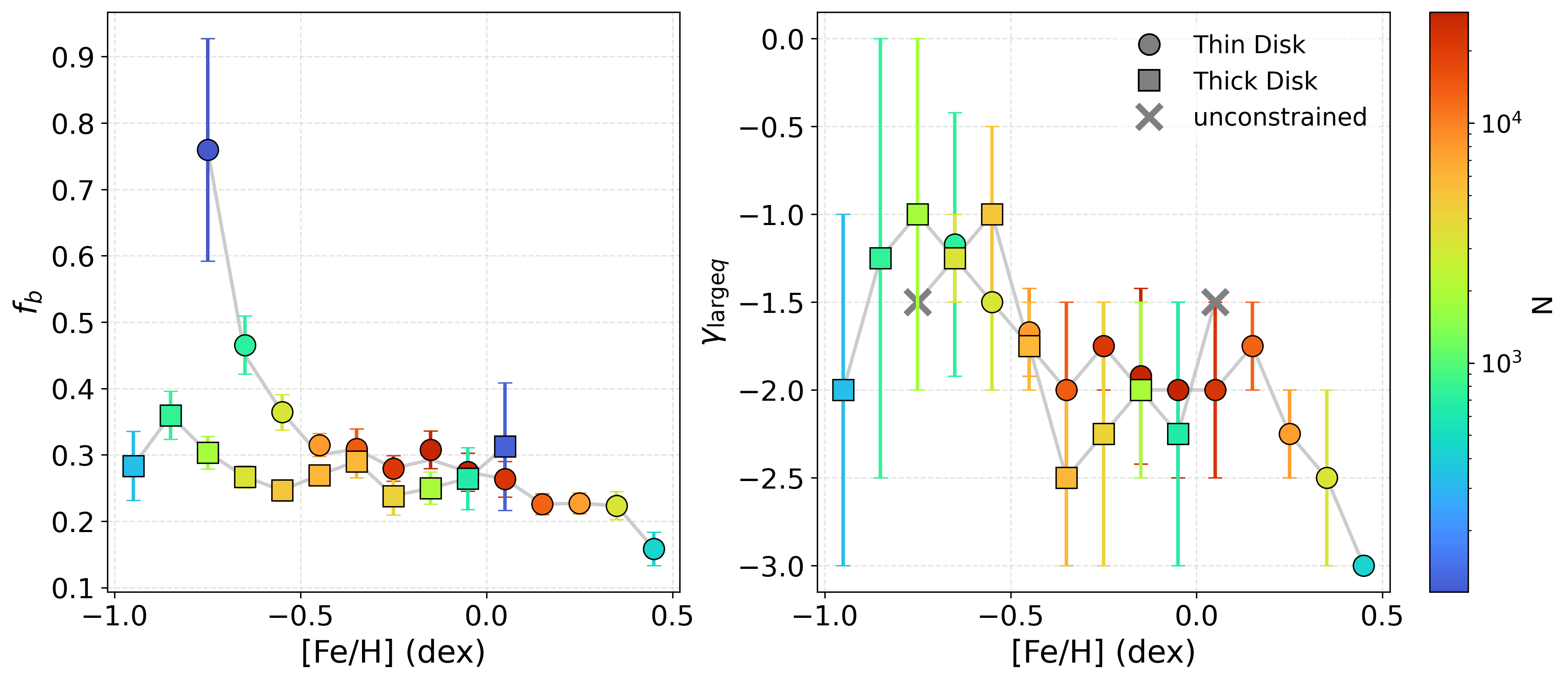}
    \caption{Binary fraction $f_b$ (left) and $\gamma_{\mathrm{large}q}$ (right) in [Fe/H] space. The thin-disk and thick-disk subsamples are marked with circular and square symbols, respectively.
    In the right panel, the errorbar of each point shows the lower limit and upper limit of its compatible group. The gray cross without an errorbar indicates that the $\gamma_{\mathrm{large}q}$ of the bin is unconstrained. All bins with size $N~<~100$ are excluded when analyzing.
    At [Fe/H] $\approx\ -0.65,\,-0.45$, and $-0.15$~dex, the $\gamma_{\mathrm{large}q}$  is identical for the thin and thick disks; the thin-disk points are offset slightly upward for visual clarity.}
    \label{fig:fb_q_feh_combine}
\end{figure*}

Figure\,\ref{fig:slot_result_example} displays an example of the fitted distribution of $(BP-G)$ color residuals. The complete result of the fitting result for each bin can be found in the online figure set of Figure~\ref{fig:slot_result_example}, while the results for different assumed  $\gamma_{\mathrm{large}q}$ models in each bin are presented in Appendix~\ref{sec:AP_fullresults}. The results of $\gamma_{\mathrm{large}q}$ and $f_b$ for the thin-disk and thick-disk subsamples in different [Fe/H] bins are shown in Figure\,\ref{fig:fb_q_feh_combine}. 

\figsetstart
\figsetnum{10}
\figsettitle{Complete SLOT fitting results for the whole binarity-estimating sample}

\figsetgrpstart
\figsetgrpnum{10.1}
\figsetgrptitle{Thin-disk subsample}
\figsetplot{Appendix_Fitting_Grid_THIN.png}
\figsetgrpnote{Complete SLOT fitting results for the thin-disk subsample in metallicity space.}
\figsetgrpend

\figsetgrpstart
\figsetgrpnum{10.2}
\figsetgrptitle{Thick-disk subsample}
\figsetplot{Appendix_Fitting_Grid_THICK.png}
\figsetgrpnote{Complete SLOT fitting results for the thick-disk subsample in metallicity space.}
\figsetgrpend
\figsetend

The left panel in Figure~\ref{fig:fb_q_feh_combine} maps the derived binary fraction $f_{\rm b}$ in the metallicity space. 
Within the thin-disk subsample, $f_b$ declines systematically with increasing [Fe/H]: it drops from $\sim$0.4--0.5 at [Fe/H]$\lesssim-0.5$~dex to $\sim$0.25--0.30 around solar metallicity, and further decreases to $\sim$0.15 in the most metal-rich bin at [Fe/H]$\sim0.45$~dex. The decline is steepest at the metal-poor end and becomes shallower toward solar metallicity. The apparently high binary fraction in the most metal-poor thin-disk bin carries a large uncertainty due to the limited number of stars in that bin. 
The uncertainty on $f_{\rm b}$ scales with the bin size: $\sim 0.13$ for $N = 100$, $\sim 0.03$ for $N = 1000$, and $\sim 0.02$ for $N = 10,000$.
For the thick-disk subsample, it shows no clear trend of $f_b$ with metallicity, with $f_b$ remaining roughly constant at 0.25--0.35 over its [Fe/H] coverage. 
The averaged binary fractions of the thin-disk subsample is slightly higher than that of the thick-disk samples, with $f_{\rm b,thin} = 0.34 \pm 0.04$ and $f_{\rm b,thick} = 0.28 \pm 0.04$.

The right panel in Figure\,\ref{fig:fb_q_feh_combine} displays the values of $\gamma_{\mathrm{large}q}$ for different [Fe/H]. 
No significant difference in $\gamma_{\mathrm{large}q}$ is observed between the thin-disk and thick-disk populations.
Metal-rich subsamples generally favor steeper slopes (more negative $\gamma_{\mathrm{large}q}$), implying a relative deficit of binary systems with intermediate mass ratios ($0.4 \lesssim q \lesssim 0.9$).
Although the SLOT method provides only modest precision on the $q$ distribution, the trends are coherent: bins near solar metallicity predominantly yield $\gamma_{\mathrm{large}q} \approx -2.0$, while those at [Fe/H] $\sim -0.5$\,dex cluster around $-1.5$. We therefore choose these two values of $\gamma_{\mathrm{large}q}$ in the subsequent analysis of the dependence of binarity on $\alpha$-process elements abundances.

We performed sensitivity tests for the fixed mass-ratio parameters \(\gamma_{{\rm small}q}\), \(q_{\rm split}\), and \(\mathcal{F}_{\rm twin}\), using three thin-disk [Fe/H] bins (centers \(-0.35, 0.05, 0.25\)~dex) and three thick-disk bins (\(-0.65, -0.45, -0.15\)~dex). We took the adopted parameter set \((\gamma_{{\rm small}q},q_{\rm split},\mathcal{F}_{\rm twin})=(0.2,0.3,0.1)\) in Section~\ref{sec:q_selection} as the baseline set, and repeated the analysis for six additional runs in which only one parameter was varied at a time: $\gamma_{{\rm small}q}=-0.2$ or 0.6, $q_{\rm split}=0.2$ or 0.4, $\mathcal{F}_{\rm twin}$= 0.05 or 0.2. The derived binary fraction is only weakly sensitive to $\gamma_{{\rm small}q}$ and $\mathcal{F}_{\rm twin}$ (typical $|\Delta f_{\rm b}|\sim0.03$ in both disks), and more strongly to \(q_{\rm split}\) (typical $|\Delta f_{\rm b}|\,\sim\,0.12$ in the thin disk and $\sim\,0.12$ in the thick disk). 
In every combination, \(f_{\rm b}\) decreases with increasing [Fe/H] in the thin disk, and the thick disk shows no obvious trend. The \(\gamma_{{\rm large}q}\) is affected by \(q_{\rm split}\): at fixed bin, changing \(q_{\rm split}\) from 0.2 to 0.4 typically shifts \(\gamma_{{\rm large}q}\) by 0.75/1.00 for thin-disk/thick-disk stars; whereas the variations of \(\gamma_{{\rm small}q}\) and \(\mathcal{F}_{\rm twin}\) barely affect \(\gamma_{{\rm large}q}\). This is because \(q_{\rm split}\) and \(\gamma_{{\rm large}q}\) both directly control the binary mass-ratio range to which the SLOT method is most sensitive. The reported $\gamma_{\mathrm{large}q}$--[Fe/H] dependence is reproduced in five of the six additional runs, with the exception of the run using $q_{\rm split}~=~0.4$, which shows weakly dependence. We therefore suggest that the $\gamma_{\mathrm{large}q}$--[Fe/H] dependence is most likely negative, and it should be interpreted qualitatively rather than quantitatively.

\begin{figure}
    \centering
    \includegraphics[width=0.98\columnwidth]{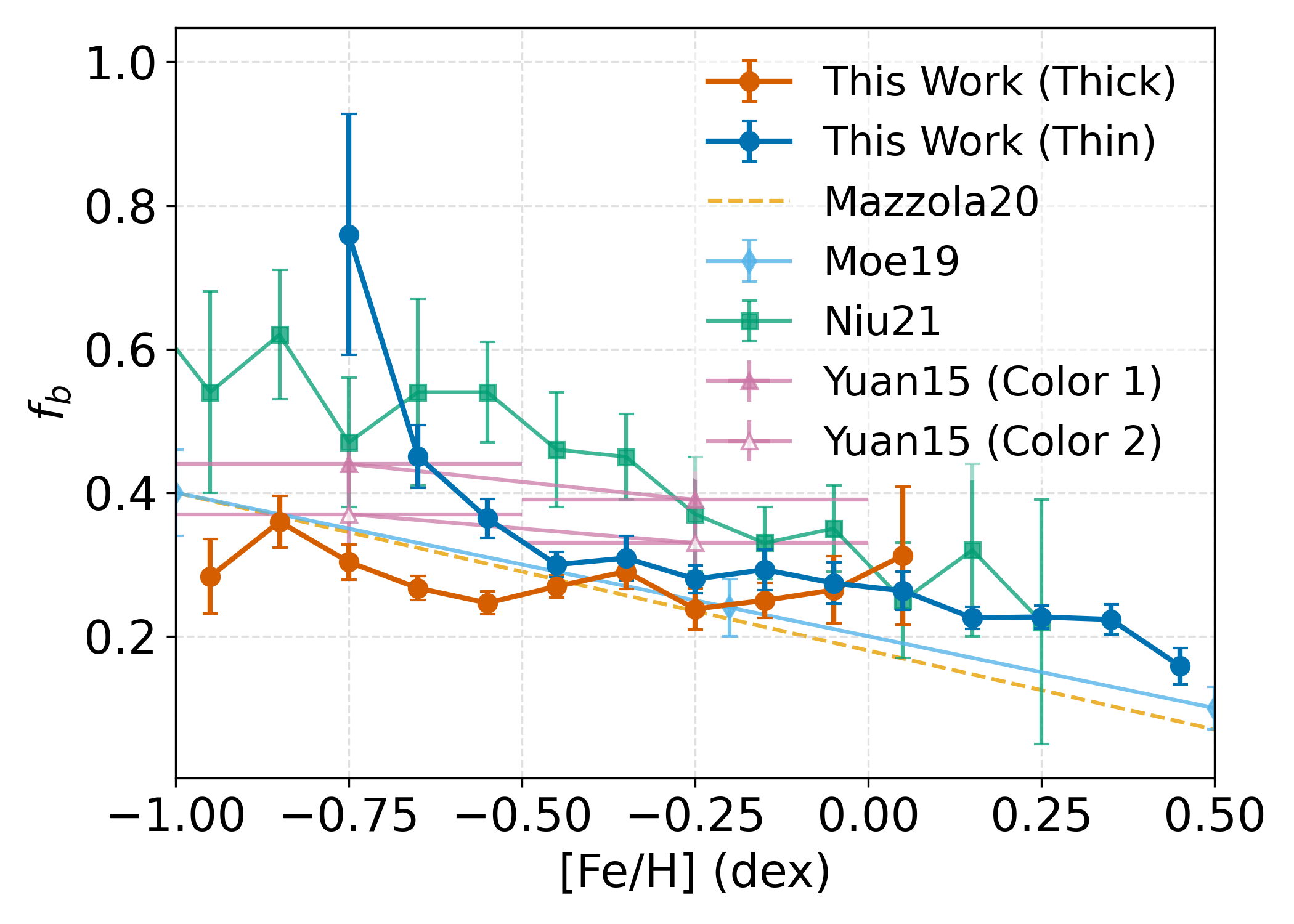}
    \caption{Binary fraction of FGK dwarf stars as a function of [Fe/H]. We compare our work with (1) \citet{Yuan2015slot}, for which we show results in two SDSS $g-i$ color bins: $0.6~\leq~g-i~<~0.9$ mag (Color 1 on the figure) and $0.9~\leq~g-i~<~1.2$ mag (Color 2 on the figure), (2) \citet{Moe2019STclosebinary}, (3) \citet{Mazzola2020closebinary}, (4) \citet{Niuzx2021FGK}, for which the thin- and thick-disk results are averaged. The Gaia $BP-RP$ color of our sample ranges from $0.8$ to $1.4$ mag, which translates to a SDSS $g-i$ color range of approximately $0.542$ to $1.366$ mag\footnote{\url{https://gea.esac.esa.int/archive/documentation/GDR3/Data_processing/chap_cu5pho/cu5pho_sec_photSystem/cu5pho_ssec_photRelations.html}}. }
    \label{fig:fb_feh_compile}
\end{figure}

We also repeated the $\gamma_{{\rm large}q}$ measurement in two $BP-RP$ bins, 0.8--1.1 and 1.1--1.4~mag. For the bluer bin, the derived [Fe/H] dependence agrees with that of the whole sample within the 1$\sigma$ uncertainties. The redder bin mainly contains thin-disk stars and covers a narrower metallicity range ($-0.6\,\lesssim $[Fe/H]$\,\lesssim \,0.3$~dex). Over most of this range, $\gamma_{{\rm large}q}$ lies between $-2.5$ and $-1$, and the $\gamma_{{\rm large}q}$--[Fe/H] relation appears weaker than in the full sample; the exception is the [Fe/H]$\approx0.2$ dex bin, where the $\gamma_{{\rm large}q}$ is poorly constrained, with values spanning  $-2.0\sim0.0$. The result of the redder bin is likely dominated by the larger photometric uncertainties and its narrow [Fe/H] range.

We compare our results for the metallicity dependence of binarity with several recent studies of Galactic field stars in Figure~\ref{fig:fb_feh_compile}. Among these, \citet{Yuan2015slot} and \citet{Niuzx2021FGK} employed the same SLOT method used here, \citet{Mazzola2020closebinary} relied on RV variations, and \citet{Moe2019STclosebinary} compiled results from multiple surveys using a combination of techniques such as eclipsing binaries and RV variations. The overall declining trend of $f_b$ with increasing [Fe/H] is qualitatively consistent with these earlier works. Quantitatively, our binary fractions are overall lower than those reported by \citet{Niuzx2021FGK}, although the qualitative trend is similar. This difference likely arises because \citet{Niuzx2021FGK} adopted a single power-law mass-ratio distribution, which predicts a higher fraction of binaries producing small color residuals and therefore leads to a systematically higher $f_b$. \citet{Niuzx2021FGK} also reported that the metallicity dependence of the binary fraction is less pronounced for thick-disk stars than for thin-disk stars. Our binary fractions are closer to those reported by \citet{Yuan2015slot}, and are systematically higher than those derived by \citet{Moe2019STclosebinary} and \citet{Mazzola2020closebinary}, whose analyses focused more closely on short-period binary systems in the thin disk. This offset is expected, as the SLOT method is capable of covering a broader range of orbital periods $P$ \citep[see Figure~\ref{fig:resolved_frac_uncounted} in our work and Figure~1 in][]{Moe2017PandQ}

\subsection{Binarity and $\alpha$-process Element Abundances} \label{sec:alpha_abund}

\begin{figure*}[ht!]
    \centering
    \includegraphics[width=\textwidth]{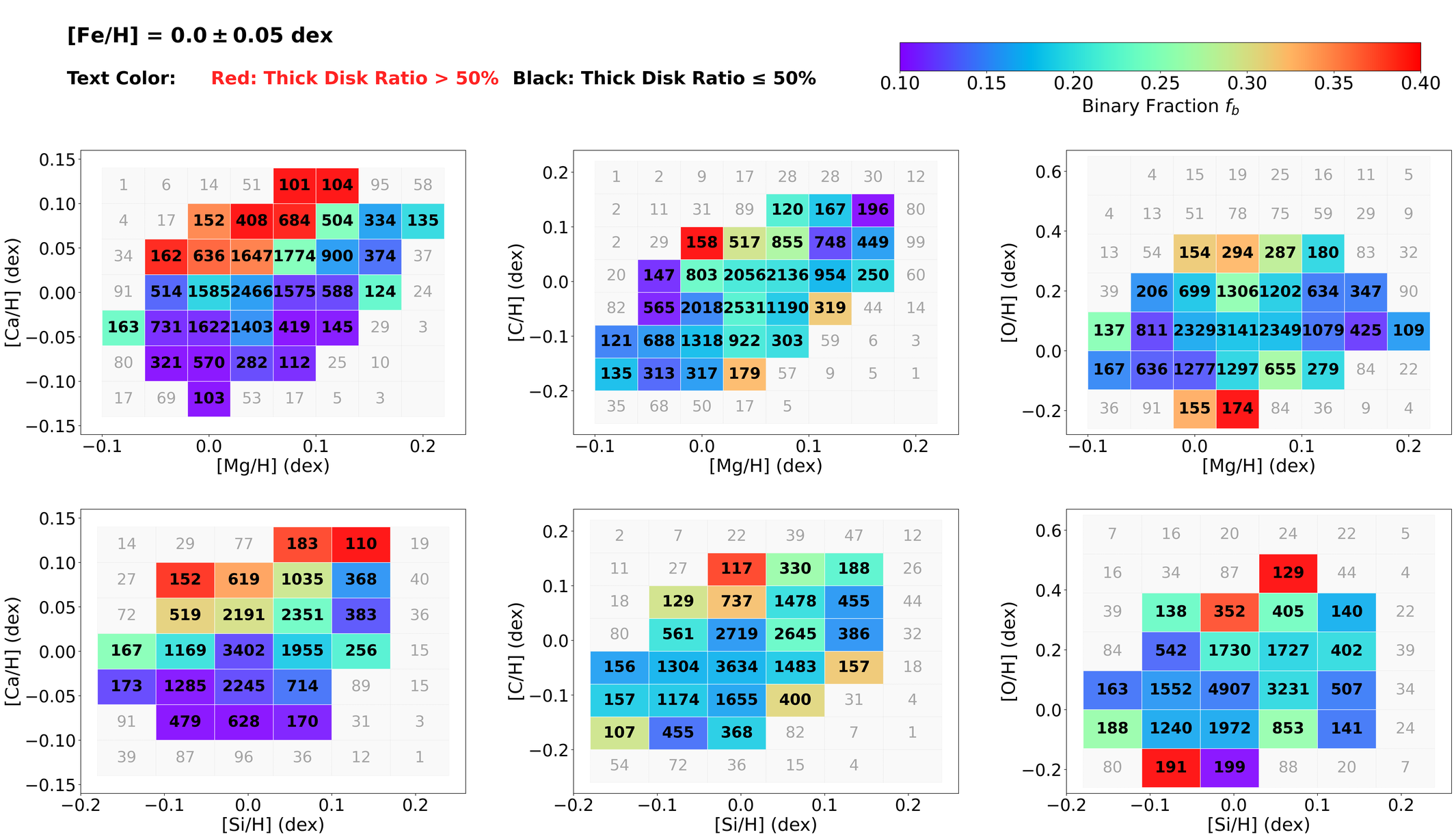}\\[1ex]
    \includegraphics[width=\textwidth]{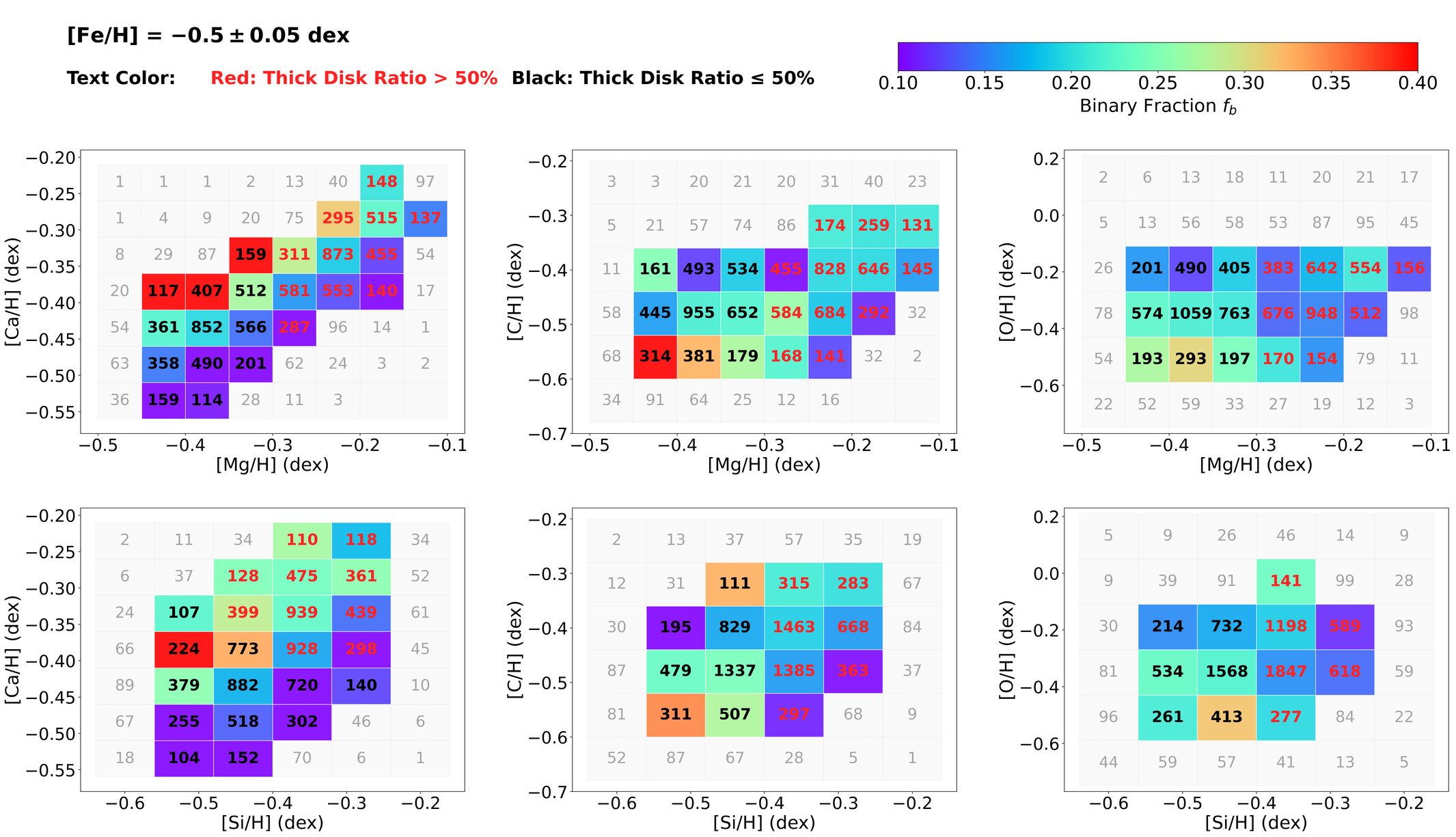}
    \caption{%
        Binary fractions of two subsamples in two-dimensional planes of $\alpha$-element abundance, including [Mg/H], [Si/H], [Ca/H], [C/H], and [O/H]. The upper panel illustrates the results of the solar-metallicity subsample ([Fe/H]$~=~0~\pm~0.05$~dex), and the lower panel illustrates the results of the moderately metal-poor subsample ([Fe/H]$~=~-0.5~\pm~0.05$~dex). The color of each bin indicates the derived binary fraction, and the number within each patch denotes the bin size. Red numbers shown indicate bins in which the fraction of thick-disk stars exceeds $50\%$. All bins with size $N~<~100$ are excluded when analyzing and are colored with light gray on the figure.
        The data used to create this figure are available online as Data behind the Figure.
    }
    \label{fig:fb_alpha}
\end{figure*}

Previous studies of the correlation between binarity and $\alpha$-element abundances have yielded conclusions that are not entirely consistent \citep{Mazzola2020closebinary, Niuzx2021FGK}. Apart from differences in sample selection, the two analyses differ in how $\alpha$ enhancement is quantified: \citet{Mazzola2020closebinary} examined the individual abundances [Mg/H], [Si/H], and [O/H], whereas \citet{Niuzx2021FGK} adopted a weighted mean of [Mg/Fe], [Si/Fe], [Ca/Fe], and [Ti/Fe]. In this work, we build upon the approach of \citet{Niuzx2021FGK} but extend the analysis by investigating the binary fraction as a function of five individual $\alpha$-process elements: Mg, Si, Ca, C, and O. Titanium is not considered here owing to its larger measurement uncertainties (private communication).

We construct two subsamples at different metallicities: a solar-metallicity subsample ([Fe/H] $= 0.00 \pm 0.05$~dex) and a moderately metal-poor subsample ([Fe/H] $= -0.50 \pm 0.05$~dex). For each $\alpha$ element, denoted X, we divide the subsample into bins whose width equal to the 75th percentile of the [X/Fe] measurement uncertainties. The solar-metallicity sample is dominated by thin-disk stars, whereas the metal-poor sample contains a mixture of thin- and thick-disk populations. The binary fraction in each bin is determined via the SLOT method on a two-dimensional grid of the selected abundance pairs.

The results are presented in Figures~\ref{fig:fb_alpha}. In both subsamples, the binary fraction declines with increasing [Si/H] and [Mg/H], with the exception of [Mg/H] in the solar-metallicity bin, where the trend is less pronounced.
In the moderately metal-poor subsample, as [Mg/H] increases from $\sim-0.4$~dex to $\sim-0.15$~dex, the binary fraction decreases from $\sim~0.3$ to $\sim~0.15$. The same trend also appears as [Si/H] increases from $\sim-0.5$~dex to $\sim-0.3$~dex, where the binary fraction decreases from $\sim~0.25$ to $\sim~0.15$. 
In the solar-metallicity subsample, when [Si/H] increases from $\sim-0.15$~dex to $\sim0.15$~dex, the binary fraction decreases from $0.25$ to $0.15$. For [Mg/H], the binary fraction peaks at about $0.25$ when [Mg/H]$\sim0.04$~dex.
This overall behavior is broadly consistent with the findings of \citet{Mazzola2020closebinary}.

Interestingly, Calcium exhibits a significant positive correlation with the binary fraction at fixed [Si/H] or [Mg/H]. 
In the moderately metal-poor subsample, an increase in [Ca/H] of $0.1$\,dex raises $f_b$ by $\Delta f_b\,\approx\,+0.15$. This effect is stronger than the opposite trends produced by Si and Mg: varying [Si/H] or [Mg/H] by the same amount changes $f_b$ by only $\Delta f_b\,\approx\,-0.05$ and $\Delta f_b\,\approx\,-0.1$, respectively. A similar pattern is seen in the solar-metallicity subsample, where $\Delta{\rm [Ca/H]}\,=\,0.1$\,dex yields $\Delta f_b\,\approx\,0.15$, whereas $\Delta{\rm [Si/H]}\,=\,0.1$\,dex produces $\Delta f_b\,\approx\,0.1$. The combined influence of Ca, Si, and Mg may account for the tentative positive correlation between $f_b$ and [$\alpha$/H] reported by \citet{Niuzx2021FGK}.

We note that among bins dominated by thick-disk stars (primarily in the metal-poor subsample), the binary fraction shows no significant correlation  with the $\alpha$-element abundances. 
Across the abundance planes shown in Figure~\ref{fig:fb_alpha}, the scatter of $f_b$ varies by among most thick-disk bins is about 0.2. In contrast, thin-disk-dominated bins exhibit a larger scatter of about 0.4. This result is consistent with \citet{Niuzx2021FGK} and highlights the contrasting sensitivity of binarity to chemical composition in the thin and thick disks.

We use 2D abundance grids to decompose the effect of $\alpha$-elements. Nevertheless, the possible effects from $T_{{\rm eff}}$, age, S/N, or thick/thin-disk membership across abundance bins are not controlled. The reported $f_{\rm b}$–abundance trends should be interpreted as statistical correlations rather than evidence for a direct causal effect of each element on binary formation.

\subsection{Binarity and Stellar Ages}\label{sec:age_and_fb}

\begin{figure}[ht!]
    \centering
    \includegraphics[width=0.98\columnwidth]{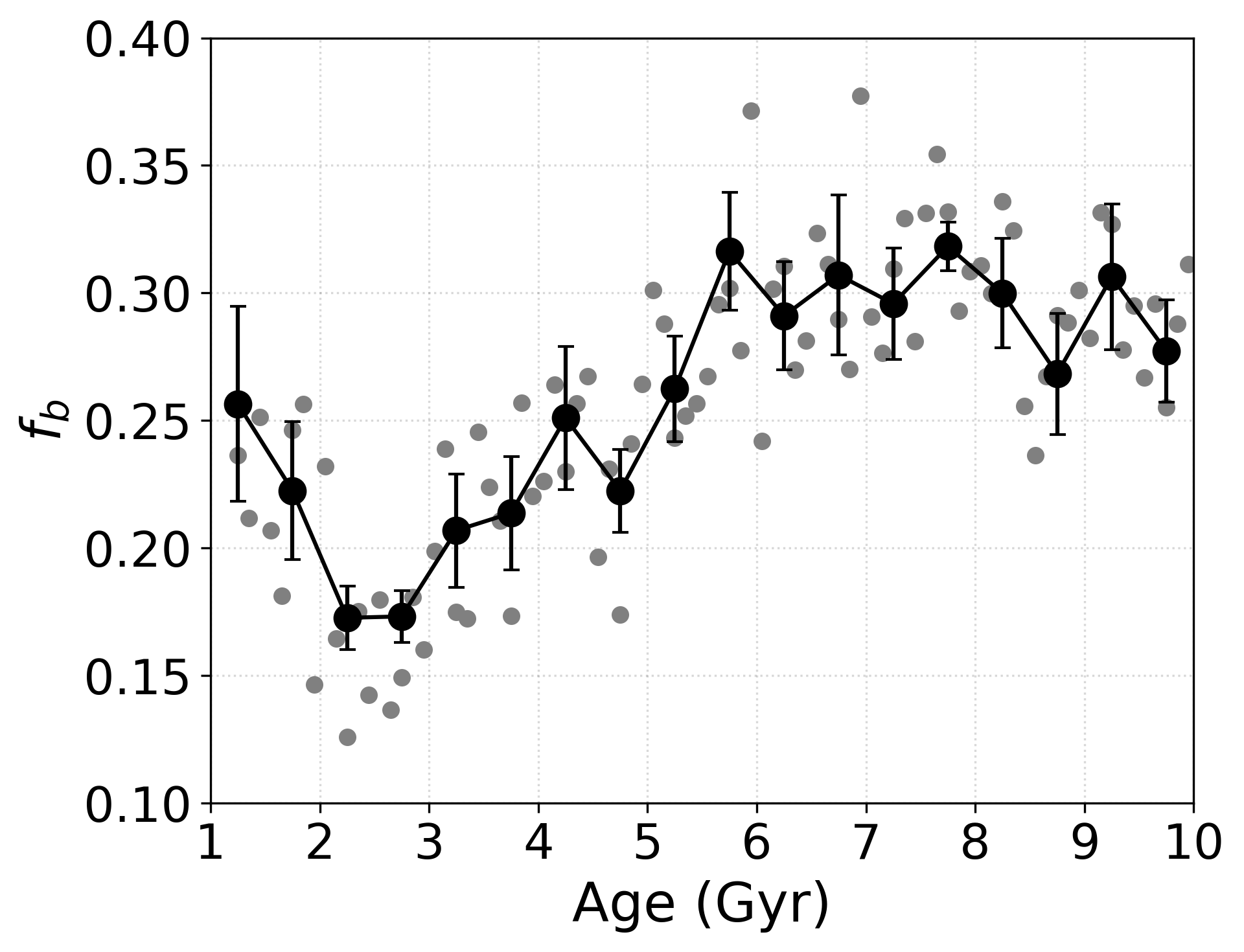}
    \caption{
    Binary fractions of the stars within an age range of $1$--$10$~Gyr.
    The gray dots in the background denote the estimated binary fractions in a smaller bin width of $0.1$~Gyr, displaying the scattering of our binarity estimation.
    The black dots in the foreground denote the estimated binary fractions in a larger bin width of $0.5$ Gyr, demonstrating the binary-fraction trend more clearly. 
    \label{fig:fb_age}
    }
\end{figure}

The stellar age estimates recently presented by \citet{Wangjh2025Ages} allow us to explore the relationship between binary fraction and stellar age. We restrict the sample to sources with ages between 1 and 10\,Gyr and bin them uniformly in logarithmic age with a width of $0.1$\,dex. For each bin, we measure the binary fraction following the same approach described in Section\,\ref{sec:q_selection}. The resulting binary fractions are shown in Figure~\ref{fig:fb_age}.

Overall, we observe only modest variation across the age bins.
The first two age bins are quite younger than the characteristic age of the Galactic thin disk ($\sim 2.5$\,Gyr; \citealt{2022Xiang}), and consequently contain relatively few stars and are subject to large uncertainties. We remain cautious about the relatively higher binary fraction among these youngest bins.
For ages between 2 and 6 Gyr, where the sample is dominated by thin-disk stars, $f_b$ increases steadily from $\sim 0.17$ at 2 Gyr at an average rate of  0.03-0.04 per Gyr. The growth slows beyond 6 Gyr, and $f_b$ reaches its maximum of $\sim 0.31$ at $\sim8$~Gyr. At older ages, where the fraction of thick-disk stars increases, $f_b$ declines slightly from the peak value and exhibits no further significant increase with age.

We note that this result requires careful interpretation, given the nature of the age estimates we used largely reflect a chemical clock, encoding the temporal evolution of abundance patterns in the Galactic disk \citep{Wangjh2025Ages}. Consequently, the correlations between binary fraction and age would essentially trace a relationship between binarity and the integrated chemical enrichment history. The modest and non-monotonic variation seen in Figure~\ref{fig:fb_age} likely arises from the combined effects of the underlying age-[Fe/H]-[$\alpha$/Fe] correlations and the contrasting binary-fraction trends with individual $\alpha$ elements discussed in Section~\ref{sec:alpha_abund}.

\section{Discussion} \label{sec:discussion}

\subsection{Implication} \label{sec:implication}

Despite extensive investigation, the detailed shape and parameter values of the mass-ratio distribution for the field binary systems remain subject to uncertainties (e.g., \citealp{Moe2017PandQ,ElBadry2019twinbinary,2022Li}). While a correlation between the stellar initial mass function (IMF) and metallicity has been indicated \citep[e.g.,][]{2023Li, 2012Marks}, and the IMF is naturally linked to the binary $q$ distribution, only \citet{LiuC2019SmokingGun} has systematically explored the relationship between the power-law index $\gamma$ and metallicity. It is therefore natural to expect that metallicity, through its influence on the thermodynamic and dynamical conditions of star formation, should also play a role in shaping the $q$ distribution of binaries. Motivated by this, we measure the $\gamma$ and investigate its dependence on metallicity.

\citet{LiuC2019SmokingGun} studies the $q$ distribution of nearby solar-type main-sequence stars and fits it with a single power law, $p(q) \propto q^{\gamma}$. They found that among low-$\gamma$ groups ($\gamma\,<\,1.2$), $\gamma$ decreases with increasing metallicity, implying that relatively metal-rich populations contain fewer high-$q$ binaries. In this work, we adopt the more flexible broken power-law model of \citet{Moe2017PandQ} and allow the high-$q$ power-law index $\gamma_{{\rm large}q}$ to vary. We find that $\gamma_{{\rm large}q}$ exhibits a trend with [Fe/H]: in the most metal-poor bins of thick-/thin- disk subsample ($-1 < {\rm [Fe/H]} < -0.7$ dex), $\gamma_{{\rm large}q}$ lies around $-1.25$; at moderately metal-poor bin ($-0.5 < {\rm [Fe/H]} < 0$ dex), $\gamma_{\rm large\,q} \sim -2.25$; and at metallicities above solar ($0.2 < {\rm [Fe/H]} < 0.5$ dex), it drops further to $\sim -2.5$.

This trend is qualitatively consistent with the anticorrelation reported by \citet{LiuC2019SmokingGun} for their low-$\gamma$ groups. During the star-formation phase, metal-poor environments tend to produce more massive stars \citep{2023Li}. When molecular cloud cores are optically thick, the more rapid cooling enabled by low metallicity enhances fragmentation of the protostellar disk, facilitating the formation of close binaries \citep{Moe2019STclosebinary, Bate2019StarsMetallicitySim}. In this process, disk fragmentation itself may favor the production of binary fragments with relatively high mass ratios. Alternatively, close binaries at early evolutionary stages are prone to mass transfer, with the secondary accreting mass from the primary, which systematically elevates their mass ratios. After star formation, dynamical disruption in clustered environments further reshapes the $q$ distribution. In higher-metallicity environments, stronger stellar winds and radiation feedback from massive stars could more efficiently reduce the local stellar density, allowing low-$q$ binaries with smaller binding energies to survive more easily. 

We also note that the mass-ratio distribution may depend on the orbital period 
\citep{Moe2017PandQ}. Consequently, differences in the period distribution among different subsamples or [Fe/H] bins could affect the inferred $q$ distribution. Nevertheless, the distance distributions of the thin- and thick-disk subsamples, as well as those across different [Fe/H] bins, are similar (see Figure~\ref{fig:stellarparams}); the accessible period ranges are therefore expected to be comparable.  This should suppress the impact of period-distribution differences on our results.

Our results provide independent evidence that the binary mass-ratio distribution correlates with metallicity, in the sense that the fraction of low-mass-ratio binaries is systematically suppressed in metal-poor populations. This trend is likely real, although an unbiased quantitative characterization of it and its underlying physical origin require further investigations.

The relationship between the binary fraction and $\alpha$-process element abundances remains relatively unexplored. The two most recent studies, by \citet{Niuzx2021FGK} and \citet{Mazzola2020closebinary}, reached partially conflicting conclusions. We note that the two works adopted different definitions of $\alpha$ enhancement: \citet{Niuzx2021FGK} used a weighted mean of several $\alpha$ elements, while \citet{Mazzola2020closebinary} focused on individual abundances. This discrepancy suggests that the net observed trend may reflect a mixture of effects from different elements. To disentangle these contributions, we have performed a cross-analysis of the binary fraction as a function of the individual abundances. Consistent with \citet{Mazzola2020closebinary}, we find an anticorrelation between $f_{\rm b}$ and both Mg and Si abundances. Moreover, we find a positive correlation between $f_b$ and the Ca abundance, which is independent of variations in Mg, Si, and Fe. This positive effect is stronger than the negative effects from Mg and Si, which likely accounts for the net positive correlation between $f_b$ and the averaged [$\alpha$/H] reported by \citet{Niuzx2021FGK}.

Mg and Si are the principal constituents of silicate dust grains \citep{Zhukovska2018FeSidust}, and their abundances can influence the opacity and cooling rates in protostellar disks, thereby affecting the conditions for disk fragmentation and binary formation \citep{2014Tanaka,Moe2019STclosebinary}. There are few studies investigating the role of Ca in the protostellar stage. Our results suggest a distinct behavior of Ca and it may indicate a different channel than Mg and Si. A comprehensive understanding of the role of calcium in shaping the binary fraction needs further astrochemical investigations.

Furthermore, we have examined the correlation between $f_b$ and [$\alpha$/H] separately for the thin-disk and thick-disk populations. As reported by both \citet{Mazzola2020closebinary} and \citet{Niuzx2021FGK}, the binary fraction among the high-$\alpha$ thick-disk stars shows no significant dependence on $\alpha$-element abundances. Our results are fully consistent with these findings. This independence is also reflected in the age-$f_b$ relation: at ages $\log(\tau/\mathrm{yr}) \gtrsim 9.8$, where the thick-disk population becomes dominant, the binary fraction remains nearly constant. The strong correlation between $\alpha$ abundance and age also breaks down in the thick-disk regime \citep{2020Huang}.

Several physical processes may account for this behavior. First, older stars may have experienced more extensive dynamical processing \citep{2011Marks}. Such processing could erase the birth signature of the binarity and decorrelate it from chemical composition.
In addition, over timescales of several Gyr, the short-period FGK binaries can initiate mass transfer and subsequently merge \citep{2008Gazeas}, producing an apparently single remnants, including WDs (e.g., \citealp{Temmink2020deceiving}). Because this depletion accumulates with time, older populations are expected to have lost a larger fraction of their binaries. This effect could contribute to the slowdown of the $f_b$ growth beyond $\sim$6 Gyr, its maximum at around 8 Gyr, and its mild decline at ages of 8–10 Gyr.
Second, the SLOT method responds to the flux of unresolved companions in the Gaia XP bands, from short-period unresolved binaries up to those at the angular-resolution limit of Gaia (see Section~\ref{sec:potential_caveats}).
The thick-disk binary population may contain a higher fraction of wide binary systems formed via core fragmentation rather than close binary systems formed via disk fragmentation; in the core-fragmentation channel, the binary fraction is no longer correlate strongly to the metallicity \citep{Bate2019StarsMetallicitySim,2023Guszejnov}. 
A third possibility relates to the debated formation channels of the Galactic thin and thick disks (e.g. \citealt{2023Spitoni,2021Johnson,2019Clarke}). Our results tentatively suggest that thin-disk binary formation occurred more gradually, allowing the chemical enrichment history to imprint itself on the binary population, whereas thick-disk binary formation may have proceeded more rapidly \citep{2022Xiang,2022Conroy}, yielding a more uniform binary fraction that demonstrates weaker correlation with the chemical enrichment history.

\subsection{Potential Caveats} \label{sec:potential_caveats}

\begin{figure}[htbp]
    \centering
    \includegraphics[width=1.0\columnwidth]{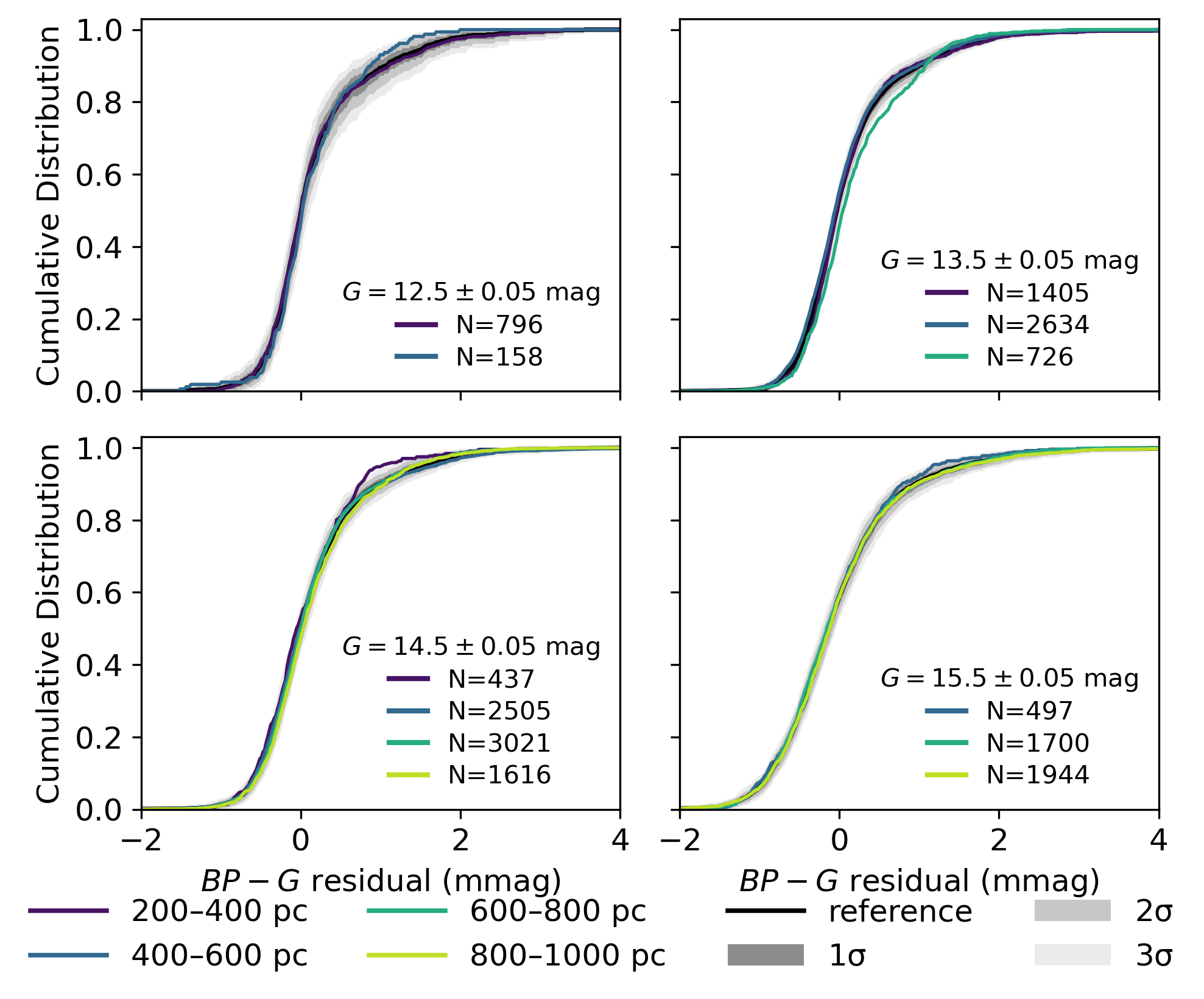}
    \caption{CDFs of the $BP-G$ residuals of the binarity-estimation sample in four representative subsamples, as labeled in the lower right corner of each panel. 
In each panel, the colored curves show CDFs for different distance bins (200--400~pc, 400--600~pc, 600--800~pc, and 800–1000~pc), and the black curve shows the reference CDF constructed from all stars in that $G$-magnitude subsample. Grey shaded regions, from dark to light, denote the $1\sigma$, $2\sigma$ and $3\sigma$ uncertainty envelopes obtained from random resamplings of the reference subsample.
    \label{fig:rescdf}
    }
\end{figure}

To investigate the possible bias caused by distance-dependent incompleteness, we construct subsamples with limited $G$-magnitude ranges to control photometric errors, and divide each subsample into distance bins (200--400, 400--600, 600--800, and 800--1000~pc). We then compare the $BP-G$ residual distributions across distance bins to examine whether the binary contribution varies with distance. Figure~\ref{fig:rescdf} shows the cumulative distribution functions (CDFs) for four representative $G$-magnitude subsamples. Most CDFs lie within the $3\sigma$ envelopes; the few deviations beyond $3\sigma$ occur in subsamples with very small sample sizes and mostly at residual levels outside the sensitive range of the SLOT method (1--2 mmag; see Figure~\ref{fig:resgaussian_scol}).
We also carried out two additional tests. Within 400\,pc ($28{,}475$ sources) , the sample is mostly thin-disk star with relatively higher metallicity (see Figure\,\ref{fig:stellarparams}). This yields $f_b\,=\,0.35\,\pm\,0.02$/$0.22\,\pm\,0.01$ and $\gamma_{\mathrm{large}q}\,=\,-2.0\,\pm 0.5$/$-2.5\,\pm\,0.5$ at [Fe/H]\,=\,$-0.25$/0.25\,dex, and the $f_b$–[Fe/H]/age as well as $\gamma_{\mathrm{large}q}$-[Fe/H] relations are preserved within 1$\sigma$ uncertainty. Similarly, raising the vertical cut to $|Z|\,>\,0.3\,$kpc  ($142{,}499$ sources) gives  $f_b\,=\,0.31\,\pm\,0.03$/$0.26\,\pm\,0.03$ and $\gamma_{\mathrm{large}q}\,=\,-1.50\,\pm 0.50$/$-2.00\,\pm\,0.50$ at [Fe/H]\,=\,$-0.5$/0.0\,dex, with the same trends preserved. Therefore, the distance-dependent selection effect is negligible.

\begin{figure}[htbp]
    \centering
    \includegraphics[width=0.98\columnwidth]{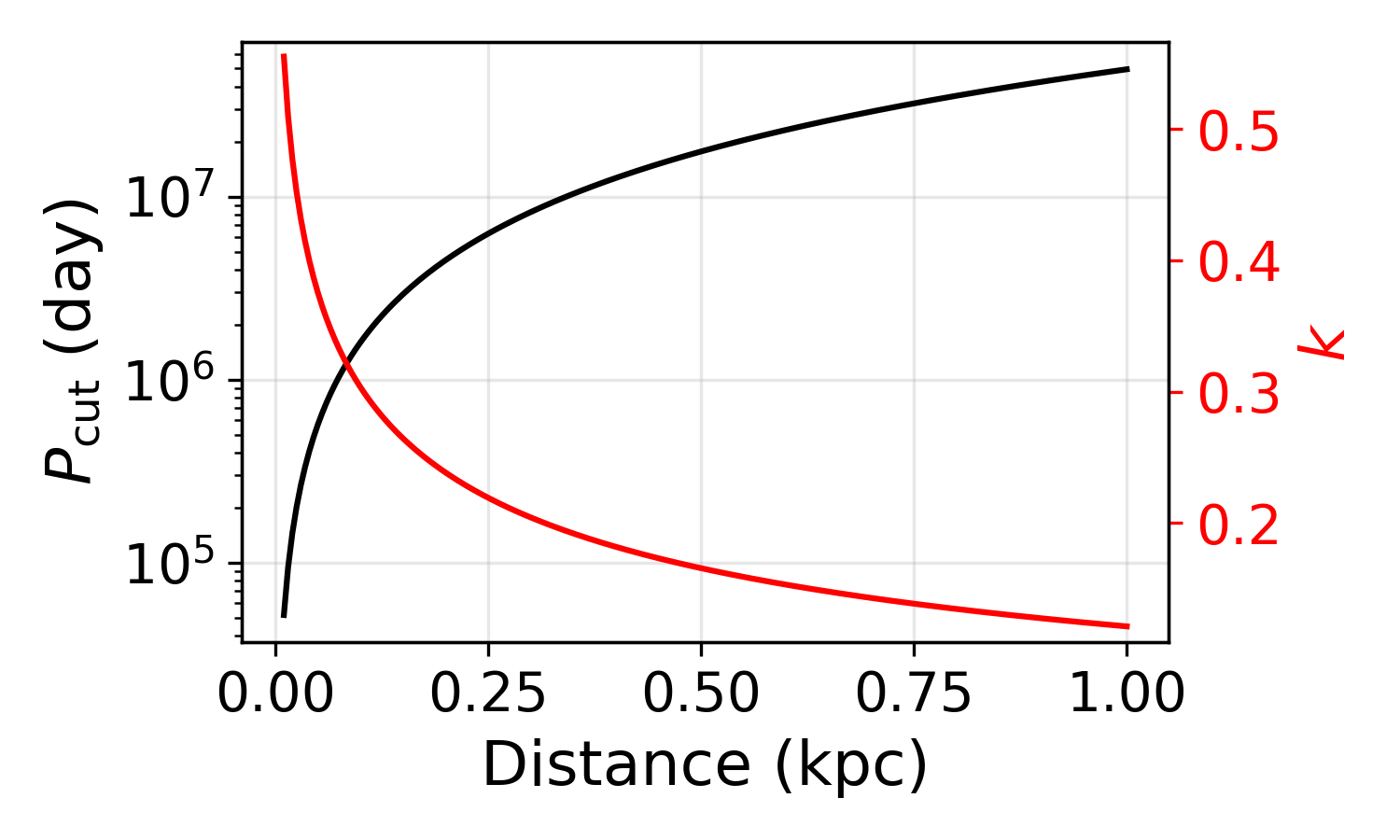}
    \caption{
    The cutoff orbital periods $P_{\rm cut}$ (which serves as the upper limits in our estimation) and the fraction $k$ of resolved binary systems that are misclassified as single stars by Gaia, both as functions of distance.
    \label{fig:resolved_frac_uncounted}
    }
\end{figure}

The SLOT method identifies unresolved photometrically detectable binary fraction. However, although the SLOT method does not require an explicit orbital-period distribution when modeling unresolved sources, the measured $f_{\rm b}$ is still limited by Gaia's angular resolution, the distance distribution of the sample, and the sensitivity of the color residuals to the companion mass. Gaia has an angular resolution of $\sim 2\arcsec$ \citep{Arenou2018GaiaDR2}, meaning that binaries with angular separations $\theta > 2\arcsec$ are likely to be recognized as two separate single stars. This leads to an incompleteness in our binary fraction estimates for wide systems. Following \citet{Niuzx2021FGK}, the corrected binary fraction can be expressed as

\begin{equation}\label{eq:fb_corr}
f_{b,\mathrm{corr}} = f_b^{\mathrm{SLOT}} \times \frac{1}{1 - k},
\end{equation}

where $k$ is the fraction of resolved systems missed by our analysis, shown as a function of distance in Figure~\ref{fig:resolved_frac_uncounted}. To estimate $k$, we adopt the log-normal period distribution of \citet{Raghavan2010survey}, with $\mu_{\log P/{\rm day}} = 5.03$ and $\sigma_{\log P/{\rm day}} = 2.28$, and assume a total system mass of $1\,M_{\odot}$.

We compute $\Delta f_b = f_{b,\mathrm{corr}} - f_b^{\mathrm{SLOT}}$ for the distance distribution within each bin. The resulting $\Delta f_b$ values are nearly constant across all bins, with a median of 0.07 and a small standard deviation of 0.01. We report $f_b^{\mathrm{SLOT}}$ throughout this work. 

Note that the adopted log-normal period distribution was measured from field stars within 25~pc, mostly thin-disk, and its applicability to the thick-disk population is not observationally constrained. 
The near-constancy of  $\Delta f_b$  is derived under this single, shared period distribution, and it should not be interpreted as evidence that population-specific period distributions would yield identical corrections.

In addition, we also explore the fraction of the binaries in our sample that are sufficiently close to undergo mass transfer. We ran a Monte Carlo simulation with $N=100,000$ draws. Primary masses $M_1$ are sampled from our sample for binarity estimation, which ranges $0.51$--$1.17~M_{\odot}$. Mass ratios are drawn from Figure\,\ref{fig:fb_q_feh_combine}. Orbital periods are drawn from the log-normal distribution of \citet{Raghavan2010survey}, and circular orbits are assumed. For each simulated system, we compute the Roche lobe radius of the primary star $R_{L1}$ based on Equation\,(2) in \citet{Eggleton1983RL} and compare it to the maximum stellar radius $R_{\rm max}$, and the system is flagged as interacting if $R_{L1} < R_{\rm max}$. For the maximum stellar radius $R_{\rm max}$, we adopt the radii of the tips of the red-giant branch (TRGBs), which has very weak dependence on the initial mass \citep{Hurley2000stellarevolution}. This guarantees Case~A and Case~B mass transfer \citep{Kippenhahn1967masstransfer}. The simulation yields an interacting fraction of about $17.8\%$. Varying $R_{\rm max}$ between 150 and 220~$R_{\odot}$ changes the result by only $\sim \pm 2\%$. The period distribution of \citet{Raghavan2010survey} may not apply to thick-disk stars. This estimate should therefore be taken as a rough order-of-magnitude guide rather than a precise prediction.

Our mock catalogs contain only single and binary systems, yet the asymmetric color residuals exploited by the SLOT method can, in principle, also arise from higher-order multiples. In practice, however, the impact of such systems on our results is negligible. The contribution of a tertiary component to the $BP-G$ residual is estimated to be less than $1$\,mmag. This shift is comparable to the precision of Gaia XP synthetic photometry and within the intrinsic width of the stellar locus; at this level, fitting the residual distribution with binaries or with higher-order multiples yields indistinguishable results. 
Furthermore, only $\sim 25\%$ of multiple systems are of order higher than two \citep{Raghavan2010survey}, which further limits any potential bias. We therefore treat all multiples as binaries in our model, as the additional contribution from higher-order components is negligible at the precision of our measurement.

Figure\,\ref{fig:contour_total} exhibits two ``gaps'' at primary masses below $0.7\,M_{\odot}$.
We speculate that several factors may contribute to these ``gaps''. One is that our training sample lacks low-mass stars at [Fe/H] $\ge 0$~dex (as shown in Figure\,\ref{fig:stellarparams}), potentially resulting in an imperfect fit of the stellar loci. Another is that the masses adopted here are derived from the PARSEC color-mass relation, which may be subject to grid effects.
Thus, our method may underestimate the binary effect on the color residuals and consequently overestimate the binary fraction for those low-mass stars.

The stellar parameters we employed in this work assume that one spectrum is contributed by one single star, neglecting contributions of companion stars in multiple systems. Therefore systematic biases are introduced in spectral fitting. \citet{ElBadry2018Signatures} evaluates such effect in different types of spectra. In LAMOST-like spectra, the typical systematic errors are $\sim 100$ K for $T_{\rm eff}$, $\sim 0.05$ dex for $\log g$, [Fe/H] and [$\alpha$/Fe] within our temperature ranges \citep[see Figure\,2 in][]{ElBadry2018Signatures}, which are ignorable compared to the accuracy of the stellar parameters in our sample.
There have been successful attempts on the decomposition of binary-star spectra. \citet{Lijd2025MSbinary} apply data-driven approach to modeling the XP spectra and construct a catalog of likely main-sequence binary stars systems. They derive photometric metallicities based on both the single-star model and the binary-star model, respectively. However, our sample share only thousands of sources with their catalog. Therefore, we still employ the stellar parameters derived from single-star models from \citet{Zhangm2025elements}. We expect future researches could bring more accurate stellar parameters of better spectra decomposition and test our results.

\section{Summary} \label{sec:summary}

In this work, we investigate the binary fraction and mass-ratio distribution for 181,011 FGK dwarfs within 1\,kpc using the SLOT method. Our sample is constructed from Gaia DR3 and LAMOST DR9, incorporating carefully recalibrated and dereddened Gaia synthetic photometry \citep{Zhangry2023dered, Huangbw2024xp}, geometric distances \citep{BJ2021distances}, and stellar parameters derived from data-driven approaches \citep{Zhangm2025elements, Wangjh2025Ages}. We also construct a dedicated training sample to fit the metallicity-dependent $(G_{\rm BP}-G)$ color locus and $M_G$ absolute magnitude locus. The SLOT method is implemented as detailed in Section\,\ref{sec:slot_overview}.

We examine the binary fraction across the metallicity space, in multi-dimensional $\alpha$-element abundance spaces, and as a function of stellar age. The mass-ratio distribution is investigated in the metallicity space. In addition to confirming well-established trends, such as the decline of the binary fraction with increasing [Fe/H], we report the following results:

\begin{enumerate}
    \item We find that the power-law index of the mass-ratio distribution likely depends on metallicity. Higher [Fe/H] yields a more negative $\gamma_{\rm large\,q}$, indicating a mass-ratio distribution increasingly skewed toward low $q$ (i.e., a deficit of near-equal-mass binaries). This trend is qualitatively consistent with the anticorrelation reported by \citet{LiuC2019SmokingGun} for their low-$\gamma$ group.
    \item Among $\alpha$-process elements, [Mg/H] and [Si/H] are anticorrelated with the binary fraction in the thin-disk-dominated subsamples; these trends, however, vanish in thick-disk-dominated subsamples.
    \item A positive correlation between $f_{\rm b}$ and [Ca/H] is found in thin-disk-dominated subsamples. This effect is stronger than the opposite trends produced by Mg and Si. 
    \item We examine $f_{\rm b}$ as a function of stellar age over the range $1$--$10$\,Gyr. The binary fraction increases modestly toward older ages, whereas no significant variation is observed at the regime of the oldest ages.
\end{enumerate}

Taken together, these results highlight the complex, nonlinear relationships among the binary fraction, the mass-ratio distribution, and multiple chemical abundance dimensions. Such complexity suggests that the binary populations we observe today have been shaped by a variety of physical processes, and that the relative importance of these processes differs markedly between the thin- and thick-disk populations. We anticipate that future studies targeting the mechanisms of binary formation and disruption will provide a more fundamental understanding of these trends.

\begin{acknowledgments}

We thank the reviewer for the suggestions on this article.

We sincerely acknowledge Prof. Maosheng Xiang for discussion and instruction about our work.

This work made use of the data from LAMOST (Large Sky Area Multi-Object Fiber Spectroscopic Telescope, also known as the Guoshoujing Telescope) (\url{https://cstr.cn/31118.02.LAMOST}). LAMOST is a Chinese national mega-science facility, operated by National Astronomical Observatories, Chinese Academy of Sciences.

This work is supported by the National Natural Science Foundation of China (NSFC), grants No. 12588202 and 12303039.
J.F.L. acknowledges support from the New Cornerstone Science Foundation through the New Cornerstone Investigator Program and the XPLORER PRIZE.

\end{acknowledgments}

\facilities{Gaia, LAMOST}

\software{Python, TOPCAT \citep{Taylor2005TOPCAT}}

\appendix

\twocolumngrid

\section{Fitting Results for the Stellar Loci}\label{sec:AP_stellarlocifitting}

This appendix provides the fitting results for both $BP-RP\,=\,f(BP-RP,\,{\rm [Fe/H]})$ and $M_{G}\,=\,f(BP-RP,\,{\rm [Fe/H]})$, as described in Section~\ref{sec:stellar_loci}. Table~\ref{tab:bpg_coeff} lists the polynomial coefficients for the form in Equation~\eqref{eq:bpg}. In this parameterization, $BP-RP$ is in mag, while the fitted $BP-G$ is given in mmag. As discussed in the main text, the color-dependent term $f_2$ is obtained via interpolation rather than a global polynomial fit; its numerical values are tabulated in Table~\ref{tab:bpgf2}. Table~\ref{tab:mg_coeff} lists the polynomial coefficients for $M_{G}\,=\,f(BP-RP,\,{\rm [Fe/H]})$.

\begin{deluxetable}{cccc}[ht!]
\tablecaption{Fitting coefficients of the metallicity-dependent color-color loci for dwarf stars. \label{tab:bpg_coeff}}
\tablewidth{0pt}
\tabletypesize{\normalsize}
\setlength{\tabcolsep}{12pt}
\tablehead{
\colhead{$a$} & \colhead{Value} & \colhead{$c$} & \colhead{Value}
}
\startdata
$a_1$ & $-0.232$ & $c_1$ & $10.668$ \\
$a_2$ & $0.681$  & $c_2$ & $61.407$ \\
$a_3$ & $-0.634$ & $c_3$ & $-0.195$ \\
$a_4$ & $-0.506$ &  &  \\
\enddata
\end{deluxetable}

\begin{deluxetable}{cccccc}[ht!]
    \tabletypesize{\normalsize}
    \tablewidth{0pt}
    \tablecaption{Interpolated values of the color-dependent term $f_2$ \label{tab:bpgf2}}  
    \tablehead{
    \colhead{\textit{Color1}} & \colhead{\textit{Color2}} & \colhead{\textit{Color1}} & 
    \colhead{\textit{Color2}} & \colhead{\textit{Color1}} & \colhead{\textit{Color2}}
    }
    \startdata
    0.30 & 3.211 & 1.30 & 0.581 & 2.30 & 1.216 \\
    0.35 & 0.522 & 1.35 & 0.610 & 2.35 & 1.251 \\
    0.40 & 0.144 & 1.40 & 0.638 & 2.40 & 1.289 \\
    0.45 & 0.164 & 1.45 & 0.667 & 2.45 & 1.324 \\
    0.50 & 0.184 & 1.50 & 0.697 & 2.50 & 1.360 \\
    0.55 & 0.205 & 1.55 & 0.726 & 2.55 & 1.395 \\
    0.60 & 0.227 & 1.60 & 0.757 & 2.60 & 1.430 \\
    0.65 & 0.249 & 1.65 & 0.788 & 2.65 & 1.465 \\
    0.70 & 0.271 & 1.70 & 0.818 & 2.70 & 1.500 \\
    0.75 & 0.295 & 1.75 & 0.849 & 2.75 & 1.535 \\
    0.80 & 0.319 & 1.80 & 0.881 & 2.80 & 1.570 \\
    0.85 & 0.343 & 1.85 & 0.912 & 2.85 & 1.605 \\
    0.90 & 0.368 & 1.90 & 0.944 & 2.90 & 1.640 \\
    0.95 & 0.393 & 1.95 & 0.977 & 2.95 & 1.675 \\
    1.00 & 0.419 & 2.00 & 1.010 & 3.00 & 1.710 \\
    1.05 & 0.445 & 2.05 & 1.043 & 3.05 & 1.745 \\
    1.10 & 0.472 & 2.10 & 1.077 & 3.10 & 1.780 \\
    1.15 & 0.499 & 2.15 & 1.111 & 3.15 & 1.814 \\
    1.20 & 0.526 & 2.20 & 1.146 & 3.20 & 1.849 \\
    1.25 & 0.553 & 2.25 & 1.181 & 3.25 & 1.884 \\
    \enddata
    \tablecomments{\textit{Color1} and \textit{Color2} correspond to $BP-RP$ and $BP-G$, respectively, both in mag.}
\end{deluxetable}

\begin{deluxetable}{crrcrr}[ht!]
\tablecaption{Fitting coefficients of the metallicity-dependent color-absolute-magnitude loci. \label{tab:mg_coeff}}
\tablewidth{0pt}
\tabletypesize{\normalsize}
\tablehead{
\colhead{Term} & \colhead{Blue} & \colhead{Red} & \colhead{Term} & \colhead{Blue} & \colhead{Red}
}
\startdata
$x^4$    & $3.002$   & -         & $x^2$ & $18.320$  & $0.794$ \\
$x^3y$   & $-1.460$  & -         & $xy$  & $-5.913$  & $1.627$ \\
$x^2y^2$ & $-0.191$  & -         & $y^2$ & $-1.678$  & $0.286$ \\
$xy^3$   & $0.410$   & -         & $x$   & $-5.074$  & $0.895$ \\
$y^4$    & $0.808$   & -         & $y$   & $1.134$   & $-1.852$ \\
$x^3$    & $-13.073$ & $-0.111$  & $C$   & $2.373$   & $4.352$ \\
$x^2y$   & $5.299$   & $-0.587$ \\
$xy^2$   & $1.537$   & $-0.066$ \\
$y^3$    & $0.137$   & $0.065$ \\
\enddata
\tablecomments{$x$ denotes the $BP-RP$ color in mag; $y$ denotes the metallicity [Fe/H]. The fitted $M_{G}$ is given in mag.}
\end{deluxetable}

\section{The systematic zero-point error} \label{sec:AP_bpg_syserr}

We empirically derive the systematic zero-point error $\sigma_{\rm ZP,sys}$ using RGB stars, following the approach of \citet{Niuzx2021FGK}. The RGB sample is constructed from the catalog of \citet{Li2022giantparams}, crossmatched with Gaia DR3, and cleaned with the same quality constraints as in \citet{Niuzx2021correction}. This RGB sample contains 57,535 sources. The metallicity-dependent color–color locus is fitted with the functional form given by \citet{Huangbw2024xp}:

\begin{equation}
\begin{split}
f_{{\rm RGB,}1} &= (u_1x^3 + u_2x^2+u^3x+u_4)*(y^3 -3ky^2 + 3k^2y)\\
f_{{\rm RGB,}2} &= v_1x^5 + v_2x^4 + v_3x^3 + v_4x^2 + v_5x + v_6
\end{split}
\end{equation}

where $x$ denotes the $BP-RP$ color in mag and $y$ denotes the [Fe/H] in dex, and $k~=~-5.25$. The fitted $BP-G$ color has the unit of mmag. The fitting coefficients are listed in Table\,\ref{tab:RGBcolorloci}.

\begin{deluxetable}{cccc}[ht!]
\tablecaption{Fitting coefficients of the metallicity-dependent color-color loci for RGB stars. \label{tab:RGBcolorloci}}
\tablewidth{0pt}
\tabletypesize{\normalsize}
\setlength{\tabcolsep}{12pt}
\tablehead{
\colhead{$u$} & \colhead{Value} & \colhead{$v$} & \colhead{Value}
}
\startdata
$u_1$ & $0.202$ & $v_1$ & $-13.500$ \\
$u_2$ & $-0.807$  & $v_2$ & $82.302$ \\
$u_3$ & $0.994$ & $v_3$ & $-210.301$ \\
$u_4$ & $-0.331$ & $v_4$ & $357.937$ \\
 & & $v_5$ & $170.972$ \\
 & & $v_6$ & $30.934$ \\
\enddata
\end{deluxetable}

\begin{figure}[htbp]
    \centering
    \includegraphics[width=0.98\columnwidth]{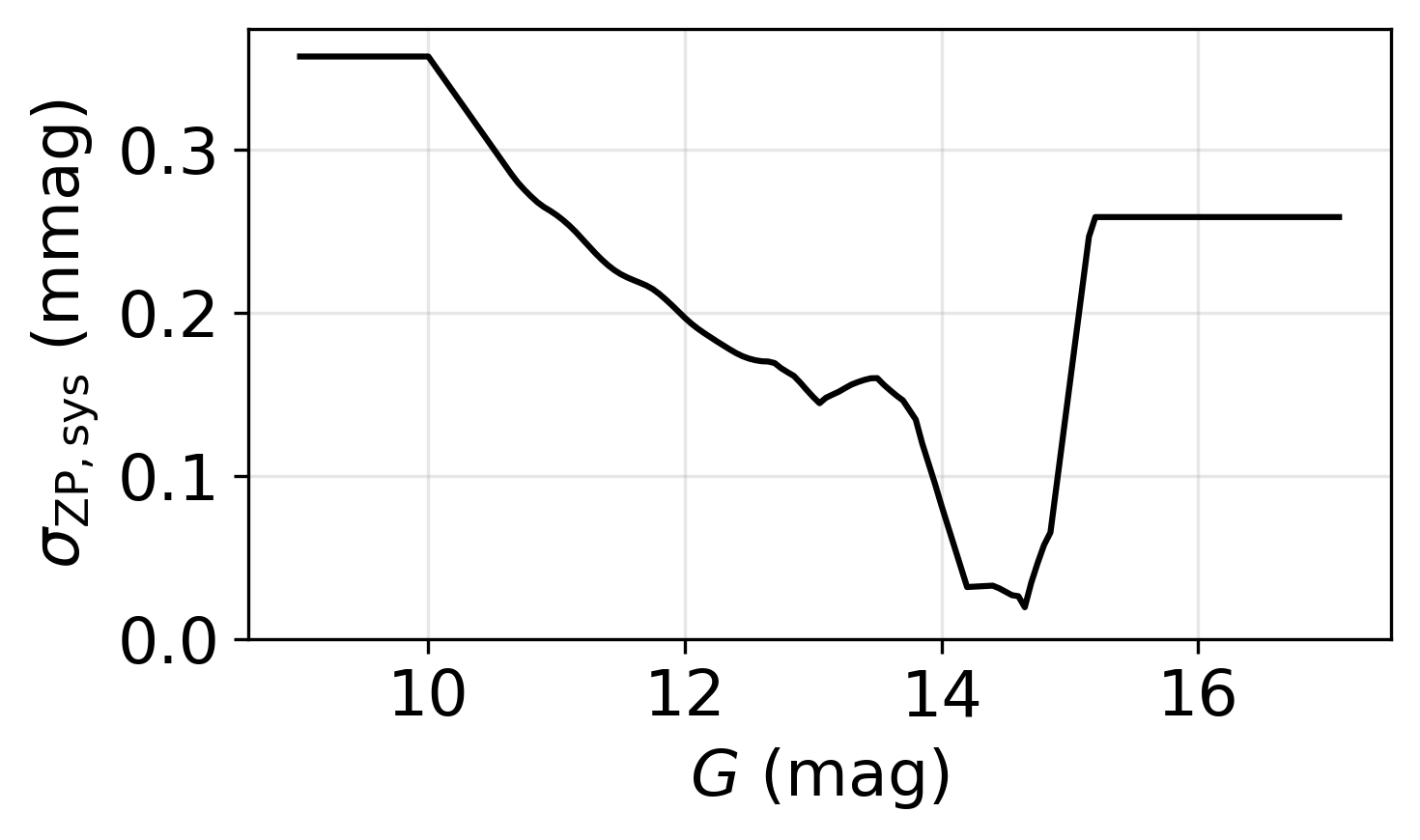}
    \caption{
    The systematic zero-point error in $BP-RP$ color as a function of $G$ magnitude.
    \label{fig:bpg_syserr}
    }
\end{figure}

Because the observing configurations of the Gaia satellite depends on source brightness, $\sigma_{\rm ZP,sys}$ is naturally expected to correlate with $G$ magnitude. We therefore divide the RGB sample into $G$-mag bins, and in each bin we generate a mock Single Set following the procedure of Figure~\ref{fig:sim_flowchart} and Section~\ref{sec:method}, and determine $\sigma_{\rm ZP,sys}$ by fitting the observed color residuals. The resulting optimized values are shown in Figure~\ref{fig:bpg_syserr}. 
With the calibrated Gaia XP synthetic photometry the systematic error is below 0.5\,mmag, which is significantly smaller than the range of $0.5$–$1.5$\,mmag, obtained from Gaia EDR3 broad‑band photometry \citep{Niuzx2021FGK}.

\section{Complete fitting results for all metallicity Bins}\label{sec:AP_fullresults}

This appendix provides the complete fitting results and model-averaging diagnostics that underlie the results presented in Section~\ref{sec:results}. For each [Fe/H] bin and each assumed $\gamma_{\mathrm{large}q}$, we list the fitted $f_b$, $\mu$, and reduced $\chi^2$, together with the models retained in the compatible group. The complete result are provided as a machine-readable table in the online journal. The first 10 rows are shown in Table~\ref{tab:SLOT_allmodels}.

\begin{deluxetable*}{lcrrrrc}
\tablecaption{Fitting results for all model grid values\label{tab:SLOT_allmodels}}
\tablehead{
\colhead{Disk} & \colhead{[Fe/H] bin} & \colhead{$\gamma_{{\rm large}q}$} &
\colhead{$f_b$} & \colhead{$\mu$} & \colhead{$\chi^2_{\rm red}$} & \colhead{Compatible group}
}
\startdata
Thin & -0.75 & -3.0 & 0.82 & -0.863 & 1.219 & 1 \\
Thin & -0.75 & -2.5 & 0.79 & -0.884 & 1.250 & 1 \\
Thin & -0.75 & -2.0 & 0.78 & -0.896 & 1.302 & 1 \\
Thin & -0.75 & -1.5 & 0.77 & -0.899 & 1.275 & 1 \\
Thin & -0.75 & -1.0 & 0.73 & -0.895 & 1.293 & 1 \\
Thin & -0.75 & -0.5 & 0.69 & -0.864 & 1.334 & 1 \\
Thin & -0.75 &  0.0 & 0.74 & -0.926 & 1.427 & 1 \\
Thin & -0.65 & -3.0 & 0.53 & -0.409 & 1.222 & 0 \\
Thin & -0.65 & -2.5 & 0.51 & -0.410 & 1.067 & 0 \\
Thin & -0.65 & -2.0 & 0.48 & -0.404 & 1.023 & 1 \\
\enddata
\tablecomments{The last column is 1 if the $\gamma_{\rm large,q}$ value belongs to the compatible group and 0 otherwise (see Section~\ref{sec:results}). The complete version of this table (168 rows) is available in machine-readable form in the online journal.}
\end{deluxetable*}

\bibliography{GaiaXP_fb_FGK}{}

@ARTICLE{Niuzx2021FGK,
       author = {{Niu}, Zexi and {Yuan}, Haibo and {Wang}, Song and {Liu}, Jifeng},
        title = "{Binary Fractions of G and K Dwarf Stars Based on Gaia EDR3 and LAMOST DR5: Impacts of the Chemical Abundances}",
      journal = {\apj},
         year = 2021,
        month = dec,
       volume = {922},
       number = {2},
          eid = {211},
        pages = {211},
          doi = {10.3847/1538-4357/ac2573},
archivePrefix = {arXiv},
       eprint = {2109.04031},
 primaryClass = {astro-ph.SR},
       adsurl = {https://ui.adsabs.harvard.edu/abs/2021ApJ...922..211N}
}

@ARTICLE{2008Gazeas,
       author = {{Gazeas}, K. and {St{\c{e}}pie{\'n}}, K.},
        title = "{Angular momentum and mass evolution of contact binaries}",
      journal = {\mnras},
         year = 2008,
        month = nov,
       volume = {390},
       number = {4},
        pages = {1577-1586},
          doi = {10.1111/j.1365-2966.2008.13844.x},
archivePrefix = {arXiv},
       eprint = {0803.0212},
 primaryClass = {astro-ph},
       adsurl = {https://ui.adsabs.harvard.edu/abs/2008MNRAS.390.1577G}
}

@ARTICLE{Niuzx2023calibration,
       author = {{Niu}, Zexi and {Yuan}, Haibo and {Liu}, Jifeng},
        title = "{Internal Calibration of LAMOST and Gaia DR3 GSP-Spec Stellar Abundances}",
      journal = {\apj},
         year = 2023,
        month = jun,
       volume = {950},
       number = {2},
          eid = {104},
        pages = {104},
          doi = {10.3847/1538-4357/accf8b},
archivePrefix = {arXiv},
       eprint = {2304.13951},
 primaryClass = {astro-ph.SR},
       adsurl = {https://ui.adsabs.harvard.edu/abs/2023ApJ...950..104N}
}

@ARTICLE{Niuzx2021correction,
       author = {{Niu}, Zexi and {Yuan}, Haibo and {Liu}, Jifeng},
        title = "{Correction to the Photometric Colors of the Gaia Data Release 2 with the Stellar Color Regression Method}",
      journal = {\apj},
         year = 2021,
        month = mar,
       volume = {909},
       number = {1},
          eid = {48},
        pages = {48},
          doi = {10.3847/1538-4357/abdbac},
archivePrefix = {arXiv},
       eprint = {2101.04290},
 primaryClass = {astro-ph.SR},
       adsurl = {https://ui.adsabs.harvard.edu/abs/2021ApJ...909...48N}
}

@ARTICLE{GaiaCollab2023Summary,
       author = {{Gaia Collaboration} and {Vallenari}, A. and {Brown}, A.~G.~A. and {Prusti}, T. and {de Bruijne}, J.~H.~J. and {Arenou}, F. and {Babusiaux}, C. and {Biermann}, M. and {Creevey}, O.~L. and {Ducourant}, C. and {Evans}, D.~W. and {Eyer}, L. and {Guerra}, R. and {Hutton}, A. and {Jordi}, C. and {Klioner}, S.~A. and {Lammers}, U.~L. and {Lindegren}, L. and {Luri}, X. and {Mignard}, F. and {Panem}, C. and {Pourbaix}, D. and {Randich}, S. and {Sartoretti}, P. and {Soubiran}, C. and {Tanga}, P. and {Walton}, N.~A. and {Bailer-Jones}, C.~A.~L. and {Bastian}, U. and {Drimmel}, R. and {Jansen}, F. and {Katz}, D. and {Lattanzi}, M.~G. and {van Leeuwen}, F. and {Bakker}, J. and {Cacciari}, C. and {Casta{\~n}eda}, J. and {De Angeli}, F. and {Fabricius}, C. and {Fouesneau}, M. and {Fr{\'e}mat}, Y. and {Galluccio}, L. and {Guerrier}, A. and {Heiter}, U. and {Masana}, E. and {Messineo}, R. and {Mowlavi}, N. and {Nicolas}, C. and {Nienartowicz}, K. and {Pailler}, F. and {Panuzzo}, P. and {Riclet}, F. and {Roux}, W. and {Seabroke}, G.~M. and {Sordo}, R. and {Th{\'e}venin}, F. and {Gracia-Abril}, G. and {Portell}, J. and {Teyssier}, D. and {Altmann}, M. and {Andrae}, R. and {Audard}, M. and {Bellas-Velidis}, I. and {Benson}, K. and {Berthier}, J. and {Blomme}, R. and {Burgess}, P.~W. and {Busonero}, D. and {Busso}, G. and {C{\'a}novas}, H. and {Carry}, B. and {Cellino}, A. and {Cheek}, N. and {Clementini}, G. and {Damerdji}, Y. and {Davidson}, M. and {de Teodoro}, P. and {Nu{\~n}ez Campos}, M. and {Delchambre}, L. and {Dell'Oro}, A. and {Esquej}, P. and {Fern{\'a}ndez-Hern{\'a}ndez}, J. and {Fraile}, E. and {Garabato}, D. and {Garc{\'\i}a-Lario}, P. and {Gosset}, E. and {Haigron}, R. and {Halbwachs}, J.-L. and {Hambly}, N.~C. and {Harrison}, D.~L. and {Hern{\'a}ndez}, J. and {Hestroffer}, D. and {Hodgkin}, S.~T. and {Holl}, B. and {Jan{\ss}en}, K. and {Jevardat de Fombelle}, G. and {Jordan}, S. and {Krone-Martins}, A. and {Lanzafame}, A.~C. and {L{\"o}ffler}, W. and {Marchal}, O. and {Marrese}, P.~M. and {Moitinho}, A. and {Muinonen}, K. and {Osborne}, P. and {Pancino}, E. and {Pauwels}, T. and {Recio-Blanco}, A. and {Reyl{\'e}}, C. and {Riello}, M. and {Rimoldini}, L. and {Roegiers}, T. and {Rybizki}, J. and {Sarro}, L.~M. and {Siopis}, C. and {Smith}, M. and {Sozzetti}, A. and {Utrilla}, E. and {van Leeuwen}, M. and {Abbas}, U. and {{\'A}brah{\'a}m}, P. and {Abreu Aramburu}, A. and {Aerts}, C. and {Aguado}, J.~J. and {Ajaj}, M. and {Aldea-Montero}, F. and {Altavilla}, G. and {{\'A}lvarez}, M.~A. and {Alves}, J. and {Anders}, F. and {Anderson}, R.~I. and {Anglada Varela}, E. and {Antoja}, T. and {Baines}, D. and {Baker}, S.~G. and {Balaguer-N{\'u}{\~n}ez}, L. and {Balbinot}, E. and {Balog}, Z. and {Barache}, C. and {Barbato}, D. and {Barros}, M. and {Barstow}, M.~A. and {Bartolom{\'e}}, S. and {Bassilana}, J.-L. and {Bauchet}, N. and {Becciani}, U. and {Bellazzini}, M. and {Berihuete}, A. and {Bernet}, M. and {Bertone}, S. and {Bianchi}, L. and {Binnenfeld}, A. and {Blanco-Cuaresma}, S. and {Blazere}, A. and {Boch}, T. and {Bombrun}, A. and {Bossini}, D. and {Bouquillon}, S. and {Bragaglia}, A. and {Bramante}, L. and {Breedt}, E. and {Bressan}, A. and {Brouillet}, N. and {Brugaletta}, E. and {Bucciarelli}, B. and {Burlacu}, A. and {Butkevich}, A.~G. and {Buzzi}, R. and {Caffau}, E. and {Cancelliere}, R. and {Cantat-Gaudin}, T. and {Carballo}, R. and {Carlucci}, T. and {Carnerero}, M.~I. and {Carrasco}, J.~M. and {Casamiquela}, L. and {Castellani}, M. and {Castro-Ginard}, A. and {Chaoul}, L. and {Charlot}, P. and {Chemin}, L. and {Chiaramida}, V. and {Chiavassa}, A. and {Chornay}, N. and {Comoretto}, G. and {Contursi}, G. and {Cooper}, W.~J. and {Cornez}, T. and {Cowell}, S. and {Crifo}, F. and {Cropper}, M. and {Crosta}, M. and {Crowley}, C. and {Dafonte}, C. and {Dapergolas}, A. and {David}, M. and {David}, P. and {de Laverny}, P. and {De Luise}, F. and {De March}, R.},
        title = "{Gaia Data Release 3. Summary of the content and survey properties}",
      journal = {\aap},
         year = 2023,
        month = jun,
       volume = {674},
          eid = {A1},
        pages = {A1},
          doi = {10.1051/0004-6361/202243940},
archivePrefix = {arXiv},
       eprint = {2208.00211},
 primaryClass = {astro-ph.GA},
       adsurl = {https://ui.adsabs.harvard.edu/abs/2023A&A...674A...1G}
}

@ARTICLE{Yuan2015stellarloci1,
       author = {{Yuan}, Haibo and {Liu}, Xiaowei and {Xiang}, Maosheng and {Huang}, Yang and {Chen}, Bingqiu},
        title = "{Stellar Loci. I. Metallicity Dependence and Intrinsic Widths}",
      journal = {\apj},
         year = 2015,
        month = feb,
       volume = {799},
       number = {2},
          eid = {134},
        pages = {134},
          doi = {10.1088/0004-637X/799/2/134},
archivePrefix = {arXiv},
       eprint = {1412.1232},
 primaryClass = {astro-ph.SR},
       adsurl = {https://ui.adsabs.harvard.edu/abs/2015ApJ...799..134Y}
}

@ARTICLE{Yuan2015slot,
       author = {{Yuan}, Haibo and {Liu}, Xiaowei and {Xiang}, Maosheng and {Huang}, Yang and {Chen}, Bingqiu and {Wu}, Yue and {Hou}, Yonghui and {Zhang}, Yong},
        title = "{Stellar Loci II. A Model-free Estimate of the Binary Fraction for Field FGK Stars}",
      journal = {\apj},
         year = 2015,
        month = feb,
       volume = {799},
       number = {2},
          eid = {135},
        pages = {135},
          doi = {10.1088/0004-637X/799/2/135},
archivePrefix = {arXiv},
       eprint = {1412.1233},
 primaryClass = {astro-ph.SR},
       adsurl = {https://ui.adsabs.harvard.edu/abs/2015ApJ...799..135Y}
}

@ARTICLE{Huangbw2024xp,
       author = {{Huang}, Bowen and {Yuan}, Haibo and {Xiang}, Maosheng and {Huang}, Yang and {Xiao}, Kai and {Xu}, Shuai and {Zhang}, Ruoyi and {Yang}, Lin and {Niu}, Zexi and {Gu}, Hongrui},
        title = "{A Comprehensive Correction of the Gaia DR3 XP Spectra}",
      journal = {\apjs},
         year = 2024,
        month = mar,
       volume = {271},
       number = {1},
          eid = {13},
        pages = {13},
          doi = {10.3847/1538-4365/ad18b1},
archivePrefix = {arXiv},
       eprint = {2401.12006},
 primaryClass = {astro-ph.IM},
       adsurl = {https://ui.adsabs.harvard.edu/abs/2024ApJS..271...13H}
}

@ARTICLE{Huangbw2025loci,
       author = {{Huang}, Bowen and {Yuan}, Haibo and {Xu}, Shuai and {Xiao}, Kai and {Xiang}, Maosheng and {Huang}, Yang and {Beers}, Timothy C.},
        title = "{Stellar Loci. VIII. Photometric Metallicities for 100 Million Stars Based on Synthetic Gaia Colors}",
      journal = {\apjs},
         year = 2025,
        month = mar,
       volume = {277},
       number = {1},
          eid = {7},
        pages = {7},
          doi = {10.3847/1538-4365/ada9e6},
archivePrefix = {arXiv},
       eprint = {2410.19895},
 primaryClass = {astro-ph.SR},
       adsurl = {https://ui.adsabs.harvard.edu/abs/2025ApJS..277....7H}
}

@ARTICLE{Moe2017PandQ,
       author = {{Moe}, Maxwell and {Di Stefano}, Rosanne},
        title = "{Mind Your Ps and Qs: The Interrelation between Period (P) and Mass-ratio (Q) Distributions of Binary Stars}",
      journal = {\apjs},
         year = 2017,
        month = jun,
       volume = {230},
       number = {2},
          eid = {15},
        pages = {15},
          doi = {10.3847/1538-4365/aa6fb6},
archivePrefix = {arXiv},
       eprint = {1606.05347},
 primaryClass = {astro-ph.SR},
       adsurl = {https://ui.adsabs.harvard.edu/abs/2017ApJS..230...15M}
}

@ARTICLE{Schlegel1998extin,
       author = {{Schlegel}, David J. and {Finkbeiner}, Douglas P. and {Davis}, Marc},
        title = "{Maps of Dust Infrared Emission for Use in Estimation of Reddening and Cosmic Microwave Background Radiation Foregrounds}",
      journal = {\apj},
         year = 1998,
        month = jun,
       volume = {500},
       number = {2},
        pages = {525-553},
          doi = {10.1086/305772},
archivePrefix = {arXiv},
       eprint = {astro-ph/9710327},
 primaryClass = {astro-ph},
       adsurl = {https://ui.adsabs.harvard.edu/abs/1998ApJ...500..525S}
}

@ARTICLE{Zhangry2023dered,
       author = {{Zhang}, Ruoyi and {Yuan}, Haibo},
        title = "{Empirical Temperature- and Extinction-dependent Extinction Coefficients for the GALEX, Pan-STARRS 1, Gaia, SDSS, 2MASS, and WISE Passbands}",
      journal = {\apjs},
         year = 2023,
        month = jan,
       volume = {264},
       number = {1},
          eid = {14},
        pages = {14},
          doi = {10.3847/1538-4365/ac9dfa},
archivePrefix = {arXiv},
       eprint = {2210.15918},
 primaryClass = {astro-ph.GA},
       adsurl = {https://ui.adsabs.harvard.edu/abs/2023ApJS..264...14Z}
}

@ARTICLE{BJ2021distances,
       author = {{Bailer-Jones}, C.~A.~L. and {Rybizki}, J. and {Fouesneau}, M. and {Demleitner}, M. and {Andrae}, R.},
        title = "{Estimating Distances from Parallaxes. V. Geometric and Photogeometric Distances to 1.47 Billion Stars in Gaia Early Data Release 3}",
      journal = {\aj},
         year = 2021,
        month = mar,
       volume = {161},
       number = {3},
          eid = {147},
        pages = {147},
          doi = {10.3847/1538-3881/abd806},
archivePrefix = {arXiv},
       eprint = {2012.05220},
 primaryClass = {astro-ph.SR},
       adsurl = {https://ui.adsabs.harvard.edu/abs/2021AJ....161..147B}
}

@ARTICLE{Zhangm2025elements,
       author = {{Zhang}, Meng and {Xiang}, Maosheng and {Ting}, Yuan-Sen and {Amarsi}, Anish Mayur and {Zhang}, Hua-Wei and {Shi}, Jianrong and {Yuan}, Haibo and {Li}, Haining and {Wang}, Jiahui and {Wu}, Yaqian and {Wu}, Tianmin and {Mou}, Lanya and {Yan}, Hong-Liang and {Liu}, Jifeng},
        title = "{Homogeneous Stellar Atmospheric Parameters and 22 Elemental Abundances for FGK Stars Derived from LAMOST Low-resolution Spectra with DD-PAYNE}",
      journal = {\apjs},
         year = 2025,
        month = jul,
       volume = {279},
       number = {1},
          eid = {5},
        pages = {5},
          doi = {10.3847/1538-4365/add016},
archivePrefix = {arXiv},
       eprint = {2506.02763},
 primaryClass = {astro-ph.SR},
       adsurl = {https://ui.adsabs.harvard.edu/abs/2025ApJS..279....5Z}
}

@ARTICLE{Wangjh2025Ages,
       author = {{Wang}, Jia-Hui and {Xiang}, Maosheng and {Zhang}, Meng and {Xie}, Ji-Wei and {Ge}, Jian and {Zhang}, Jinghua and {Mou}, Lanya and {Liu}, Ji-Feng},
        title = "{Spectroscopic Ages for 4 Million Main-sequence Dwarf Stars from LAMOST DR10 Estimated with a Data-driven Approach}",
      journal = {\apjs},
         year = 2025,
        month = sep,
       volume = {280},
       number = {1},
          eid = {13},
        pages = {13},
          doi = {10.3847/1538-4365/aded16},
archivePrefix = {arXiv},
       eprint = {2508.03019},
 primaryClass = {astro-ph.SR},
       adsurl = {https://ui.adsabs.harvard.edu/abs/2025ApJS..280...13W}
}

@ARTICLE{Qiud2024Mdwarfs,
       author = {{Qiu}, Dan and {Li}, Jiadong and {Zhang}, Bo and {Liu}, Chao and {Tian}, Haijun and {Niu}, Zexi},
        title = "{Calibration of metallicity of LAMOST M dwarf stars using FGK+M wide binaries}",
      journal = {\mnras},
         year = 2024,
        month = feb,
       volume = {527},
       number = {4},
        pages = {11866-11881},
          doi = {10.1093/mnras/stad3950},
archivePrefix = {arXiv},
       eprint = {2312.12827},
 primaryClass = {astro-ph.SR},
       adsurl = {https://ui.adsabs.harvard.edu/abs/2024MNRAS.52711866Q}
}

@article{Mazzola2020closebinary,
    author = {Mazzola, Christine N and Badenes, Carles and Moe, Maxwell and Koposov, Sergey E and Kounkel, Marina and Kratter, Kaitlin and Covey, Kevin and Walker, Matthew G and Thompson, Todd A and Andrews, Brett and Freeman, Peter E and Anguiano, Borja and Carlberg, Joleen K and De Lee, Nathan M and Frinchaboy, Peter M and Lewis, Hannah M and Majewski, Steven and Nidever, David and Nitschelm, Christian and Price-Whelan, Adrian M and Roman-Lopes, Alexandre and Stassun, Keivan G and Troup, Nicholas W},
    title = {The close binary fraction as a function of stellar parameters in APOGEE: a strong anticorrelation with α abundances},
    journal = {Monthly Notices of the Royal Astronomical Society},
    volume = {499},
    number = {2},
    pages = {1607-1626},
    year = {2020},
    month = {09},
    issn = {0035-8711},
    doi = {10.1093/mnras/staa2859},
    url = {https://doi.org/10.1093/mnras/staa2859},
    eprint = {https://academic.oup.com/mnras/article-pdf/499/2/1607/33890920/staa2859.pdf},
}

@ARTICLE{2022Li,
       author = {{Li}, Jiangdan and {Li}, Jiao and {Liu}, Chao and {Li}, Chunqian and {Guo}, Yanjun and {Wang}, Luqian and {Chen}, Xuefei and {Xing}, Lifeng and {Hou}, Yonghui and {Han}, Zhanwen},
        title = "{Mass-ratio Distribution of Binaries from the LAMOST-MRS Survey}",
      journal = {\apj},
         year = 2022,
        month = jul,
       volume = {933},
       number = {2},
          eid = {119},
        pages = {119},
          doi = {10.3847/1538-4357/ac731d},
archivePrefix = {arXiv},
       eprint = {2205.11734},
 primaryClass = {astro-ph.SR},
       adsurl = {https://ui.adsabs.harvard.edu/abs/2022ApJ...933..119L}
}

@ARTICLE{ElBadry2019twinbinary,
       author = {{El-Badry}, Kareem and {Rix}, Hans-Walter and {Tian}, Haijun and {Duch{\^e}ne}, Gaspard and {Moe}, Maxwell},
        title = "{Discovery of an equal-mass `twin' binary population reaching 1000 + au separations}",
      journal = {\mnras},
         year = 2019,
        month = nov,
       volume = {489},
       number = {4},
        pages = {5822-5857},
          doi = {10.1093/mnras/stz2480},
archivePrefix = {arXiv},
       eprint = {1906.10128},
 primaryClass = {astro-ph.SR},
       adsurl = {https://ui.adsabs.harvard.edu/abs/2019MNRAS.489.5822E}
}

@ARTICLE{2020Huang,
       author = {{Huang}, Yang and {Sch{\"o}nrich}, Ralph and {Zhang}, Huawei and {Wu}, Yaqian and {Chen}, Bingqiu and {Wang}, Haifeng and {Xiang}, Maosheng and {Wang}, Chun and {Yuan}, Haibo and {Li}, Xinyi and {Sun}, Weixiang and {Li}, Ji and {Liu}, Xiaowei},
        title = "{Mapping the Galactic Disk with the LAMOST and Gaia Red Clump Sample. I. Precise Distances, Masses, Ages, and 3D Velocities of {\ensuremath{\sim}}140,000 Red Clump Stars}",
      journal = {\apjs},
         year = 2020,
        month = aug,
       volume = {249},
       number = {2},
          eid = {29},
        pages = {29},
          doi = {10.3847/1538-4365/ab994f},
archivePrefix = {arXiv},
       eprint = {2006.02686},
 primaryClass = {astro-ph.SR},
       adsurl = {https://ui.adsabs.harvard.edu/abs/2020ApJS..249...29H}
}

@ARTICLE{2005Goodwin,
       author = {{Goodwin}, S.~P. and {Kroupa}, P.},
        title = "{Limits on the primordial stellar multiplicity}",
      journal = {\aap},
         year = 2005,
        month = aug,
       volume = {439},
       number = {2},
        pages = {565-569},
          doi = {10.1051/0004-6361:20052654},
archivePrefix = {arXiv},
       eprint = {astro-ph/0505470},
 primaryClass = {astro-ph},
       adsurl = {https://ui.adsabs.harvard.edu/abs/2005A&A...439..565G}
}

@ARTICLE{1995Kroupa,
       author = {{Kroupa}, Pavel},
        title = "{The dynamical properties of stellar systems in the Galactic disc}",
      journal = {\mnras},
         year = 1995,
        month = dec,
       volume = {277},
        pages = {1507},
          doi = {10.1093/mnras/277.4.1507},
archivePrefix = {arXiv},
       eprint = {astro-ph/9508084},
 primaryClass = {astro-ph},
       adsurl = {https://ui.adsabs.harvard.edu/abs/1995MNRAS.277.1507K}
}

@ARTICLE{2026Niu,
       author = {{Niu}, Zexi and {Sun}, Ning-Chen and {Zapartas}, Emmanouil and {Souropanis}, Dimitris and {Cui}, Yingzhen and {Maund}, Justyn R. and {Andrews}, Jeff J. and {Briel}, Max M. and {Fraser}, Morgan and {Gossage}, Seth and {Kruckow}, Matthias U. and {Liotine}, Camille and {Liu}, Zhengwei and {Podsiadlowski}, Philipp and {Srivastava}, Philipp M. and {Teng}, Elizabeth and {Wang}, Xiaofeng and {Yang}, Yi and {Liu}, Jifeng},
        title = "{A binary merger product as the direct progenitor of a Type II-P supernova}",
      journal = {Science Bulletin},
         year = 2026,
        month = mar,
       volume = {71},
       number = {5},
        pages = {1023-1033},
          doi = {10.1016/j.scib.2026.01.036},
       adsurl = {https://ui.adsabs.harvard.edu/abs/2026SciBu..71.1023N}
}

@ARTICLE{2023Guszejnov,
       author = {{Guszejnov}, D{\'a}vid and {Raju}, Aman N. and {Offner}, Stella S.~R. and {Grudi{\'c}}, Michael Y. and {Faucher-Gigu{\`e}re}, Claude-Andr{\'e} and {Hopkins}, Philip F. and {Rosen}, Anna L.},
        title = "{Effects of the environment on the multiplicity properties of stars in the STARFORGE simulations}",
      journal = {\mnras},
         year = 2023,
        month = jan,
       volume = {518},
       number = {3},
        pages = {4693-4712},
          doi = {10.1093/mnras/stac3268},
archivePrefix = {arXiv},
       eprint = {2208.02844},
 primaryClass = {astro-ph.GA},
       adsurl = {https://ui.adsabs.harvard.edu/abs/2023MNRAS.518.4693G}
}

@ARTICLE{2011Marks,
       author = {{Marks}, Michael and {Kroupa}, Pavel},
        title = "{Dynamical population synthesis: constructing the stellar single and binary contents of galactic field populations}",
      journal = {\mnras},
         year = 2011,
        month = nov,
       volume = {417},
       number = {3},
        pages = {1702-1714},
          doi = {10.1111/j.1365-2966.2011.19519.x},
archivePrefix = {arXiv},
       eprint = {1109.2896},
 primaryClass = {astro-ph.GA},
       adsurl = {https://ui.adsabs.harvard.edu/abs/2011MNRAS.417.1702M}
}

@ARTICLE{2012Marks,
       author = {{Marks}, Michael and {Kroupa}, Pavel and {Dabringhausen}, J{\"o}rg and {Pawlowski}, Marcel S.},
        title = "{Evidence for top-heavy stellar initial mass functions with increasing density and decreasing metallicity}",
      journal = {\mnras},
         year = 2012,
        month = may,
       volume = {422},
       number = {3},
        pages = {2246-2254},
          doi = {10.1111/j.1365-2966.2012.20767.x},
archivePrefix = {arXiv},
       eprint = {1202.4755},
 primaryClass = {astro-ph.GA},
       adsurl = {https://ui.adsabs.harvard.edu/abs/2012MNRAS.422.2246M}
}

@ARTICLE{2023Li,
       author = {{Li}, Jiadong and {Liu}, Chao and {Zhang}, Zhi-Yu and {Tian}, Hao and {Fu}, Xiaoting and {Li}, Jiao and {Yan}, Zhi-Qiang},
        title = "{Stellar initial mass function varies with metallicity and time}",
      journal = {\nat},
         year = 2023,
        month = jan,
       volume = {613},
       number = {7944},
        pages = {460-462},
          doi = {10.1038/s41586-022-05488-1},
archivePrefix = {arXiv},
       eprint = {2301.07029},
 primaryClass = {astro-ph.GA},
       adsurl = {https://ui.adsabs.harvard.edu/abs/2023Natur.613..460L}
}

@ARTICLE{2022Conroy,
       author = {{Conroy}, Charlie and {Weinberg}, David H. and {Naidu}, Rohan P. and {Buck}, Tobias and {Johnson}, James W. and {Cargile}, Phillip and {Bonaca}, Ana and {Caldwell}, Nelson and {Chandra}, Vedant and {Han}, Jiwon Jesse and {Johnson}, Benjamin D. and {Speagle}, Joshua S. and {Ting}, Yuan-Sen and {Woody}, Turner and {Zaritsky}, Dennis},
        title = "{Birth of the Galactic Disk Revealed by the H3 Survey}",
      journal = {arXiv e-prints},
         year = 2022,
        month = apr,
          eid = {arXiv:2204.02989},
        pages = {arXiv:2204.02989},
          doi = {10.48550/arXiv.2204.02989},
archivePrefix = {arXiv},
       eprint = {2204.02989},
 primaryClass = {astro-ph.GA},
       adsurl = {https://ui.adsabs.harvard.edu/abs/2022arXiv220402989C}
}

@ARTICLE{2019Clarke,
       author = {{Clarke}, Adam J. and {Debattista}, Victor P. and {Nidever}, David L. and {Loebman}, Sarah R. and {Simons}, Raymond C. and {Kassin}, Susan and {Du}, Min and {Ness}, Melissa and {Fisher}, Deanne B. and {Quinn}, Thomas R. and {Wadsley}, James and {Freeman}, Ken C. and {Popescu}, Cristina C.},
        title = "{The imprint of clump formation at high redshift - I. A disc {\ensuremath{\alpha}}-abundance dichotomy}",
      journal = {\mnras},
         year = 2019,
        month = apr,
       volume = {484},
       number = {3},
        pages = {3476-3490},
          doi = {10.1093/mnras/stz104},
archivePrefix = {arXiv},
       eprint = {1901.00931},
 primaryClass = {astro-ph.GA},
       adsurl = {https://ui.adsabs.harvard.edu/abs/2019MNRAS.484.3476C}
}

@ARTICLE{2021Johnson,
       author = {{Johnson}, James W. and {Weinberg}, David H. and {Vincenzo}, Fiorenzo and {Bird}, Jonathan C. and {Loebman}, Sarah R. and {Brooks}, Alyson M. and {Quinn}, Thomas R. and {Christensen}, Charlotte R. and {Griffith}, Emily J.},
        title = "{Stellar migration and chemical enrichment in the milky way disc: a hybrid model}",
      journal = {\mnras},
         year = 2021,
        month = dec,
       volume = {508},
       number = {3},
        pages = {4484-4511},
          doi = {10.1093/mnras/stab2718},
archivePrefix = {arXiv},
       eprint = {2103.09838},
 primaryClass = {astro-ph.GA},
       adsurl = {https://ui.adsabs.harvard.edu/abs/2021MNRAS.508.4484J}
}

@ARTICLE{2023Spitoni,
       author = {{Spitoni}, E. and {Recio-Blanco}, A. and {de Laverny}, P. and {Palicio}, P.~A. and {Kordopatis}, G. and {Schultheis}, M. and {Contursi}, G. and {Poggio}, E. and {Romano}, D. and {Matteucci}, F.},
        title = "{Beyond the two-infall model. I. Indications for a recent gas infall with Gaia DR3 chemical abundances}",
      journal = {\aap},
         year = 2023,
        month = feb,
       volume = {670},
          eid = {A109},
        pages = {A109},
          doi = {10.1051/0004-6361/202244349},
archivePrefix = {arXiv},
       eprint = {2206.12436},
 primaryClass = {astro-ph.GA},
       adsurl = {https://ui.adsabs.harvard.edu/abs/2023A&A...670A.109S}
}

@ARTICLE{2014Tanaka,
       author = {{Tanaka}, Kei E.~I. and {Omukai}, Kazuyuki},
        title = "{Gravitational instability in protostellar discs at low metallicities}",
      journal = {\mnras},
         year = 2014,
        month = apr,
       volume = {439},
       number = {2},
        pages = {1884-1896},
          doi = {10.1093/mnras/stu069},
archivePrefix = {arXiv},
       eprint = {1401.2993},
 primaryClass = {astro-ph.GA},
       adsurl = {https://ui.adsabs.harvard.edu/abs/2014MNRAS.439.1884T}
}

@ARTICLE{2022Xiang,
       author = {{Xiang}, Maosheng and {Rix}, Hans-Walter},
        title = "{A time-resolved picture of our Milky Way's early formation history}",
      journal = {\nat},
         year = 2022,
        month = mar,
       volume = {603},
       number = {7902},
        pages = {599-603},
          doi = {10.1038/s41586-022-04496-5},
archivePrefix = {arXiv},
       eprint = {2203.12110},
 primaryClass = {astro-ph.GA},
       adsurl = {https://ui.adsabs.harvard.edu/abs/2022Natur.603..599X}
}

@misc{Offner2022OriginEvoMultiplestar,
      title={The Origin and Evolution of Multiple Star Systems}, 
      author={Stella S. R. Offner and Maxwell Moe and Kaitlin M. Kratter and Sarah I. Sadavoy and Eric L. N. Jensen and John J. Tobin},
      year={2022},
      eprint={2203.10066},
      archivePrefix={arXiv},
      primaryClass={astro-ph.SR},
      url={https://arxiv.org/abs/2203.10066}, 
}

@article{LiuC2019SmokingGun,
    author = {Liu, Chao},
    title = {Smoking gun of the dynamical processing of solar-type field binary stars},
    journal = {Monthly Notices of the Royal Astronomical Society},
    volume = {490},
    number = {1},
    pages = {550-565},
    year = {2019},
    month = {10},
    issn = {0035-8711},
    doi = {10.1093/mnras/stz2274},
    url = {https://academic.oup.com/mnras/article/490/1/550/5581703},
    eprint = {https://academic.oup.com/mnras/article-pdf/490/1/550/30105380/stz2274.pdf},
}

@article{Bate2019StarsMetallicitySim,
    author = {Bate, Matthew R},
    title = {The statistical properties of stars and their dependence on metallicity},
    journal = {Monthly Notices of the Royal Astronomical Society},
    volume = {484},
    number = {2},
    pages = {2341-2361},
    year = {2019},
    month = {01},
    issn = {0035-8711},
    doi = {10.1093/mnras/stz103},
    url = {https://doi.org/10.1093/mnras/stz103},
    eprint = {https://academic.oup.com/mnras/article-pdf/484/2/2341/27635804/stz103.pdf},
}

@article{Arenou2018GaiaDR2,
	author = {{Arenou, F.} and {Luri, X.} and {Babusiaux, C.} and {Fabricius, C.} and {Helmi, A.} and {Muraveva, T.} and {Robin, A. C.} and {Spoto, F.} and {Vallenari, A.} and {Antoja, T.} and {Cantat-Gaudin, T.} and {Jordi, C.} and {Leclerc, N.} and {Reylé, C.} and {Romero-Gómez, M.} and {Shih, I.-C.} and {Soria, S.} and {Barache, C.} and {Bossini, D.} and {Bragaglia, A.} and {Breddels, M. A.} and {Fabrizio, M.} and {Lambert, S.} and {Marrese, P. M.} and {Massari, D.} and {Moitinho, A.} and {Robichon, N.} and {Ruiz-Dern, L.} and {Sordo, R.} and {Veljanoski, J.} and {Eyer, L.} and {Jasniewicz, G.} and {Pancino, E.} and {Soubiran, C.} and {Spagna, A.} and {Tanga, P.} and {Turon, C.} and {Zurbach, C.}},
	title = {Gaia Data Release 2 - Catalogue validation},
	DOI= "10.1051/0004-6361/201833234",
	url= "https://doi.org/10.1051/0004-6361/201833234",
	journal = {A\&A},
	year = 2018,
	volume = 616,
	pages = "A17",
}

@article{Raghavan2010survey,
doi = {10.1088/0067-0049/190/1/1},
url = {https://doi.org/10.1088/0067-0049/190/1/1},
year = {2010},
month = {aug},
publisher = {The American Astronomical Society},
volume = {190},
number = {1},
pages = {1},
author = {Raghavan, Deepak and McAlister, Harold A. and Henry, Todd J. and Latham, David W. and Marcy, Geoffrey W. and Mason, Brian D. and Gies, Douglas R. and White, Russel J. and ten Brummelaar, Theo A.},
title = {A SURVEY OF STELLAR FAMILIES: MULTIPLICITY OF SOLAR-TYPE STARS},
journal = {The Astrophysical Journal Supplement Series}
}

@article{GaiaCollab2016mission,
	author = {{Gaia Collaboration} and {Prusti, T.} and {de Bruijne, J. H. J.} and {Brown, A. G. A.} and {Vallenari, A.} and {Babusiaux, C.} and {Bailer-Jones, C. A. L.} and {Bastian, U.} and {Biermann, M.} and {Evans, D. W.} and {Eyer, L.} and {Jansen, F.} and {Jordi, C.} and {Klioner, S. A.} and {Lammers, U.} and {Lindegren, L.} and {Luri, X.} and {Mignard, F.} and {Milligan, D. J.} and {Panem, C.} and {Poinsignon, V.} and {Pourbaix, D.} and {Randich, S.} and {Sarri, G.} and {Sartoretti, P.} and {Siddiqui, H. I.} and {Soubiran, C.} and {Valette, V.} and {van Leeuwen, F.} and {Walton, N. A.} and {Aerts, C.} and {Arenou, F.} and {Cropper, M.} and {Drimmel, R.} and {Høg, E.} and {Katz, D.} and {Lattanzi, M. G.} and {O’Mullane, W.} and {Grebel, E. K.} and {Holland, A. D.} and {Huc, C.} and {Passot, X.} and {Bramante, L.} and {Cacciari, C.} and {Castañeda, J.} and {Chaoul, L.} and {Cheek, N.} and {De Angeli, F.} and {Fabricius, C.} and {Guerra, R.} and {Hernández, J.} and {Jean-Antoine-Piccolo, A.} and {Masana, E.} and {Messineo, R.} and {Mowlavi, N.} and {Nienartowicz, K.} and {Ordóñez-Blanco, D.} and {Panuzzo, P.} and {Portell, J.} and {Richards, P. J.} and {Riello, M.} and {Seabroke, G. M.} and {Tanga, P.} and {Thévenin, F.} and {Torra, J.} and {Els, S. G.} and {Gracia-Abril, G.} and {Comoretto, G.} and {Garcia-Reinaldos, M.} and {Lock, T.} and {Mercier, E.} and {Altmann, M.} and {Andrae, R.} and {Astraatmadja, T. L.} and {Bellas-Velidis, I.} and {Benson, K.} and {Berthier, J.} and {Blomme, R.} and {Busso, G.} and {Carry, B.} and {Cellino, A.} and {Clementini, G.} and {Cowell, S.} and {Creevey, O.} and {Cuypers, J.} and {Davidson, M.} and {De Ridder, J.} and {de Torres, A.} and {Delchambre, L.} and {Dell’Oro, A.} and {Ducourant, C.} and {Frémat, Y.} and {García-Torres, M.} and {Gosset, E.} and {Halbwachs, J.-L.} and {Hambly, N. C.} and {Harrison, D. L.} and {Hauser, M.} and {Hestroffer, D.} and {Hodgkin, S. T.} and {Huckle, H. E.} and {Hutton, A.} and {Jasniewicz, G.} and {Jordan, S.} and {Kontizas, M.} and {Korn, A. J.} and {Lanzafame, A. C.} and {Manteiga, M.} and {Moitinho, A.} and {Muinonen, K.} and {Osinde, J.} and {Pancino, E.} and {Pauwels, T.} and {Petit, J.-M.} and {Recio-Blanco, A.} and {Robin, A. C.} and {Sarro, L. M.} and {Siopis, C.} and {Smith, M.} and {Smith, K. W.} and {Sozzetti, A.} and {Thuillot, W.} and {van Reeven, W.} and {Viala, Y.} and {Abbas, U.} and {Abreu Aramburu, A.} and {Accart, S.} and {Aguado, J. J.} and {Allan, P. M.} and {Allasia, W.} and {Altavilla, G.} and {Álvarez, M. A.} and {Alves, J.} and {Anderson, R. I.} and {Andrei, A. H.} and {Anglada Varela, E.} and {Antiche, E.} and {Antoja, T.} and {Antón, S.} and {Arcay, B.} and {Atzei, A.} and {Ayache, L.} and {Bach, N.} and {Baker, S. G.} and {Balaguer-Núñez, L.} and {Barache, C.} and {Barata, C.} and {Barbier, A.} and {Barblan, F.} and {Baroni, M.} and {Barrado y Navascués, D.} and {Barros, M.} and {Barstow, M. A.} and {Becciani, U.} and {Bellazzini, M.} and {Bellei, G.} and {Bello García, A.} and {Belokurov, V.} and {Bendjoya, P.} and {Berihuete, A.} and {Bianchi, L.} and {Bienaymé, O.} and {Billebaud, F.} and {Blagorodnova, N.} and {Blanco-Cuaresma, S.} and {Boch, T.} and {Bombrun, A.} and {Borrachero, R.} and {Bouquillon, S.} and {Bourda, G.} and {Bouy, H.} and {Bragaglia, A.} and {Breddels, M. A.} and {Brouillet, N.} and {Brüsemeister, T.} and {Bucciarelli, B.} and {Budnik, F.} and {Burgess, P.} and {Burgon, R.} and {Burlacu, A.} and {Busonero, D.} and {Buzzi, R.} and {Caffau, E.} and {Cambras, J.} and {Campbell, H.} and {Cancelliere, R.} and {Cantat-Gaudin, T.} and {Carlucci, T.} and {Carrasco, J. M.} and {Castellani, M.} and {Charlot, P.} and {Charnas, J.} and {Charvet, P.} and {Chassat, F.} and {Chiavassa, A.} and {Clotet, M.} and {Cocozza, G.} and {Collins, R. S.} and {Collins, P.} and {Costigan, G.} and {Crifo, F.} and {Cross, N. J. G.} and {Crosta, M.} and {Crowley, C.} and {Dafonte, C.} and {Damerdji, Y.} and {Dapergolas, A.} and {David, P.} and {David, M.} and {De Cat, P.} and {de Felice, F.} and {de Laverny, P.} and {De Luise, F.} and {De March, R.} and {de Martino, D.} and {de Souza, R.} and {Debosscher, J.} and {del Pozo, E.} and {Delbo, M.} and {Delgado, A.} and {Delgado, H. E.} and {di Marco, F.} and {Di Matteo, P.} and {Diakite, S.} and {Distefano, E.} and {Dolding, C.} and {Dos Anjos, S.} and {Drazinos, P.} and {Durán, J.} and {Dzigan, Y.} and {Ecale, E.} and {Edvardsson, B.} and {Enke, H.} and {Erdmann, M.} and {Escolar, D.} and {Espina, M.} and {Evans, N. W.} and {Eynard Bontemps, G.} and {Fabre, C.} and {Fabrizio, M.} and {Faigler, S.} and {Falcão, A. J.} and {Farràs Casas, M.} and {Faye, F.} and {Federici, L.} and {Fedorets, G.} and {Fernández-Hernández, J.} and {Fernique, P.} and {Fienga, A.} and {Figueras, F.} and {Filippi, F.} and {Findeisen, K.} and {Fonti, A.} and {Fouesneau, M.} and {Fraile, E.} and {Fraser, M.} and {Fuchs, J.} and {Furnell, R.} and {Gai, M.} and {Galleti, S.} and {Galluccio, L.} and {Garabato, D.} and {García-Sedano, F.} and {Garé, P.} and {Garofalo, A.} and {Garralda, N.} and {Gavras, P.} and {Gerssen, J.} and {Geyer, R.} and {Gilmore, G.} and {Girona, S.} and {Giuffrida, G.} and {Gomes, M.} and {González-Marcos, A.} and {González-Núñez, J.} and {González-Vidal, J. J.} and {Granvik, M.} and {Guerrier, A.} and {Guillout, P.} and {Guiraud, J.} and {Gúrpide, A.} and {Gutiérrez-Sánchez, R.} and {Guy, L. P.} and {Haigron, R.} and {Hatzidimitriou, D.} and {Haywood, M.} and {Heiter, U.} and {Helmi, A.} and {Hobbs, D.} and {Hofmann, W.} and {Holl, B.} and {Holland, G.} and {Hunt, J. A. S.} and {Hypki, A.} and {Icardi, V.} and {Irwin, M.} and {Jevardat de Fombelle, G.} and {Jofré, P.} and {Jonker, P. G.} and {Jorissen, A.} and {Julbe, F.} and {Karampelas, A.} and {Kochoska, A.} and {Kohley, R.} and {Kolenberg, K.} and {Kontizas, E.} and {Koposov, S. E.} and {Kordopatis, G.} and {Koubsky, P.} and {Kowalczyk, A.} and {Krone-Martins, A.} and {Kudryashova, M.} and {Kull, I.} and {Bachchan, R. K.} and {Lacoste-Seris, F.} and {Lanza, A. F.} and {Lavigne, J.-B.} and {Le Poncin-Lafitte, C.} and {Lebreton, Y.} and {Lebzelter, T.} and {Leccia, S.} and {Leclerc, N.} and {Lecoeur-Taibi, I.} and {Lemaitre, V.} and {Lenhardt, H.} and {Leroux, F.} and {Liao, S.} and {Licata, E.} and {Lindstrøm, H. E. P.} and {Lister, T. A.} and {Livanou, E.} and {Lobel, A.} and {Löffler, W.} and {López, M.} and {Lopez-Lozano, A.} and {Lorenz, D.} and {Loureiro, T.} and {MacDonald, I.} and {Magalhães Fernandes, T.} and {Managau, S.} and {Mann, R. G.} and {Mantelet, G.} and {Marchal, O.} and {Marchant, J. M.} and {Marconi, M.} and {Marie, J.} and {Marinoni, S.} and {Marrese, P. M.} and {Marschalkó, G.} and {Marshall, D. J.} and {Martín-Fleitas, J. M.} and {Martino, M.} and {Mary, N.} and {Matijevič, G.} and {Mazeh, T.} and {McMillan, P. J.} and {Messina, S.} and {Mestre, A.} and {Michalik, D.} and {Millar, N. R.} and {Miranda, B. M. H.} and {Molina, D.} and {Molinaro, R.} and {Molinaro, M.} and {Molnár, L.} and {Moniez, M.} and {Montegriffo, P.} and {Monteiro, D.} and {Mor, R.} and {Mora, A.} and {Morbidelli, R.} and {Morel, T.} and {Morgenthaler, S.} and {Morley, T.} and {Morris, D.} and {Mulone, A. F.} and {Muraveva, T.} and {Musella, I.} and {Narbonne, J.} and {Nelemans, G.} and {Nicastro, L.} and {Noval, L.} and {Ordénovic, C.} and {Ordieres-Meré, J.} and {Osborne, P.} and {Pagani, C.} and {Pagano, I.} and {Pailler, F.} and {Palacin, H.} and {Palaversa, L.} and {Parsons, P.} and {Paulsen, T.} and {Pecoraro, M.} and {Pedrosa, R.} and {Pentikäinen, H.} and {Pereira, J.} and {Pichon, B.} and {Piersimoni, A. M.} and {Pineau, F.-X.} and {Plachy, E.} and {Plum, G.} and {Poujoulet, E.} and {Prša, A.} and {Pulone, L.} and {Ragaini, S.} and {Rago, S.} and {Rambaux, N.} and {Ramos-Lerate, M.} and {Ranalli, P.} and {Rauw, G.} and {Read, A.} and {Regibo, S.} and {Renk, F.} and {Reylé, C.} and {Ribeiro, R. A.} and {Rimoldini, L.} and {Ripepi, V.} and {Riva, A.} and {Rixon, G.} and {Roelens, M.} and {Romero-Gómez, M.} and {Rowell, N.} and {Royer, F.} and {Rudolph, A.} and {Ruiz-Dern, L.} and {Sadowski, G.} and {Sagristà Sellés, T.} and {Sahlmann, J.} and {Salgado, J.} and {Salguero, E.} and {Sarasso, M.} and {Savietto, H.} and {Schnorhk, A.} and {Schultheis, M.} and {Sciacca, E.} and {Segol, M.} and {Segovia, J. C.} and {Segransan, D.} and {Serpell, E.} and {Shih, I-C.} and {Smareglia, R.} and {Smart, R. L.} and {Smith, C.} and {Solano, E.} and {Solitro, F.} and {Sordo, R.} and {Soria Nieto, S.} and {Souchay, J.} and {Spagna, A.} and {Spoto, F.} and {Stampa, U.} and {Steele, I. A.} and {Steidelmüller, H.} and {Stephenson, C. A.} and {Stoev, H.} and {Suess, F. F.} and {Süveges, M.} and {Surdej, J.} and {Szabados, L.} and {Szegedi-Elek, E.} and {Tapiador, D.} and {Taris, F.} and {Tauran, G.} and {Taylor, M. B.} and {Teixeira, R.} and {Terrett, D.} and {Tingley, B.} and {Trager, S. C.} and {Turon, C.} and {Ulla, A.} and {Utrilla, E.} and {Valentini, G.} and {van Elteren, A.} and {Van Hemelryck, E.} and {van Leeuwen, M.} and {Varadi, M.} and {Vecchiato, A.} and {Veljanoski, J.} and {Via, T.} and {Vicente, D.} and {Vogt, S.} and {Voss, H.} and {Votruba, V.} and {Voutsinas, S.} and {Walmsley, G.} and {Weiler, M.} and {Weingrill, K.} and {Werner, D.} and {Wevers, T.} and {Whitehead, G.} and {Wyrzykowski, Ł.} and {Yoldas, A.} and {Žerjal, M.} and {Zucker, S.} and {Zurbach, C.} and {Zwitter, T.} and {Alecu, A.} and {Allen, M.} and {Allende Prieto, C.} and {Amorim, A.} and {Anglada-Escudé, G.} and {Arsenijevic, V.} and {Azaz, S.} and {Balm, P.} and {Beck, M.} and {Bernstein, H.-H.} and {Bigot, L.} and {Bijaoui, A.} and {Blasco, C.} and {Bonfigli, M.} and {Bono, G.} and {Boudreault, S.} and {Bressan, A.} and {Brown, S.} and {Brunet, P.-M.} and {Bunclark, P.} and {Buonanno, R.} and {Butkevich, A. G.} and {Carret, C.} and {Carrion, C.} and {Chemin, L.} and {Chéreau, F.} and {Corcione, L.} and {Darmigny, E.} and {de Boer, K. S.} and {de Teodoro, P.} and {de Zeeuw, P. T.} and {Delle Luche, C.} and {Domingues, C. D.} and {Dubath, P.} and {Fodor, F.} and {Frézouls, B.} and {Fries, A.} and {Fustes, D.} and {Fyfe, D.} and {Gallardo, E.} and {Gallegos, J.} and {Gardiol, D.} and {Gebran, M.} and {Gomboc, A.} and {Gómez, A.} and {Grux, E.} and {Gueguen, A.} and {Heyrovsky, A.} and {Hoar, J.} and {Iannicola, G.} and {Isasi Parache, Y.} and {Janotto, A.-M.} and {Joliet, E.} and {Jonckheere, A.} and {Keil, R.} and {Kim, D.-W.} and {Klagyivik, P.} and {Klar, J.} and {Knude, J.} and {Kochukhov, O.} and {Kolka, I.} and {Kos, J.} and {Kutka, A.} and {Lainey, V.} and {LeBouquin, D.} and {Liu, C.} and {Loreggia, D.} and {Makarov, V. V.} and {Marseille, M. G.} and {Martayan, C.} and {Martinez-Rubi, O.} and {Massart, B.} and {Meynadier, F.} and {Mignot, S.} and {Munari, U.} and {Nguyen, A.-T.} and {Nordlander, T.} and {Ocvirk, P.} and {O’Flaherty, K. S.} and {Olias Sanz, A.} and {Ortiz, P.} and {Osorio, J.} and {Oszkiewicz, D.} and {Ouzounis, A.} and {Palmer, M.} and {Park, P.} and {Pasquato, E.} and {Peltzer, C.} and {Peralta, J.} and {Péturaud, F.} and {Pieniluoma, T.} and {Pigozzi, E.} and {Poels, J.} and {Prat, G.} and {Prod’homme, T.} and {Raison, F.} and {Rebordao, J. M.} and {Risquez, D.} and {Rocca-Volmerange, B.} and {Rosen, S.} and {Ruiz-Fuertes, M. I.} and {Russo, F.} and {Sembay, S.} and {Serraller Vizcaino, I.} and {Short, A.} and {Siebert, A.} and {Silva, H.} and {Sinachopoulos, D.} and {Slezak, E.} and {Soffel, M.} and {Sosnowska, D.} and {Straižys, V.} and {ter Linden, M.} and {Terrell, D.} and {Theil, S.} and {Tiede, C.} and {Troisi, L.} and {Tsalmantza, P.} and {Tur, D.} and {Vaccari, M.} and {Vachier, F.} and {Valles, P.} and {Van Hamme, W.} and {Veltz, L.} and {Virtanen, J.} and {Wallut, J.-M.} and {Wichmann, R.} and {Wilkinson, M. I.} and {Ziaeepour, H.} and {Zschocke, S.}},
	title = {The Gaia mission⋆},
	DOI= "10.1051/0004-6361/201629272",
	url= "https://doi.org/10.1051/0004-6361/201629272",
	journal = {A\&A},
	year = 2016,
	volume = 595,
	pages = "A1",
}

@ARTICLE{Tokovinin2000origin,
       author = {{Tokovinin}, A.~A.},
        title = "{On the origin of binaries with twin components}",
      journal = {\aap},
         year = 2000,
        month = aug,
       volume = {360},
        pages = {997-1002},
       adsurl = {https://ui.adsabs.harvard.edu/abs/2000A&A...360..997T}
}

@ARTICLE{Sana2012BInteract,
       author = {{Sana}, H. and {de Mink}, S.~E. and {de Koter}, A. and {Langer}, N. and {Evans}, C.~J. and {Gieles}, M. and {Gosset}, E. and {Izzard}, R.~G. and {Le Bouquin}, J.-B. and {Schneider}, F.~R.~N.},
        title = "{Binary Interaction Dominates the Evolution of Massive Stars}",
      journal = {Science},
         year = 2012,
        month = jul,
       volume = {337},
       number = {6093},
        pages = {444},
          doi = {10.1126/science.1223344},
archivePrefix = {arXiv},
       eprint = {1207.6397},
 primaryClass = {astro-ph.SR},
       adsurl = {https://ui.adsabs.harvard.edu/abs/2012Sci...337..444S}
}

@ARTICLE{Hartkopf1996orbits,
       author = {{Hartkopf}, William I. and {Mason}, Brian D. and {McAlister}, Harold A.},
        title = "{Binary Star Orbits From Speckle Interferometry. VIII. Orbits of 37 Close Visual System}",
      journal = {\aj},
         year = 1996,
        month = jan,
       volume = {111},
        pages = {370},
          doi = {10.1086/117790},
       adsurl = {https://ui.adsabs.harvard.edu/abs/1996AJ....111..370H}
}

@ARTICLE{Eggleton1989vbs,
       author = {{Eggleton}, Peter P. and {Fitchett}, Michael J. and {Tout}, Christopher A.},
        title = "{The Distribution of Visual Binaries with Two Bright Components}",
      journal = {\apj},
         year = 1989,
        month = dec,
       volume = {347},
        pages = {998},
          doi = {10.1086/168190},
       adsurl = {https://ui.adsabs.harvard.edu/abs/1989ApJ...347..998E}
}

@ARTICLE{Badenes2018APOGEEview,
       author = {{Badenes}, Carles and {Mazzola}, Christine and {Thompson}, Todd A. and {Covey}, Kevin and {Freeman}, Peter E. and {Walker}, Matthew G. and {Moe}, Maxwell and {Troup}, Nicholas and {Nidever}, David and {Allende Prieto}, Carlos and {Andrews}, Brett and {Barb{\'a}}, Rodolfo H. and {Beers}, Timothy C. and {Bovy}, Jo and {Carlberg}, Joleen K. and {De Lee}, Nathan and {Johnson}, Jennifer and {Lewis}, Hannah and {Majewski}, Steven R. and {Pinsonneault}, Marc and {Sobeck}, Jennifer and {Stassun}, Keivan G. and {Stringfellow}, Guy S. and {Zasowski}, Gail},
        title = "{Stellar Multiplicity Meets Stellar Evolution and Metallicity: The APOGEE View}",
      journal = {\apj},
         year = 2018,
        month = feb,
       volume = {854},
       number = {2},
          eid = {147},
        pages = {147},
          doi = {10.3847/1538-4357/aaa765},
archivePrefix = {arXiv},
       eprint = {1711.00660},
 primaryClass = {astro-ph.SR},
       adsurl = {https://ui.adsabs.harvard.edu/abs/2018ApJ...854..147B}
}

@ARTICLE{2026Souropanis,
       author = {{Souropanis}, D. and {Zapartas}, E. and {Pessi}, T. and {Briel}, M.~M. and {Renzo}, M. and {Guti{\'e}rrez}, C.~P. and {Andrews}, J.~J. and {Gossage}, S. and {Kruckow}, M.~U. and {Liotine}, C. and {Srivastava}, P.~M. and {Teng}, E.},
        title = "{The power of binaries on stripped-envelope supernovae across metallicity: uniform progenitor parameter space and persistently low ejecta masses, but subtype diversity}",
      journal = {\mnras},
         year = 2026,
        month = jan,
       volume = {545},
       number = {3},
          eid = {staf2163},
        pages = {staf2163},
          doi = {10.1093/mnras/staf2163},
archivePrefix = {arXiv},
       eprint = {2508.21042},
 primaryClass = {astro-ph.SR},
       adsurl = {https://ui.adsabs.harvard.edu/abs/2026MNRAS.545f2163S}
}

@ARTICLE{2023Liu,
       author = {{Liu}, Zheng-Wei and {R{\"o}pke}, Friedrich K. and {Han}, Zhanwen},
        title = "{Type Ia Supernova Explosions in Binary Systems: A Review}",
      journal = {Research in Astronomy and Astrophysics},
         year = 2023,
        month = aug,
       volume = {23},
       number = {8},
          eid = {082001},
        pages = {082001},
          doi = {10.1088/1674-4527/acd89e},
archivePrefix = {arXiv},
       eprint = {2305.13305},
 primaryClass = {astro-ph.HE},
       adsurl = {https://ui.adsabs.harvard.edu/abs/2023RAA....23h2001L}
}

@article{ChenX2024newMil,
title = {Binary stars in the new millennium},
journal = {Progress in Particle and Nuclear Physics},
volume = {134},
pages = {104083},
year = {2024},
issn = {0146-6410},
doi = {https://doi.org/10.1016/j.ppnp.2023.104083},
url = {https://www.sciencedirect.com/science/article/pii/S0146641023000649},
author = {Xuefei Chen and Zhengwei Liu and Zhanwen Han}
}

@ARTICLE{Duchene2013multipl,
       author = {{Duch{\^e}ne}, Gaspard and {Kraus}, Adam},
        title = "{Stellar Multiplicity}",
      journal = {\araa},
         year = 2013,
        month = aug,
       volume = {51},
       number = {1},
        pages = {269-310},
          doi = {10.1146/annurev-astro-081710-102602},
archivePrefix = {arXiv},
       eprint = {1303.3028},
 primaryClass = {astro-ph.SR},
       adsurl = {https://ui.adsabs.harvard.edu/abs/2013ARA&A..51..269D}
}

@article{RM2007PCEB,
    author = {Rebassa-Mansergas, A. and Gänsicke, B. T. and Rodríguez-Gil, P. and Schreiber, M. R. and Koester, D.},
    title = {Post-common-envelope binaries from SDSS – I. 101 white dwarf main-sequence binaries with multiple Sloan Digital Sky Survey spectroscopy},
    journal = {Monthly Notices of the Royal Astronomical Society},
    volume = {382},
    number = {4},
    pages = {1377-1393},
    year = {2007},
    month = {12},
    issn = {0035-8711},
    doi = {10.1111/j.1365-2966.2007.12288.x},
    url = {https://doi.org/10.1111/j.1365-2966.2007.12288.x},
    eprint = {https://academic.oup.com/mnras/article-pdf/382/4/1377/40680636/mnras_382_4_1377.pdf},
}

@ARTICLE{ElBadry2018Signatures,
       author = {{El-Badry}, Kareem and {Rix}, Hans-Walter and {Ting}, Yuan-Sen and {Weisz}, Daniel R. and {Bergemann}, Maria and {Cargile}, Phillip and {Conroy}, Charlie and {Eilers}, Anna-Christina},
        title = "{Signatures of unresolved binaries in stellar spectra: implications for spectral fitting}",
      journal = {\mnras},
         year = 2018,
        month = feb,
       volume = {473},
       number = {4},
        pages = {5043-5049},
          doi = {10.1093/mnras/stx2758},
archivePrefix = {arXiv},
       eprint = {1709.03983},
 primaryClass = {astro-ph.SR},
       adsurl = {https://ui.adsabs.harvard.edu/abs/2018MNRAS.473.5043E}
}

@article{Roulston2020classify,
doi = {10.3847/1538-4365/aba1e7},
url = {https://doi.org/10.3847/1538-4365/aba1e7},
year = {2020},
month = {aug},
publisher = {The American Astronomical Society},
volume = {249},
number = {2},
pages = {34},
author = {Roulston, Benjamin R. and Green, Paul J. and Kesseli, Aurora Y.},
title = {Classifying Single Stars and Spectroscopic Binaries Using Optical Stellar Templates},
journal = {The Astrophysical Journal Supplement Series}
}

@ARTICLE{Xiang2021DDSp,
       author = {{Xiang}, Maosheng and {Rix}, Hans-Walter and {Ting}, Yuan-Sen and {Zari}, Eleonora and {El-Badry}, Kareem and {Yuan}, Hai-Bo and {Cui}, Wen-Yuan},
        title = "{Data-driven Spectroscopic Estimates of Absolute Magnitude, Distance, and Binarity: Method and Catalog of 16,002 O- and B-type Stars from LAMOST}",
      journal = {\apjs},
         year = 2021,
        month = mar,
       volume = {253},
       number = {1},
          eid = {22},
        pages = {22},
          doi = {10.3847/1538-4365/abd6ba},
archivePrefix = {arXiv},
       eprint = {2008.10637},
 primaryClass = {astro-ph.SR},
       adsurl = {https://ui.adsabs.harvard.edu/abs/2021ApJS..253...22X}
}

@article{Moe2013MWandMC,
doi = {10.1088/0004-637X/778/2/95},
url = {https://doi.org/10.1088/0004-637X/778/2/95},
year = {2013},
month = {nov},
publisher = {The American Astronomical Society},
volume = {778},
number = {2},
pages = {95},
author = {Moe, Maxwell and Di Stefano, Rosanne},
title = {THE CLOSE BINARY PROPERTIES OF MASSIVE STARS IN THE MILKY WAY AND LOW-METALLICITY MAGELLANIC CLOUDS},
journal = {The Astrophysical Journal}
}

@article{TianZ2018lmdr4,
doi = {10.1088/1674-4527/18/5/52},
url = {https://doi.org/10.1088/1674-4527/18/5/52},
year = {2018},
month = {may},
publisher = {National Astronomical Observatories, CAS and IOP Publishing Ltd.},
volume = {18},
number = {5},
pages = {052},
author = {Tian, Zhi-Jia and Liu, Xiao-Wei and Yuan, Hai-Bo and Chen, Bing-Qiu and Xiang, Mao-Sheng and Huang, Yang and Wang, Chun and Zhang, Hua-Wei and Guo, Jin-Cheng and Ren, Juan-Juan and Huo, Zhi-Ying and Yang, Yong and Zhang, Meng and Bi, Shao-Lan and Yang, Wu-Ming and Liu, Kang and Zhang, Xian-Fei and Li, Tan-Da and Wu, Ya-Qian and Zhang, Jing-Hua},
title = {Binary Star Fractions from the LAMOST DR4},
journal = {Research in Astronomy and Astrophysics}
}

@ARTICLE{Pawlak2016OGLEMCs,
       author = {{Pawlak}, M. and {Soszy{\'n}ski}, I. and {Udalski}, A. and {Szyma{\'n}ski}, M.~K. and {Wyrzykowski}, {\L}. and {Ulaczyk}, K. and {Poleski}, R. and {Pietrukowicz}, P. and {Koz{\l}owski}, S. and {Skowron}, D.~M. and {Skowron}, J. and {Mr{\'o}z}, P. and {Hamanowicz}, A.},
        title = "{The OGLE Collection of Variable Stars. Eclipsing Binaries in the Magellanic System}",
      journal = {\actaa},
         year = 2016,
        month = dec,
       volume = {66},
       number = {4},
        pages = {421-432},
          doi = {10.48550/arXiv.1612.06394},
archivePrefix = {arXiv},
       eprint = {1612.06394},
 primaryClass = {astro-ph.SR},
       adsurl = {https://ui.adsabs.harvard.edu/abs/2016AcA....66..421P}
}

@ARTICLE{Soszynski2016OGLEbulge,
       author = {{Soszy{\'n}ski}, I. and {Pawlak}, M. and {Pietrukowicz}, P. and {Udalski}, A. and {Szyma{\'n}ski}, M.~K. and {Wyrzykowski}, {\L}. and {Ulaczyk}, K. and {Poleski}, R. and {Koz{\l}owski}, S. and {Skowron}, D.~M. and {Skowron}, J. and {Mr{\'o}z}, P. and {Hamanowicz}, A.},
        title = "{The OGLE Collection of Variable Stars. Over 450 000 Eclipsing and Ellipsoidal Binary Systems Toward the Galactic Bulge}",
      journal = {\actaa},
         year = 2016,
        month = dec,
       volume = {66},
       number = {4},
        pages = {405-420},
          doi = {10.48550/arXiv.1701.03105},
archivePrefix = {arXiv},
       eprint = {1701.03105},
 primaryClass = {astro-ph.SR},
       adsurl = {https://ui.adsabs.harvard.edu/abs/2016AcA....66..405S}
}

@ARTICLE{Jayasinghe2019ASASSN,
       author = {{Jayasinghe}, T. and {Stanek}, K.~Z. and {Kochanek}, C.~S. and {Shappee}, B.~J. and {Holoien}, T.~W.-S. and {Thompson}, Todd A. and {Prieto}, J.~L. and {Dong}, Subo and {Pawlak}, M. and {Pejcha}, O. and {Shields}, J.~V. and {Pojmanski}, G. and {Otero}, S. and {Hurst}, N. and {Britt}, C.~A. and {Will}, D.},
        title = "{The ASAS-SN catalogue of variable stars III: variables in the southern TESS continuous viewing zone}",
      journal = {\mnras},
         year = 2019,
        month = may,
       volume = {485},
       number = {1},
        pages = {961-971},
          doi = {10.1093/mnras/stz444},
archivePrefix = {arXiv},
       eprint = {1901.00009},
 primaryClass = {astro-ph.SR},
       adsurl = {https://ui.adsabs.harvard.edu/abs/2019MNRAS.485..961J}
}

@article{ElBadry2019wb,
    author = {El-Badry, Kareem and Rix, Hans-Walter},
    title = {The wide binary fraction of solar-type stars: emergence of metallicity dependence at a \&lt; 200 au},
    journal = {Monthly Notices of the Royal Astronomical Society: Letters},
    volume = {482},
    number = {1},
    pages = {L139-L144},
    year = {2019},
    month = {01},
    issn = {1745-3925},
    doi = {10.1093/mnrasl/sly206},
    url = {https://doi.org/10.1093/mnrasl/sly206},
    eprint = {https://academic.oup.com/mnrasl/article-pdf/482/1/L139/56977189/mnrasl_482_1_l139.pdf},
}

@article{ChenX2020ZTF,
doi = {10.3847/1538-4365/ab9cae},
url = {https://doi.org/10.3847/1538-4365/ab9cae},
year = {2020},
month = {jul},
publisher = {The American Astronomical Society},
volume = {249},
number = {1},
pages = {18},
author = {Chen, Xiaodian and Wang, Shu and Deng, Licai and de Grijs, Richard and Yang, Ming and Tian, Hao},
title = {The Zwicky Transient Facility Catalog of Periodic Variable Stars},
journal = {The Astrophysical Journal Supplement Series}
}

@article{Heinze2018ATLAS,
doi = {10.3847/1538-3881/aae47f},
url = {https://doi.org/10.3847/1538-3881/aae47f},
year = {2018},
month = {nov},
publisher = {The American Astronomical Society},
volume = {156},
number = {5},
pages = {241},
author = {Heinze, A. N. and Tonry, J. L. and Denneau, L. and Flewelling, H. and Stalder, B. and Rest, A. and Smith, K. W. and Smartt, S. J. and Weiland, H.},
title = {A First Catalog of Variable Stars Measured by the Asteroid Terrestrial-impact Last Alert System (ATLAS)},
journal = {The Astronomical Journal}
}

@ARTICLE{Sana2014southernmassive,
       author = {{Sana}, H. and {Le Bouquin}, J.-B. and {Lacour}, S. and {Berger}, J.-P. and {Duvert}, G. and {Gauchet}, L. and {Norris}, B. and {Olofsson}, J. and {Pickel}, D. and {Zins}, G. and {Absil}, O. and {de Koter}, A. and {Kratter}, K. and {Schnurr}, O. and {Zinnecker}, H.},
        title = "{Southern Massive Stars at High Angular Resolution: Observational Campaign and Companion Detection}",
      journal = {\apjs},
         year = 2014,
        month = nov,
       volume = {215},
       number = {1},
          eid = {15},
        pages = {15},
          doi = {10.1088/0067-0049/215/1/15},
archivePrefix = {arXiv},
       eprint = {1409.6304},
 primaryClass = {astro-ph.SR},
       adsurl = {https://ui.adsabs.harvard.edu/abs/2014ApJS..215...15S}
}

@INPROCEEDINGS{OT2024DYNAMO,
       author = {{Ordo{\~n}ez-Toro}, J. and {Dzib}, S.~A. and {Loinard}, L.},
        title = "{Dynamical Masses of Young Stellar Multiple Systems with the VLBA (DYNAMO-VLBA)}",
    booktitle = {Proceedings of the 16th EVN Symposium},
         year = 2024,
       editor = {{Ros}, E. and {Benke}, P. and {Dzib}, S.~A. and {Rottmann}, I. and {Zensus}, J.~A.},
        month = sep,
        pages = {35-38},
          doi = {10.48550/arXiv.2511.04642},
archivePrefix = {arXiv},
       eprint = {2511.04642},
 primaryClass = {astro-ph.SR},
       adsurl = {https://ui.adsabs.harvard.edu/abs/2024evn..conf...35O}
}

@article{Hartman2020SUPERWIDE,
doi = {10.3847/1538-4365/ab79a6},
url = {https://doi.org/10.3847/1538-4365/ab79a6},
year = {2020},
month = {apr},
publisher = {The American Astronomical Society},
volume = {247},
number = {2},
pages = {66},
author = {Hartman, Zachary D. and Lépine, Sébastien},
title = {The SUPERWIDE Catalog: A Catalog of 99,203 Wide Binaries Found in Gaia and Supplemented by the SUPERBLINK High Proper Motion Catalog},
journal = {The Astrophysical Journal Supplement Series}
}

@article{Drake2017catalina,
    author = {Drake, A. J. and Djorgovski, S. G. and Catelan, M. and Graham, M. J. and Mahabal, A. A. and Larson, S. and Christensen, E. and Torrealba, G. and Beshore, E. and McNaught, R. H. and Garradd, G. and Belokurov, V. and Koposov, S. E.},
    title = {The Catalina Surveys Southern periodic variable star catalogue},
    journal = {Monthly Notices of the Royal Astronomical Society},
    volume = {469},
    number = {3},
    pages = {3688-3712},
    year = {2017},
    month = {08},
    issn = {0035-8711},
    doi = {10.1093/mnras/stx1085},
    url = {https://doi.org/10.1093/mnras/stx1085},
    eprint = {https://academic.oup.com/mnras/article-pdf/469/3/3688/18211203/stx1085.pdf},
}

@article{Moe2019STclosebinary,
doi = {10.3847/1538-4357/ab0d88},
url = {https://doi.org/10.3847/1538-4357/ab0d88},
year = {2019},
month = {apr},
publisher = {The American Astronomical Society},
volume = {875},
number = {1},
pages = {61},
author = {Moe, Maxwell and Kratter, Kaitlin M. and Badenes, Carles},
title = {The Close Binary Fraction of Solar-type Stars Is Strongly Anticorrelated with Metallicity},
journal = {The Astrophysical Journal}
}

@BOOK{Warner1995cataclysmic,
       author = {{Warner}, Brian},
        title = "{Cataclysmic variable stars}",
         year = 1995,
       volume = {28},
       adsurl = {https://ui.adsabs.harvard.edu/abs/1995cvs..book.....W}
}

@ARTICLE{Metzger2020KNe,
       author = {{Metzger}, Brian D.},
        title = "{Kilonovae}",
      journal = {Living Reviews in Relativity},
         year = 2020,
        month = dec,
       volume = {23},
       number = {1},
          eid = {1},
        pages = {1},
          doi = {10.1007/s41114-019-0024-0},
archivePrefix = {arXiv},
       eprint = {1910.01617},
 primaryClass = {astro-ph.HE},
       adsurl = {https://ui.adsabs.harvard.edu/abs/2020LRR....23....1M}
}

@ARTICLE{2014Berger,
       author = {{Berger}, Edo},
        title = "{Short-Duration Gamma-Ray Bursts}",
      journal = {\araa},
         year = 2014,
        month = aug,
       volume = {52},
        pages = {43-105},
          doi = {10.1146/annurev-astro-081913-035926},
archivePrefix = {arXiv},
       eprint = {1311.2603},
 primaryClass = {astro-ph.HE},
       adsurl = {https://ui.adsabs.harvard.edu/abs/2014ARA&A..52...43B}
}

@ARTICLE{Tetarenko2016BHXB,
       author = {{Tetarenko}, B.~E. and {Sivakoff}, G.~R. and {Heinke}, C.~O. and {Gladstone}, J.~C.},
        title = "{WATCHDOG: A Comprehensive All-sky Database of Galactic Black Hole X-ray Binaries}",
      journal = {\apjs},
         year = 2016,
        month = feb,
       volume = {222},
       number = {2},
          eid = {15},
        pages = {15},
          doi = {10.3847/0067-0049/222/2/15},
archivePrefix = {arXiv},
       eprint = {1512.00778},
 primaryClass = {astro-ph.HE},
       adsurl = {https://ui.adsabs.harvard.edu/abs/2016ApJS..222...15T}
}

@ARTICLE{Arcones2023eleorigin,
       author = {{Arcones}, Almudena and {Thielemann}, Friedrich-Karl},
        title = "{Origin of the elements}",
      journal = {\aapr},
         year = 2023,
        month = dec,
       volume = {31},
       number = {1},
          eid = {1},
        pages = {1},
          doi = {10.1007/s00159-022-00146-x},
       adsurl = {https://ui.adsabs.harvard.edu/abs/2023A&ARv..31....1A}
}

@ARTICLE{Abbott2023GWTC3,
       author = {{Abbott}, R. and {Abbott}, T.~D. and {Acernese}, F. and {Ackley}, K. and {Adams}, C. and {Adhikari}, N. and {Adhikari}, R.~X. and {Adya}, V.~B. and {Affeldt}, C. and {Agarwal}, D. and {Agathos}, M. and {Agatsuma}, K. and {Aggarwal}, N. and {Aguiar}, O.~D. and {Aiello}, L. and {Ain}, A. and {Ajith}, P. and {Akcay}, S. and {Akutsu}, T. and {Albanesi}, S. and {Allocca}, A. and {Altin}, P.~A. and {Amato}, A. and {Anand}, C. and {Anand}, S. and {Ananyeva}, A. and {Anderson}, S.~B. and {Anderson}, W.~G. and {Ando}, M. and {Andrade}, T. and {Andres}, N. and {Andri{\'c}}, T. and {Angelova}, S.~V. and {Ansoldi}, S. and {Antelis}, J.~M. and {Antier}, S. and {Appert}, S. and {Arai}, Koji and {Arai}, Koya and {Arai}, Y. and {Araki}, S. and {Araya}, A. and {Araya}, M.~C. and {Areeda}, J.~S. and {Ar{\`e}ne}, M. and {Aritomi}, N. and {Arnaud}, N. and {Arogeti}, M. and {Aronson}, S.~M. and {Arun}, K.~G. and {Asada}, H. and {Asali}, Y. and {Ashton}, G. and {Aso}, Y. and {Assiduo}, M. and {Aston}, S.~M. and {Astone}, P. and {Aubin}, F. and {Austin}, C. and {Babak}, S. and {Badaracco}, F. and {Bader}, M.~K.~M. and {Badger}, C. and {Bae}, S. and {Bae}, Y. and {Baer}, A.~M. and {Bagnasco}, S. and {Bai}, Y. and {Baiotti}, L. and {Baird}, J. and {Bajpai}, R. and {Ball}, M. and {Ballardin}, G. and {Ballmer}, S.~W. and {Balsamo}, A. and {Baltus}, G. and {Banagiri}, S. and {Bankar}, D. and {Barayoga}, J.~C. and {Barbieri}, C. and {Barish}, B.~C. and {Barker}, D. and {Barneo}, P. and {Barone}, F. and {Barr}, B. and {Barsotti}, L. and {Barsuglia}, M. and {Barta}, D. and {Bartlett}, J. and {Barton}, M.~A. and {Bartos}, I. and {Bassiri}, R. and {Basti}, A. and {Bawaj}, M. and {Bayley}, J.~C. and {Baylor}, A.~C. and {Bazzan}, M. and {B{\'e}csy}, B. and {Bedakihale}, V.~M. and {Bejger}, M. and {Belahcene}, I. and {Benedetto}, V. and {Beniwal}, D. and {Bennett}, T.~F. and {Bentley}, J.~D. and {Benyaala}, M. and {Bergamin}, F. and {Berger}, B.~K. and {Bernuzzi}, S. and {Berry}, C.~P.~L. and {Bersanetti}, D. and {Bertolini}, A. and {Betzwieser}, J. and {Beveridge}, D. and {Bhandare}, R. and {Bhardwaj}, U. and {Bhattacharjee}, D. and {Bhaumik}, S. and {Bilenko}, I.~A. and {Billingsley}, G. and {Bini}, S. and {Birney}, R. and {Birnholtz}, O. and {Biscans}, S. and {Bischi}, M. and {Biscoveanu}, S. and {Bisht}, A. and {Biswas}, B. and {Bitossi}, M. and {Bizouard}, M.-A. and {Blackburn}, J.~K. and {Blair}, C.~D. and {Blair}, D.~G. and {Blair}, R.~M. and {Bobba}, F. and {Bode}, N. and {Boer}, M. and {Bogaert}, G. and {Boldrini}, M. and {Bonavena}, L.~D. and {Bondu}, F. and {Bonilla}, E. and {Bonnand}, R. and {Booker}, P. and {Boom}, B.~A. and {Bork}, R. and {Boschi}, V. and {Bose}, N. and {Bose}, S. and {Bossilkov}, V. and {Boudart}, V. and {Bouffanais}, Y. and {Bozzi}, A. and {Bradaschia}, C. and {Brady}, P.~R. and {Bramley}, A. and {Branch}, A. and {Branchesi}, M. and {Brandt}, J. and {Brau}, J.~E. and {Breschi}, M. and {Briant}, T. and {Briggs}, J.~H. and {Brillet}, A. and {Brinkmann}, M. and {Brockill}, P. and {Brooks}, A.~F. and {Brooks}, J. and {Brown}, D.~D. and {Brunett}, S. and {Bruno}, G. and {Bruntz}, R. and {Bryant}, J. and {Bulik}, T. and {Bulten}, H.~J. and {Buonanno}, A. and {Buscicchio}, R. and {Buskulic}, D. and {Buy}, C. and {Byer}, R.~L. and {Davies}, G.~S. Cabourn and {Cadonati}, L. and {Cagnoli}, G. and {Cahillane}, C. and {Bustillo}, J. Calder{\'o}n and {Callaghan}, J.~D. and {Callister}, T.~A. and {Calloni}, E. and {Cameron}, J. and {Camp}, J.~B. and {Canepa}, M. and {Canevarolo}, S. and {Cannavacciuolo}, M. and {Cannon}, K.~C. and {Cao}, H. and {Cao}, Z. and {Capocasa}, E. and {Capote}, E. and {Carapella}, G. and {Carbognani}, F.},
        title = "{GWTC-3: Compact Binary Coalescences Observed by LIGO and Virgo during the Second Part of the Third Observing Run}",
      journal = {Physical Review X},
         year = 2023,
        month = oct,
       volume = {13},
       number = {4},
          eid = {041039},
        pages = {041039},
          doi = {10.1103/PhysRevX.13.041039},
archivePrefix = {arXiv},
       eprint = {2111.03606},
 primaryClass = {gr-qc},
       adsurl = {https://ui.adsabs.harvard.edu/abs/2023PhRvX..13d1039A}
}

@article{HanZ2020binarypopsyn,
doi = {10.1088/1674-4527/20/10/161},
url = {https://doi.org/10.1088/1674-4527/20/10/161},
year = {2020},
month = {oct},
publisher = {National Astronomical Observatories, CAS and IOP Publishing Ltd.},
volume = {20},
number = {10},
pages = {161},
author = {Han, Zhan-Wen and Ge, Hong-Wei and Chen, Xue-Fei and Chen, Hai-Liang},
title = {Binary Population Synthesis},
journal = {Research in Astronomy and Astrophysics}
}

@article{Li2022giantparams,
    author = {Li, Zhuohan and Zhao, Gang and Chen, Yuqin and Liang, Xilong and Zhao, Jingkun},
    title = {The stellar parameters and elemental abundances from low-resolution spectra – I. 1.2 million giants from LAMOST DR8},
    journal = {Monthly Notices of the Royal Astronomical Society},
    volume = {517},
    number = {4},
    pages = {4875-4891},
    year = {2022},
    month = {12},
    issn = {0035-8711},
    doi = {10.1093/mnras/stac1959},
    url = {https://doi.org/10.1093/mnras/stac1959},
    eprint = {https://academic.oup.com/mnras/article-pdf/517/4/4875/46854744/stac1959.pdf},
}

@ARTICLE{Zhukovska2018FeSidust,
       author = {{Zhukovska}, Svitlana and {Henning}, Thomas and {Dobbs}, Clare},
        title = "{Iron and Silicate Dust Growth in the Galactic Interstellar Medium: Clues from Element Depletions}",
      journal = {\apj},
         year = 2018,
        month = apr,
       volume = {857},
       number = {2},
          eid = {94},
        pages = {94},
          doi = {10.3847/1538-4357/aab438},
archivePrefix = {arXiv},
       eprint = {1803.01929},
 primaryClass = {astro-ph.GA},
       adsurl = {https://ui.adsabs.harvard.edu/abs/2018ApJ...857...94Z}
}

@misc{FM2013emcee,
       author = {{Foreman-Mackey}, Daniel and {Conley}, Alex and {Meierjurgen Farr}, Will and {Hogg}, David W. and {Lang}, Dustin and {Marshall}, Phil and {Price-Whelan}, Adrian and {Sanders}, Jeremy and {Zuntz}, Joe},
        title = "{emcee: The MCMC Hammer}",
     publisher = {Astrophysics Source Code Library, record ascl:1303.002},
         year = 2013,
        month = mar,
          eid = {ascl:1303.002},
archivePrefix = {ascl},
       eprint = {1303.002},
       adsurl = {https://ui.adsabs.harvard.edu/abs/2013ascl.soft03002F}
}

@article{Bressan2012PARSEC,
    author = {Bressan, Alessandro and Marigo, Paola and Girardi, Léo and Salasnich, Bernardo and Dal Cero, Claudia and Rubele, Stefano and Nanni, Ambra},
    title = {PARSEC: stellar tracks and isochrones with the PAdova and TRieste Stellar Evolution Code},
    journal = {Monthly Notices of the Royal Astronomical Society},
    volume = {427},
    number = {1},
    pages = {127-145},
    year = {2012},
    month = {11},
    issn = {0035-8711},
    doi = {10.1111/j.1365-2966.2012.21948.x},
    url = {https://doi.org/10.1111/j.1365-2966.2012.21948.x},
    eprint = {https://academic.oup.com/mnras/article-pdf/427/1/127/18231042/427-1-127.pdf},
}

@ARTICLE{Lijd2025MSbinary,
       author = {{Li}, Jiadong and {Rix}, Hans-Walter and {Ting}, Yuan-Sen and {M{\"u}ller-Horn}, Johanna and {El-Badry}, Kareem and {Liu}, Chao and {Seeburger}, Rhys and {Green}, Gregory M. and {Zhang}, Xiangyu},
        title = "{Millions of main-sequence binary stars from Gaia BP/RP spectra}",
      journal = {\aap},
         year = 2025,
        month = dec,
       volume = {704},
          eid = {A126},
        pages = {A126},
          doi = {10.1051/0004-6361/202556362},
archivePrefix = {arXiv},
       eprint = {2507.09622},
 primaryClass = {astro-ph.SR},
       adsurl = {https://ui.adsabs.harvard.edu/abs/2025A&A...704A.126L}
}

@ARTICLE{Zhaog2012LAMOSToverview,
       author = {{Zhao}, Gang and {Zhao}, Yong-Heng and {Chu}, Yao-Quan and {Jing}, Yi-Peng and {Deng}, Li-Cai},
        title = "{LAMOST spectral survey {\textemdash} An overview}",
      journal = {Research in Astronomy and Astrophysics},
         year = 2012,
        month = jul,
       volume = {12},
       number = {7},
        pages = {723-734},
          doi = {10.1088/1674-4527/12/7/002},
       adsurl = {https://ui.adsabs.harvard.edu/abs/2012RAA....12..723Z}
}

@ARTICLE{Dengl2012LAMOSTsp,
       author = {{Deng}, Li-Cai and {Newberg}, Heidi Jo and {Liu}, Chao and {Carlin}, Jeffrey L. and {Beers}, Timothy C. and {Chen}, Li and {Chen}, Yu-Qin and {Christlieb}, Norbert and {Grillmair}, Carl J. and {Guhathakurta}, Puragra and {Han}, Zhan-Wen and {Hou}, Jin-Liang and {Lee}, Hsu-Tai and {L{\'e}pine}, S{\'e}bastien and {Li}, Jing and {Liu}, Xiao-Wei and {Pan}, Kai-Ke and {Sellwood}, J.~A. and {Wang}, Bo and {Wang}, Hong-Chi and {Yang}, Fan and {Yanny}, Brian and {Zhang}, Hao-Tong and {Zhang}, Yue-Yang and {Zheng}, Zheng and {Zhu}, Zi},
        title = "{LAMOST Experiment for Galactic Understanding and Exploration (LEGUE) {\textemdash} The survey's science plan}",
      journal = {Research in Astronomy and Astrophysics},
         year = 2012,
        month = jul,
       volume = {12},
       number = {7},
        pages = {735-754},
          doi = {10.1088/1674-4527/12/7/003},
archivePrefix = {arXiv},
       eprint = {1206.3578},
 primaryClass = {astro-ph.GA},
       adsurl = {https://ui.adsabs.harvard.edu/abs/2012RAA....12..735D}
}

@ARTICLE{Cuix2012LAMOST,
       author = {{Cui}, Xiang-Qun and {Zhao}, Yong-Heng and {Chu}, Yao-Quan and {Li}, Guo-Ping and {Li}, Qi and {Zhang}, Li-Ping and {Su}, Hong-Jun and {Yao}, Zheng-Qiu and {Wang}, Ya-Nan and {Xing}, Xiao-Zheng and {Li}, Xin-Nan and {Zhu}, Yong-Tian and {Wang}, Gang and {Gu}, Bo-Zhong and {Luo}, A.-Li and {Xu}, Xin-Qi and {Zhang}, Zhen-Chao and {Liu}, Gen-Rong and {Zhang}, Hao-Tong and {Yang}, De-Hua and {Cao}, Shu-Yun and {Chen}, Hai-Yuan and {Chen}, Jian-Jun and {Chen}, Kun-Xin and {Chen}, Ying and {Chu}, Jia-Ru and {Feng}, Lei and {Gong}, Xue-Fei and {Hou}, Yong-Hui and {Hu}, Hong-Zhuan and {Hu}, Ning-Sheng and {Hu}, Zhong-Wen and {Jia}, Lei and {Jiang}, Fang-Hua and {Jiang}, Xiang and {Jiang}, Zi-Bo and {Jin}, Ge and {Li}, Ai-Hua and {Li}, Yan and {Li}, Ye-Ping and {Liu}, Guan-Qun and {Liu}, Zhi-Gang and {Lu}, Wen-Zhi and {Mao}, Yin-Dun and {Men}, Li and {Qi}, Yong-Jun and {Qi}, Zhao-Xiang and {Shi}, Huo-Ming and {Tang}, Zheng-Hong and {Tao}, Qing-Sheng and {Wang}, Da-Qi and {Wang}, Dan and {Wang}, Guo-Min and {Wang}, Hai and {Wang}, Jia-Ning and {Wang}, Jian and {Wang}, Jian-Ling and {Wang}, Jian-Ping and {Wang}, Lei and {Wang}, Shu-Qing and {Wang}, You and {Wang}, Yue-Fei and {Xu}, Ling-Zhe and {Xu}, Yan and {Yang}, Shi-Hai and {Yu}, Yong and {Yuan}, Hui and {Yuan}, Xiang-Yan and {Zhai}, Chao and {Zhang}, Jing and {Zhang}, Yan-Xia and {Zhang}, Yong and {Zhao}, Ming and {Zhou}, Fang and {Zhou}, Guo-Hua and {Zhu}, Jie and {Zou}, Si-Cheng},
        title = "{The Large Sky Area Multi-Object Fiber Spectroscopic Telescope (LAMOST)}",
      journal = {Research in Astronomy and Astrophysics},
         year = 2012,
        month = sep,
       volume = {12},
       number = {9},
        pages = {1197-1242},
          doi = {10.1088/1674-4527/12/9/003},
       adsurl = {https://ui.adsabs.harvard.edu/abs/2012RAA....12.1197C}
}

@ARTICLE{Luoa2015firstlm,
       author = {{Luo}, A.-Li and {Zhao}, Yong-Heng and {Zhao}, Gang and {Deng}, Li-Cai and {Liu}, Xiao-Wei and {Jing}, Yi-Peng and {Wang}, Gang and {Zhang}, Hao-Tong and {Shi}, Jian-Rong and {Cui}, Xiang-Qun and {Chu}, Yao-Quan and {Li}, Guo-Ping and {Bai}, Zhong-Rui and {Wu}, Yue and {Cai}, Yan and {Cao}, Shu-Yun and {Cao}, Zi-Huang and {Carlin}, Jeffrey L. and {Chen}, Hai-Yuan and {Chen}, Jian-Jun and {Chen}, Kun-Xin and {Chen}, Li and {Chen}, Xue-Lei and {Chen}, Xiao-Yan and {Chen}, Ying and {Christlieb}, Norbert and {Chu}, Jia-Ru and {Cui}, Chen-Zhou and {Dong}, Yi-Qiao and {Du}, Bing and {Fan}, Dong-Wei and {Feng}, Lei and {Fu}, Jian-Ning and {Gao}, Peng and {Gong}, Xue-Fei and {Gu}, Bo-Zhong and {Guo}, Yan-Xin and {Han}, Zhan-Wen and {He}, Bo-Liang and {Hou}, Jin-Liang and {Hou}, Yong-Hui and {Hou}, Wen and {Hu}, Hong-Zhuan and {Hu}, Ning-Sheng and {Hu}, Zhong-Wen and {Huo}, Zhi-Ying and {Jia}, Lei and {Jiang}, Fang-Hua and {Jiang}, Xiang and {Jiang}, Zhi-Bo and {Jin}, Ge and {Kong}, Xiao and {Kong}, Xu and {Lei}, Ya-Juan and {Li}, Ai-Hua and {Li}, Chang-Hua and {Li}, Guang-Wei and {Li}, Hai-Ning and {Li}, Jian and {Li}, Qi and {Li}, Shuang and {Li}, Sha-Sha and {Li}, Xin-Nan and {Li}, Yan and {Li}, Yin-Bi and {Li}, Ye-Ping and {Liang}, Yuan and {Lin}, Chien-Cheng and {Liu}, Chao and {Liu}, Gen-Rong and {Liu}, Guan-Qun and {Liu}, Zhi-Gang and {Lu}, Wen-Zhi and {Luo}, Yu and {Mao}, Yin-Dun and {Newberg}, Heidi and {Ni}, Ji-Jun and {Qi}, Zhao-Xiang and {Qi}, Yong-Jun and {Shen}, Shi-Yin and {Shi}, Huo-Ming and {Song}, Jing and {Song}, Yi-Han and {Su}, Ding-Qiang and {Su}, Hong-Jun and {Tang}, Zheng-Hong and {Tao}, Qing-Sheng and {Tian}, Yuan and {Wang}, Dan and {Wang}, Da-Qi and {Wang}, Feng-Fei and {Wang}, Guo-Min and {Wang}, Hai and {Wang}, Hong-Chi and {Wang}, Jian and {Wang}, Jia-Ning and {Wang}, Jian-Ling and {Wang}, Jian-Ping and {Wang}, Jun-Xian and {Wang}, Lei and {Wang}, Meng-Xin and {Wang}, Shou-Guan and {Wang}, Shu-Qing and {Wang}, Xia and {Wang}, Ya-Nan and {Wang}, You and {Wang}, Yue-Fei and {Wang}, You-Fen and {Wei}, Peng and {Wei}, Ming-Zhi and {Wu}, Hong and {Wu}, Ke-Fei and {Wu}, Xue-Bing and {Wu}, Yu-Zhong and {Xing}, Xiao-Zheng and {Xu}, Ling-Zhe and {Xu}, Xin-Qi and {Xu}, Yan and {Yan}, Tai-Sheng and {Yang}, De-Hua and {Yang}, Hai-Feng and {Yang}, Hui-Qin and {Yang}, Ming and {Yao}, Zheng-Qiu and {Yu}, Yong and {Yuan}, Hui and {Yuan}, Hai-Bo and {Yuan}, Hai-Long and {Yuan}, Wei-Min and {Zhai}, Chao and {Zhang}, En-Peng and {Zhang}, Hua-Wei and {Zhang}, Jian-Nan and {Zhang}, Li-Pin and {Zhang}, Wei and {Zhang}, Yong and {Zhang}, Yan-Xia and {Zhang}, Zheng-Chao and {Zhao}, Ming and {Zhou}, Fang and {Zhou}, Xu and {Zhu}, Jie and {Zhu}, Yong-Tian and {Zou}, Si-Cheng and {Zuo}, Fang},
        title = "{The first data release (DR1) of the LAMOST regular survey}",
      journal = {Research in Astronomy and Astrophysics},
         year = 2015,
        month = aug,
       volume = {15},
       number = {8},
          eid = {1095},
        pages = {1095},
          doi = {10.1088/1674-4527/15/8/002},
archivePrefix = {arXiv},
       eprint = {1505.01570},
 primaryClass = {astro-ph.GA},
       adsurl = {https://ui.adsabs.harvard.edu/abs/2015RAA....15.1095L}
}

@ARTICLE{GaiaCollab2022edr3,
       author = {{Gaia Collaboration} and {Brown}, A.~G.~A. and {Vallenari}, A. and {Prusti}, T. and {de Bruijne}, J.~H.~J. and {Babusiaux}, C. and {Biermann}, M. and {Creevey}, O.~L. and {Evans}, D.~W. and {Eyer}, L. and {Hutton}, A. and {Jansen}, F. and {Jordi}, C. and {Klioner}, S.~A. and {Lammers}, U. and {Lindegren}, L. and {Luri}, X. and {Mignard}, F. and {Panem}, C. and {Pourbaix}, D. and {Randich}, S. and {Sartoretti}, P. and {Soubiran}, C. and {Walton}, N.~A. and {Arenou}, F. and {Bailer-Jones}, C.~A.~L. and {Bastian}, U. and {Cropper}, M. and {Drimmel}, R. and {Katz}, D. and {Lattanzi}, M.~G. and {van Leeuwen}, F. and {Bakker}, J. and {Cacciari}, C. and {Casta{\~n}eda}, J. and {De Angeli}, F. and {Ducourant}, C. and {Fabricius}, C. and {Fouesneau}, M. and {Fr{\'e}mat}, Y. and {Guerra}, R. and {Guerrier}, A. and {Guiraud}, J. and {Jean-Antoine Piccolo}, A. and {Masana}, E. and {Messineo}, R. and {Mowlavi}, N. and {Nicolas}, C. and {Nienartowicz}, K. and {Pailler}, F. and {Panuzzo}, P. and {Riclet}, F. and {Roux}, W. and {Seabroke}, G.~M. and {Sordo}, R. and {Tanga}, P. and {Th{\'e}venin}, F. and {Gracia-Abril}, G. and {Portell}, J. and {Teyssier}, D. and {Altmann}, M. and {Andrae}, R. and {Bellas-Velidis}, I. and {Benson}, K. and {Berthier}, J. and {Blomme}, R. and {Brugaletta}, E. and {Burgess}, P.~W. and {Busso}, G. and {Carry}, B. and {Cellino}, A. and {Cheek}, N. and {Clementini}, G. and {Damerdji}, Y. and {Davidson}, M. and {Delchambre}, L. and {Dell'Oro}, A. and {Fern{\'a}ndez-Hern{\'a}ndez}, J. and {Galluccio}, L. and {Garc{\'\i}a-Lario}, P. and {Garcia-Reinaldos}, M. and {Gonz{\'a}lez-N{\'u}{\~n}ez}, J. and {Gosset}, E. and {Haigron}, R. and {Halbwachs}, J.-L. and {Hambly}, N.~C. and {Harrison}, D.~L. and {Hatzidimitriou}, D. and {Heiter}, U. and {Hern{\'a}ndez}, J. and {Hestroffer}, D. and {Hodgkin}, S.~T. and {Holl}, B. and {Jan{\ss}en}, K. and {Jevardat de Fombelle}, G. and {Jordan}, S. and {Krone-Martins}, A. and {Lanzafame}, A.~C. and {L{\"o}ffler}, W. and {Lorca}, A. and {Manteiga}, M. and {Marchal}, O. and {Marrese}, P.~M. and {Moitinho}, A. and {Mora}, A. and {Muinonen}, K. and {Osborne}, P. and {Pancino}, E. and {Pauwels}, T. and {Petit}, J.-M. and {Recio-Blanco}, A. and {Richards}, P.~J. and {Riello}, M. and {Rimoldini}, L. and {Robin}, A.~C. and {Roegiers}, T. and {Rybizki}, J. and {Sarro}, L.~M. and {Siopis}, C. and {Smith}, M. and {Sozzetti}, A. and {Ulla}, A. and {Utrilla}, E. and {van Leeuwen}, M. and {van Reeven}, W. and {Abbas}, U. and {Abreu Aramburu}, A. and {Accart}, S. and {Aerts}, C. and {Aguado}, J.~J. and {Ajaj}, M. and {Altavilla}, G. and {{\'A}lvarez}, M.~A. and {{\'A}lvarez Cid-Fuentes}, J. and {Alves}, J. and {Anderson}, R.~I. and {Anglada Varela}, E. and {Antoja}, T. and {Audard}, M. and {Baines}, D. and {Baker}, S.~G. and {Balaguer-N{\'u}{\~n}ez}, L. and {Balbinot}, E. and {Balog}, Z. and {Barache}, C. and {Barbato}, D. and {Barros}, M. and {Barstow}, M.~A. and {Bartolom{\'e}}, S. and {Bassilana}, J.-L. and {Bauchet}, N. and {Baudesson-Stella}, A. and {Becciani}, U. and {Bellazzini}, M. and {Bernet}, M. and {Bertone}, S. and {Bianchi}, L. and {Blanco-Cuaresma}, S. and {Boch}, T. and {Bombrun}, A. and {Bossini}, D. and {Bouquillon}, S. and {Bragaglia}, A. and {Bramante}, L. and {Breedt}, E. and {Bressan}, A. and {Brouillet}, N. and {Bucciarelli}, B. and {Burlacu}, A. and {Busonero}, D. and {Butkevich}, A.~G. and {Buzzi}, R. and {Caffau}, E. and {Cancelliere}, R. and {C{\'a}novas}, H. and {Cantat-Gaudin}, T. and {Carballo}, R. and {Carlucci}, T. and {Carnerero}, M.~I. and {Carrasco}, J.~M. and {Casamiquela}, L. and {Castellani}, M. and {Castro-Ginard}, A. and {Castro Sampol}, P. and {Chaoul}, L. and {Charlot}, P. and {Chemin}, L. and {Chiavassa}, A. and {Cioni}, M.-R.~L. and {Comoretto}, G. and {Cooper}, W.~J. and {Cornez}, T. and {Cowell}, S. and {Crifo}, F. and {Crosta}, M. and {Crowley}, C. and {Dafonte}, C. and {Dapergolas}, A. and {David}, M. and {David}, P.},
        title = "{Gaia Early Data Release 3. Summary of the contents and survey properties}",
      journal = {\aap},
         year = 2021,
        month = may,
       volume = {649},
          eid = {A1},
        pages = {A1},
          doi = {10.1051/0004-6361/202039657},
archivePrefix = {arXiv},
       eprint = {2012.01533},
 primaryClass = {astro-ph.GA},
       adsurl = {https://ui.adsabs.harvard.edu/abs/2021A&A...649A...1G}
}

@ARTICLE{Martin2005GALEX,
       author = {{Martin}, D. Christopher and {Fanson}, James and {Schiminovich}, David and {Morrissey}, Patrick and {Friedman}, Peter G. and {Barlow}, Tom A. and {Conrow}, Tim and {Grange}, Robert and {Jelinsky}, Patrick N. and {Milliard}, Bruno and {Siegmund}, Oswald H.~W. and {Bianchi}, Luciana and {Byun}, Yong-Ik and {Donas}, Jose and {Forster}, Karl and {Heckman}, Timothy M. and {Lee}, Young-Wook and {Madore}, Barry F. and {Malina}, Roger F. and {Neff}, Susan G. and {Rich}, R. Michael and {Small}, Todd and {Surber}, Frank and {Szalay}, Alex S. and {Welsh}, Barry and {Wyder}, Ted K.},
        title = "{The Galaxy Evolution Explorer: A Space Ultraviolet Survey Mission}",
      journal = {\apjl},
         year = 2005,
        month = jan,
       volume = {619},
       number = {1},
        pages = {L1-L6},
          doi = {10.1086/426387},
archivePrefix = {arXiv},
       eprint = {astro-ph/0411302},
 primaryClass = {astro-ph},
       adsurl = {https://ui.adsabs.harvard.edu/abs/2005ApJ...619L...1M}
}

@ARTICLE{Alam2015sdss,
       author = {{Alam}, Shadab and {Albareti}, Franco D. and {Allende Prieto}, Carlos and {Anders}, F. and {Anderson}, Scott F. and {Anderton}, Timothy and {Andrews}, Brett H. and {Armengaud}, Eric and {Aubourg}, {\'E}ric and {Bailey}, Stephen and {Basu}, Sarbani and {Bautista}, Julian E. and {Beaton}, Rachael L. and {Beers}, Timothy C. and {Bender}, Chad F. and {Berlind}, Andreas A. and {Beutler}, Florian and {Bhardwaj}, Vaishali and {Bird}, Jonathan C. and {Bizyaev}, Dmitry and {Blake}, Cullen H. and {Blanton}, Michael R. and {Blomqvist}, Michael and {Bochanski}, John J. and {Bolton}, Adam S. and {Bovy}, Jo and {Shelden Bradley}, A. and {Brandt}, W.~N. and {Brauer}, D.~E. and {Brinkmann}, J. and {Brown}, Peter J. and {Brownstein}, Joel R. and {Burden}, Angela and {Burtin}, Etienne and {Busca}, Nicol{\'a}s G. and {Cai}, Zheng and {Capozzi}, Diego and {Carnero Rosell}, Aurelio and {Carr}, Michael A. and {Carrera}, Ricardo and {Chambers}, K.~C. and {Chaplin}, William James and {Chen}, Yen-Chi and {Chiappini}, Cristina and {Chojnowski}, S. Drew and {Chuang}, Chia-Hsun and {Clerc}, Nicolas and {Comparat}, Johan and {Covey}, Kevin and {Croft}, Rupert A.~C. and {Cuesta}, Antonio J. and {Cunha}, Katia and {da Costa}, Luiz N. and {Da Rio}, Nicola and {Davenport}, James R.~A. and {Dawson}, Kyle S. and {De Lee}, Nathan and {Delubac}, Timoth{\'e}e and {Deshpande}, Rohit and {Dhital}, Saurav and {Dutra-Ferreira}, Let{\'\i}cia and {Dwelly}, Tom and {Ealet}, Anne and {Ebelke}, Garrett L. and {Edmondson}, Edward M. and {Eisenstein}, Daniel J. and {Ellsworth}, Tristan and {Elsworth}, Yvonne and {Epstein}, Courtney R. and {Eracleous}, Michael and {Escoffier}, Stephanie and {Esposito}, Massimiliano and {Evans}, Michael L. and {Fan}, Xiaohui and {Fern{\'a}ndez-Alvar}, Emma and {Feuillet}, Diane and {Filiz Ak}, Nurten and {Finley}, Hayley and {Finoguenov}, Alexis and {Flaherty}, Kevin and {Fleming}, Scott W. and {Font-Ribera}, Andreu and {Foster}, Jonathan and {Frinchaboy}, Peter M. and {Galbraith-Frew}, J.~G. and {Garc{\'\i}a}, Rafael A. and {Garc{\'\i}a-Hern{\'a}ndez}, D.~A. and {Garc{\'\i}a P{\'e}rez}, Ana E. and {Gaulme}, Patrick and {Ge}, Jian and {G{\'e}nova-Santos}, R. and {Georgakakis}, A. and {Ghezzi}, Luan and {Gillespie}, Bruce A. and {Girardi}, L{\'e}o and {Goddard}, Daniel and {Gontcho}, Satya Gontcho A. and {Gonz{\'a}lez Hern{\'a}ndez}, Jonay I. and {Grebel}, Eva K. and {Green}, Paul J. and {Grieb}, Jan Niklas and {Grieves}, Nolan and {Gunn}, James E. and {Guo}, Hong and {Harding}, Paul and {Hasselquist}, Sten and {Hawley}, Suzanne L. and {Hayden}, Michael and {Hearty}, Fred R. and {Hekker}, Saskia and {Ho}, Shirley and {Hogg}, David W. and {Holley-Bockelmann}, Kelly and {Holtzman}, Jon A. and {Honscheid}, Klaus and {Huber}, Daniel and {Huehnerhoff}, Joseph and {Ivans}, Inese I. and {Jiang}, Linhua and {Johnson}, Jennifer A. and {Kinemuchi}, Karen and {Kirkby}, David and {Kitaura}, Francisco and {Klaene}, Mark A. and {Knapp}, Gillian R. and {Kneib}, Jean-Paul and {Koenig}, Xavier P. and {Lam}, Charles R. and {Lan}, Ting-Wen and {Lang}, Dustin and {Laurent}, Pierre and {Le Goff}, Jean-Marc and {Leauthaud}, Alexie and {Lee}, Khee-Gan and {Lee}, Young Sun and {Licquia}, Timothy C. and {Liu}, Jian and {Long}, Daniel C. and {L{\'o}pez-Corredoira}, Mart{\'\i}n and {Lorenzo-Oliveira}, Diego and {Lucatello}, Sara and {Lundgren}, Britt and {Lupton}, Robert H. and {Mack}, III, Claude E. and {Mahadevan}, Suvrath and {Maia}, Marcio A.~G. and {Majewski}, Steven R. and {Malanushenko}, Elena and {Malanushenko}, Viktor and {Manchado}, A. and {Manera}, Marc and {Mao}, Qingqing and {Maraston}, Claudia and {Marchwinski}, Robert C. and {Margala}, Daniel and {Martell}, Sarah L. and {Martig}, Marie and {Masters}, Karen L. and {Mathur}, Savita and {McBride}, Cameron K. and {McGehee}, Peregrine M. and {McGreer}, Ian D. and {McMahon}, Richard G. and {M{\'e}nard}, Brice and {Menzel}, Marie-Luise and {Merloni}, Andrea and {M{\'e}sz{\'a}ros}, Szabolcs and {Miller}, Adam A. and {Miralda-Escud{\'e}}, Jordi and {Miyatake}, Hironao and {Montero-Dorta}, Antonio D. and {More}, Surhud and {Morganson}, Eric and {Morice-Atkinson}, Xan and {Morrison}, Heather L. and {Mosser}, Ben{\^o}it and {Muna}, Demitri and {Myers}, Adam D. and {Nandra}, Kirpal and {Newman}, Jeffrey A. and {Neyrinck}, Mark and {Nguyen}, Duy Cuong and {Nichol}, Robert C. and {Nidever}, David L. and {Noterdaeme}, Pasquier and {Nuza}, Sebasti{\'a}n E. and {O'Connell}, Julia E. and {O'Connell}, Robert W. and {O'Connell}, Ross and {Ogando}, Ricardo L.~C. and {Olmstead}, Matthew D. and {Oravetz}, Audrey E. and {Oravetz}, Daniel J. and {Osumi}, Keisuke and {Owen}, Russell and {Padgett}, Deborah L. and {Padmanabhan}, Nikhil and {Paegert}, Martin and {Palanque-Delabrouille}, Nathalie and {Pan}, Kaike},
        title = "{The Eleventh and Twelfth Data Releases of the Sloan Digital Sky Survey: Final Data from SDSS-III}",
      journal = {\apjs},
         year = 2015,
        month = jul,
       volume = {219},
       number = {1},
          eid = {12},
        pages = {12},
          doi = {10.1088/0067-0049/219/1/12},
archivePrefix = {arXiv},
       eprint = {1501.00963},
 primaryClass = {astro-ph.IM},
       adsurl = {https://ui.adsabs.harvard.edu/abs/2015ApJS..219...12A}
}

@INPROCEEDINGS{Chambers2018Panstarrs,
       author = {{Chambers}, Kenneth and {Pan-STARRS Team}},
        title = "{The Pan-STARRS1 Surveys}",
    booktitle = {American Astronomical Society Meeting Abstracts \#231},
         year = 2018,
       series = {American Astronomical Society Meeting Abstracts},
       volume = {231},
        month = jan,
          eid = {102.01},
        pages = {102.01},
       adsurl = {https://ui.adsabs.harvard.edu/abs/2018AAS...23110201C}
}

@ARTICLE{Skrutskie20062MASS,
       author = {{Skrutskie}, M.~F. and {Cutri}, R.~M. and {Stiening}, R. and {Weinberg}, M.~D. and {Schneider}, S. and {Carpenter}, J.~M. and {Beichman}, C. and {Capps}, R. and {Chester}, T. and {Elias}, J. and {Huchra}, J. and {Liebert}, J. and {Lonsdale}, C. and {Monet}, D.~G. and {Price}, S. and {Seitzer}, P. and {Jarrett}, T. and {Kirkpatrick}, J.~D. and {Gizis}, J.~E. and {Howard}, E. and {Evans}, T. and {Fowler}, J. and {Fullmer}, L. and {Hurt}, R. and {Light}, R. and {Kopan}, E.~L. and {Marsh}, K.~A. and {McCallon}, H.~L. and {Tam}, R. and {Van Dyk}, S. and {Wheelock}, S.},
        title = "{The Two Micron All Sky Survey (2MASS)}",
      journal = {\aj},
         year = 2006,
        month = feb,
       volume = {131},
       number = {2},
        pages = {1163-1183},
          doi = {10.1086/498708},
       adsurl = {https://ui.adsabs.harvard.edu/abs/2006AJ....131.1163S}
}

@ARTICLE{Wright2010WISE,
       author = {{Wright}, Edward L. and {Eisenhardt}, Peter R.~M. and {Mainzer}, Amy K. and {Ressler}, Michael E. and {Cutri}, Roc M. and {Jarrett}, Thomas and {Kirkpatrick}, J. Davy and {Padgett}, Deborah and {McMillan}, Robert S. and {Skrutskie}, Michael and {Stanford}, S.~A. and {Cohen}, Martin and {Walker}, Russell G. and {Mather}, John C. and {Leisawitz}, David and {Gautier}, III, Thomas N. and {McLean}, Ian and {Benford}, Dominic and {Lonsdale}, Carol J. and {Blain}, Andrew and {Mendez}, Bryan and {Irace}, William R. and {Duval}, Valerie and {Liu}, Fengchuan and {Royer}, Don and {Heinrichsen}, Ingolf and {Howard}, Joan and {Shannon}, Mark and {Kendall}, Martha and {Walsh}, Amy L. and {Larsen}, Mark and {Cardon}, Joel G. and {Schick}, Scott and {Schwalm}, Mark and {Abid}, Mohamed and {Fabinsky}, Beth and {Naes}, Larry and {Tsai}, Chao-Wei},
        title = "{The Wide-field Infrared Survey Explorer (WISE): Mission Description and Initial On-orbit Performance}",
      journal = {\aj},
         year = 2010,
        month = dec,
       volume = {140},
       number = {6},
        pages = {1868-1881},
          doi = {10.1088/0004-6256/140/6/1868},
archivePrefix = {arXiv},
       eprint = {1008.0031},
 primaryClass = {astro-ph.IM},
       adsurl = {https://ui.adsabs.harvard.edu/abs/2010AJ....140.1868W}
}

@INPROCEEDINGS{Taylor2005TOPCAT,
       author = {{Taylor}, M.~B.},
        title = "{TOPCAT \& STIL: Starlink Table/VOTable Processing Software}",
    booktitle = {Astronomical Data Analysis Software and Systems XIV},
         year = 2005,
       editor = {{Shopbell}, P. and {Britton}, M. and {Ebert}, R.},
       series = {Astronomical Society of the Pacific Conference Series},
       volume = {347},
        month = dec,
        pages = {29},
       adsurl = {https://ui.adsabs.harvard.edu/abs/2005ASPC..347...29T}
}

@misc{ZhangRY2026taledustdisksmilky,
      title={A Tale of Two Dust Disks in Our Milky Way}, 
      author={Ruoyi Zhang and Haibo Yuan and Bingqiu Chen and Maosheng Xiang and Yang Huang and Xiaowei Liu and Jifeng Liu},
      year={2026},
      eprint={2601.02724},
      archivePrefix={arXiv},
      primaryClass={astro-ph.GA},
      url={https://arxiv.org/abs/2601.02724}, 
}

@ARTICLE{Eggleton1983RL,
       author = {{Eggleton}, P.~P.},
        title = "{Aproximations to the radii of Roche lobes.}",
      journal = {\apj},
         year = 1983,
        month = may,
       volume = {268},
        pages = {368-369},
          doi = {10.1086/160960},
       adsurl = {https://ui.adsabs.harvard.edu/abs/1983ApJ...268..368E}
}
\bibliographystyle{aasjournalv7}

\end{document}